\documentclass[12pt, letterpaper]{article}
\usepackage[english]{babel}
\usepackage[latin1]{inputenc}
\usepackage{graphicx}       % For å inkludere figurer
\usepackage{url}
\usepackage{hyperref}
\usepackage{amsmath,amssymb,amsfonts,bm}
\usepackage{relsize}
\usepackage{pstricks}
\usepackage{multirow}
\usepackage{rotating}
\usepackage{subfigure}
\usepackage{float}
\usepackage{fancyhdr}
\usepackage{ae}
\usepackage{natbib}
\usepackage{multicol}
\usepackage{enumerate}
\usepackage{wrapfig}
\usepackage[hang,flushmargin]{footmisc}
\usepackage{setspace}

\usepackage[explicit]{titlesec}
\usepackage[normalem]{ulem}
\titleformat{\section}
  {\large\bfseries\filcenter}{}{0em}{\thesection.\hspace*{ 0.5em}#1}
\titleformat{name=\section,numberless}{\large\bfseries\filcenter}{}{0em}
	{\MakeUppercase{#1}}

\usepackage{ccaption}
\captionnamefont{\bfseries}
\bibpunct{(}{)}{;}{a}{}{,}

\usepackage[margin=0.90in]{geometry}

\begin{document}
	
%\topmargin -22mm
	
%\numberwithin{equation}{section}
%\numberwithin{figure}{section}
%\numberwithin{table}{section}

%\hspace{1cm} \vspace{-0.cm}
{\begin{center}{\LARGE \textbf{Bayesian Detection of Changepoints in Finite-State Markov Chains for Multiple Sequences} \\}
	\vspace{1cm}
{\Large Petter \textsc{Arnesen}, Tracy \textsc{Holsclaw} and Padhraic \textsc{Smyth}	}
\let\thefootnote\relax\footnote{Petter Arnesen is a PhD student with the Department of Mathematical Sciences,
Norwegian University of Science and Technology, Trondheim 7491, Norway
(email: \textit{petterar@math.ntnu.no}).
Tracy Holsclaw is a Postdoctoral Scholar with the Department of Statistics, University of California, Irvine, CA (email: \textit{tholscla@ams.ucsc.edu}). Padhraic Smyth is a Professor with the
Department of Computer Science and the Department of Statistics, University of California, Irvine, CA (email: \textit{smyth@ics.uci.edu}).}
\end{center}}

We consider the analysis of sets of categorical sequences consisting
of piecewise homogeneous Markov segments.
The sequences are assumed to be governed by a common underlying process with segments occurring in the same order for each sequence. 
Segments are defined by a set of unobserved changepoints where the positions and number of changepoints can vary from
sequence to sequence. We propose a Bayesian framework for analyzing such
data, placing priors on the locations of the changepoints and on
the transition matrices and using Markov chain Monte Carlo (MCMC)
techniques to obtain posterior samples given the data. Experimental
results using simulated data illustrates how the methodology can be used for inference of posterior
distributions for parameters and changepoints, as well as the ability to
handle considerable variability in the locations of the changepoints
across different sequences. We also investigate the application of the
approach to sequential data from two 
applications involving monsoonal
rainfall patterns and branching patterns in trees. 

\vspace{0.5cm}

\vspace{-0.1cm}
\renewcommand{\baselinestretch}{1.9} \small\normalsize

\section{INTRODUCTION}

Finite-state Markov chains are widely used to model sequential data in applications such as
weather models \citep{gab62}, speech recognition \citep{rabiner1989}, bioinformatics \citep{durbin1998}, and more \citep{guttorp1995}. A common assumption is that the chain is homogeneous, often motivated by a desire to keep the number of model parameters tractable. In practice, however, inhomogeneity in various forms is often present. 

\begin{figure}
\vspace{-2cm}
  \centering
    \includegraphics[width=0.5\textwidth]{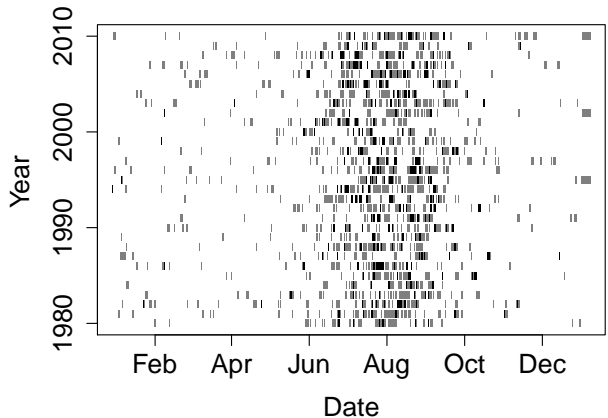}
\caption{31 years of rainfall categorized into no/light rainfall (white), medium rainfall (gray), and heavy rainfall (black).}
\label{fig:rainfallExample}
\end{figure}

In particular in this paper we investigate the problem of modeling sets of categorical-valued sequences where each sequence is assumed to be generated by an ordered set of segments, with unobserved segment boundaries that can vary from sequence to sequence. Each segment has its own Markov dynamics, representing common ``phases" for some underlying process. As a specific example, consider the modeling of rainfall at a particular location. Markov chains have a long history of use as stochastic models of rainfall  \citep{newnham1916persistence, gold1929note, cochran1938extension, wilks1999weather, chen2010daily}.
%{\color{red} Markov chains are well known to be a useful starting point as stochastic models of daily rainfall occurrence  
%\citep{newnham1916persistence, gold1929note, cochran1938extension} and a variety of methods based on Markov chains are in wide use as %precipitation models  in climate science and hydrology \citep{wilks1999weather, chen2010daily}.} 
Figure \ref{fig:rainfallExample} is an illustration of annual daily rainfall sequences for a weather station in Northern India.
% (discussed in more detail later in the paper).
The appearance of the Indian monsoon in early summer causes a visible
phase-change in the sequence of daily rainfall for each year,
motivating the type of segment-based model that is the focus of this
paper. 
%Figure \ref{fig:rainfallExample} is suggestive of three phases or
%segments, corresponding to pre-monsoon, monsoon, and post-monsoon,
%with changepoints corresponding to monsoon onset and withdrawal, and
%with a significant degree of interannual variability in the locations
%of the changepoints. 
Being able to infer monsoon onset and withdrawal dates, and to effectively handle interannual variability, is of considerable interest in agriculture modeling and climate science (e.g., \cite{joseph1994}).
Other examples of data sets characterized by multiple sequences with common segmental structure include tree branching patterns 
\citep{guedon2001}, sequences of bird songs \citep{craig1943,raftery1994}, wind-speed data 
\citep{berchtold1999}, DNA sequences  \citep{berchtold2002,fearnhead2007}, and dialog transcripts
\citep{levin2000}. 

In this paper we model a particular type of shared structure where
a categorical sequence is assumed to consist of $K+1$ {\it segments} separated by
$K$ unobserved {\it changepoints}. Within each segment the observed
data are assumed to be generated by a homogeneous finite-state Markov
chain. We will assume that each of the $K+1$ segments has its own Markov
parameters, resulting in $K+1$ transition matrices. The location of each
changepoint is modeled by a discrete distribution conditioned on the
location of the previous changepoint. We will consider the case of a
set of $L$ observed sequences (of potentially varying length) where
the locations of the $K$ changepoints can vary across the 
sequences.  The $L$ sequences share information in that the parameters
of the Markov transition matrices are global, and the order in which
the segments appear in each sequence are assumed to be the
same, corresponding to common underlying phases.
Our framework allows any segment to be skipped in a particular
sequence, and thus, $K$ is in effect the maximal number of changepoints that can occur in any
particular sequence.
%relaxing the assumption that all sequences
%must have the same number of changepoints, while continuing to draw
%strength across all sequences.

There is a large amount of
%There is a large  
prior literature on changepoint detection for sequential data, covering a variety of aspects of this problem, including online/offline estimation, single versus multiple sequences, Bayesian versus non-Bayesian inference, and methods for inferring the number of changepoints. Much of this prior work has focused on the case of independent observations and/or non-categorical data. An exception is the double-Markov chain approach proposed by \cite{berchtold1999} using the Expectation-Maximization (EM) algorithm for inference, and which was further developed within a Bayesian framework by \cite{fitzpatrick2012}. Because of the Markov assumption at the segment-transition level, the model imposes an implicit geometric distribution on segment lengths which is not always ideal in practice (we will followup on this point later in the paper). The geometric assumption can be relaxed by using a hidden semi-Markov approach for example \citep{guedon2003}, but this imposes computational restrictions, namely that the time complexity of inference (e.g., via the forward-backward algorithm, \cite{rabiner1989}) can scale as $O(T^2)$ 
where $T$ is the length of a sequence.

\cite{polansky2007} proposed a likelihood-based framework for segmentation assuming a process that switches between different Markov chains with unknown parameters, but for the case of a single sequence rather than multiple
  sequences. A restricted variant of
  the multiple sequence changepoint problem is the case where
  sequences are required to be of the same length and the changepoints
  are assumed to occur at the same location in each sequence, which can
   also be viewed as a single multivariate sequence
  \citep{lai2011, xing2012, fitzpatrick2012}.
\cite{zhang2012} relax this approach to allow subsets of sequences
 (of equal length and for the case of independent real-valued
 observations) to share changepoints. In the approach proposed here we allow each sequence to have its own changepoint locations and sequences to have different lengths. 
 
While most prior work has focused on likelihood-based inference and point estimates of changepoints, here we use full
  Bayesian inference for both changepoint locations and model
  parameters.
  %providing useful information in terms of
  %posterior uncertainty about changepoint locations (for example). 
  Our MCMC inference algorithm
  has a time complexity per iteration that is linear in the
  length $T$ of a sequence, avoiding the
  $O(T^2)$ time complexity incurred by  
   \cite{rigaill2012}.
 \cite{fearnhead2011} developed a Bayesian changepoint detection that is linear in $T$ in terms of time complexity, but for the case of a single real-valued sequence.

 Our focus on multiple sequences also provides a natural context for using cross-validation
(across sequences) for model selection, providing a practical alternative to  approaches such as the BIC criterion \citep{guedon2003,polansky2007,fitzpatrick2012,luong2013}, and related penalized likelihood approaches \citep{zhang2007,rigaill2012,cleynen2013}, which are not always appropriate or effective for changepoint problems (e.g., see the discussion in  \cite{cleynen2014}).

%In summary, our paper focuses on Bayesian changepoint estimation for multiple categorical sequences  where the sequences are assumed to %share common dynamics but can have different lengths and have changepoints in different locations.  
%
%
%In summary, the results presented here differ from prior work  by (a) using non-geometric duration distributions 
%%(in contrast to the double-Markov chain approach), 
%(b) allowing for Markov dependence within segments 
%%(unlike for instance \cite{guedon2003} where independence is assumed),  
%(c) developing a fully Bayesian inference procedure conditioned on a fixed number of changepoints, and (d) employing an MCMC inference scheme that scales linearly (rather than quadratically) in sequence length $T$.

In Section \ref{sec:model} we introduce our proposed changepoint Markov model, 
% Section \ref{sec:algorithm} describes in detail the MCMC algorithm for inferring changepoints and parameters, including a discussion of how missing data can be handled,
with results on model selection in Section \ref{sec:modelEvaluation}.
% describes in detail our model selection precedurein Section
%\ref{sec:modelEvaluation}. 
In Section \ref{sec:ex1} we present
an example of simulated data to illustrate different aspects of the
proposed approach. Section \ref{sec:apple} describes the application of our approach to a data set of annual branching of apple trees, and in 
Section \ref{sec:rain} we discuss the application of the model 
to a data set consisting of multiple years of daily rainfall in India. We conclude the paper in Section \ref{sec:discussion} with a brief discussion of the main results as well as suggestions for future directions. Discussions on MCMC sampling, how to handle missing data and additional simulated examples
%, and an application to a second real data set 
are provided in the supplementary materials.

\section{A CHANGEPOINT MARKOV MODEL FOR CATEGORICAL SEQUENCES}\label{sec:model}

In this section we  define our Bayesian model for detection of a fixed number of changepoints in multiple sequences. The MCMC sampling algorithm we construct for sampling from the resulting posterior distribution is given in Section S.1 in the supplementary materials. 

\subsection{\sc Homogeneous Markov Chains}
Given a sequence of discrete observations $\bold y=(y_1,...,y_T)$,
such that $y_j \in \{1,...,n\}$ for all $j=1,...,T$, $\bold y$ is a
realization from a finite-state Markov chain if the joint probability
of $\bold y$ can be written as
\begin{equation}
p(\bold y)=p_1(y_1)\prod_{j=2}^{T}p_j(y_j|y_{j-1},...,y_1) = \prod_{j=1}^{T}p_j(y_j|y_{j-1}),
\end{equation}
where the conditional distribution of $y_j$ given $y_{j-1},...,y_{1}$ only depends on the previous state $y_{j-1}$, and we assume some given initial state $y_0$ such that $p_1(y_1)=p_1(y_1|y_0)$.
If the transition distribution $p_j(y_j|y_{j-1})=p(y_j|y_{j-1})$ for
all $j=1,...,T$, we say that the Markov chain is homogeneous. The
transition probabilities can be organized into an $n\times n$
transition matrix $\bold Q$, where the rows indicate the value $y_{j-1}$ and the columns represent the values of $y_j$. We relax these assumptions below to allow the Markov chain to be {\it piecewise homogeneous}, allowing inhomogeneity via local segmentation of the sequence $\bm y$.

\subsection{\sc Modeling Local Segments}
 To define a piecewise homogeneous Markov chain, we divide the $T$
 observations into $K+1$ segments by introducing $K$ changepoints $
 \tau_0 = 0 \leq \tau_1 \leq ... \leq \tau_{K} \leq T$, and write
 $\bm{\tau}=(\tau_1,...,\tau_{K})$ as the vector of changepoints. If $\tau_{i-1}<\tau_{i}$ then we denote $s_i=\{\tau_{i-1}+1,...,\tau_{i}\}$ to be the $i^{th}$ segment and if $\tau_{i-1}=\tau_{i}$ (a segment of length zero) then $s_i=\emptyset$. Introducing the notation $\bm{y}_s=(y_j|j\in s)$ for $s\subseteq \{1,...,T\}$, the $i^{th}$ segment will include the data points $\bm{y}_{s_i}$. We assume the locations of the $K$ changepoints to be generated by a distribution of the form
\begin{equation}
p(\bm \tau|T,\bar{ \bm \theta})=p(\tau_1,...,\tau_{K}|T,\bar{ \bm\theta})=\prod_{i=1}^{K} p(\tau_i|\tau_{i-1},T,\bm \theta_{i}),
\end{equation}
where $p(\tau_i|\tau_{i-1},T,\bm \theta_i)$ is a discrete parametric distribution for changepoint $\tau_i$, with parameter vector $\bm \theta_i$, and where we let $\bar{\bm \theta}$ denote the collection of the parameter vectors $\bm \theta_i$ for $i=1,...,K$. In the remainder of the paper we assume the distribution for $\tau_i$ to be the negative binomial distribution truncated to the interval $(\tau_{i-1},T)$ (additional details in Section \ref{sec:multipeSeq}). The two-parameter negative binomial distribution provides additional flexibility in modeling segment lengths compared to a single parameter distribution such as the geometric distribution. This can lead to more accurate detection of changepoints, as we will see later in the paper.

\subsection{\sc The Piecewise Homogeneous Markov Chain}
We assume that within each segment the sequential observations in $\bm y$
evolve according to a fixed transition matrix, i.e., $\bm y$ is
generated by a piecewise homogeneous Markov chain within each
segment. With $K$ changepoints, we have a total of $K+1$
transition matrices $\bold Q^{(1)},...,\bold Q^{(K+1)}$, each having
size $n\times n$. In what follows we present the case where the $K+1$
transition matrices are modeled separately and independently, but it
is straightforward to constrain some of these matrices to be the same
(as in our rainfall example later in the paper) and to reduce the
parameter count accordingly. Let $\bar{\mathbf{Q}}$ denote the
collection of these $K+1$ transition matrices, and let $\bold Q^{(i)}_{k,l}$ denote the element at the $k^{th}$ row and the $l^{th}$ column of matrix $i$. Given an initial state $y_0$, the data likelihood is
\begin{equation}
p(\bold y|\bm
\tau,\bar{\mathbf{Q}},y_0)=\prod_{j=1}^Tp(y_j|y_{j-1},\bar{\mathbf{Q}},\bm
\tau)=\prod_{i=1}^{K+1}\prod_{j\in
  s_i}\bold Q_{y_{j-1},y_j}^{(i)}.
\label{eq:dataLikelihood}
\end{equation}
Assume for the moment that no segments are of length zero, i.e., no segments are skipped. If $y_j$ is
the first data point in segment $s_i$, $i>1$, then we assume it to be
distributed according to the conditional distribution of $y_j$ given
$y_{j-1}$ (the last point in segment $s_{i-1}$) using the transition
matrix for segment $s_i$, $\bold Q^{(i)}$. The extension to segments of length zero is straightforward.

We adopt a fully Bayesian approach conditioned on fixed $K$ for our
model. For priors for the transition matrices we assume each of the
rows to be independently distributed according to the Dirichlet
distribution. In particular, $\bold Q^{(i)}_{k,\cdot}\sim
Dir(\bm \alpha_{k}^{(i)})$, where $\bold Q^{(i)}_{k,\cdot}$ is the
$k^{th}$ row of the $i^{th}$ transition matrix and where $Dir(\bm
\alpha_k^{(i)})$ denotes the Dirichlet distribution with parameter
vector $\bm \alpha_k^{(i)}$ of appropriate length. We set all
elements in $\bm\alpha_k^{(i)}$ to be equal to 1,  for all $i$ and $k$ in our examples, 
rendering the prior equivalent to having seen   one transition from each category to every other category including itself.

\subsection{\sc Modeling Multiple Sequences of Variable Length}\label{sec:multipeSeq}
To handle multiple sequences, consider $L$ conditionally independent
sequences of observations, $\bm y^{(l)}$, $l=1,...,L$, with lengths
$T_1,\ldots, T_L$,  where each sequence consists of $K+1$ segments
occurring in the same order as described above. Also let $y_0^{(l)}$
denote the initial value for each of the $l$ sequences. Assuming the sequences
to be conditionally independent,  the likelihood for multiple
sequences is the product of the likelihoods  for each individual
sequence  (as defined earlier). Segments are allowed to be of zero
length (i.e., skipped), effectively allowing the number of
changepoints per sequence to differ (see the simulation studies
in Section S.3.1 and S.3.2 in the supplementary material for examples of this property). 
Let $\bm
\tau^{(l)}=(\tau_1^{(l)},...,\tau_{K}^{(l)})$ denote the
changepoints in sequence $l$.  There are a number of options for
modeling how the locations of the changepoints $\bm \tau^{(l)}$ are
related to the lengths of the sequences $T_l$. One could allow the
changepoints to have distributions defined in absolute units (e.g., of
time) and treat the total length of the sequence as a random
quantity. The approach we take here is to assume that the distribution
on changepoints (or equivalently, on segment length) is specified in
terms of  position {\it relative} to the total length of the sequence,
where we treat the observed total sequence lengths $T_l$ as fixed
quantities and condition on them. In practice the choice of
parametrization will depend on the specific nature of the application,
and for the special case where the sequences are all of the same
length, the absolute and relative approaches will be equivalent. In
particular we assume that the position of the changepoints
$\tau_i^{(l)}$ have a negative binomial distribution, with  parameters
$\bm \theta_i=(r_i,b_i) \in (0,1)\times (0,1)$, truncated to the range $(\tau_{i-1}^{(l)},T_l)$. We write
\begin{equation}
p(\tau_i^{(l)}|\tau_{i-1}^{(l)},T_l,\bm{\theta_i})\propto\frac{\Gamma(\tau_i^{(l)}+\gamma(\bm
  \theta_i,T_l))}{\tau_{i}^{(l)}! \ \Gamma(\gamma(\bm
  \theta_i,T_l))}b_i^{\gamma(\bm \theta_i,T_l)}(1-b_i)^{\tau_i^{(l)}},\ \tau_i^{(l)}=\tau_{i-1}^{(l)},...,T_l,
\end{equation}
where $\gamma(\bm \theta_i,T_l)$ is defined via the expression
$r_iT_l=\gamma(\bm \theta_i,T_l) (1-b_i)/b_i$, which corresponds to
the expected value of the negative binomial distribution without
truncation. The parameter $r_i$ will therefore be related to the
expected position of changepoint $\tau_i^{(l)}$ scaled by the length
$T_l$ of sequence $l$, while $b_i$ will be related to variance of the
distribution. This truncated distribution can be efficiently computed
as it does not include a computationally demanding normalizing
constant. We also assume apriori that the $\bm \theta_i$,
$i=1,...,K$ are independent, and  that $r_i$ and $b_i$ are
independent and uniformly distributed in the unit interval,
$U(0,1)$.

\section{MODEL SELECTION AND COMPARISON} \label{sec:modelEvaluation}

Previous work on estimating the number of changepoints has often used  formulations  such as BIC \citep{guedon2003,polansky2007,fitzpatrick2012,luong2013}. The BIC criterion, however, is not directly applicable to changepoint problems (as discussed in \cite{zhang2007}), leading to alternative penalized likelihood formulations, see for instance \cite{zhang2007} \cite{cleynen2013} and \cite{cleynen2014}. More fully Bayesian approaches have also been pursued (for example, by \cite{fearnhead2011} and \cite{rigaill2012}) but for single-sequences with real or count-valued IID observations. 

For multiple sequences,  \cite{xing2012} investigate a Bayesian approach for inferring boundaries for piecewise homogeneous Markov chains with unobserved changepoints, similar to the problem we address in this paper. However, their approach  assumes that all sequences are the same length and have changepoints in the same position, essentially restricting the approach to the case of a single sequence with a multivariate distribution.  For IID observations \cite{zhang2012} also consider the scenario of multiple equi-length sequences with aligned changepoints across sequences, using a modified BIC criterion for model selection. 

An alternative approach that could be pursued is that of  transdimensional MCMC algorithms based on a joint posterior distribution for the model and parameter space \citep{green1995}. Reversible jump MCMC (RJMCMC) algorithms \citep{green1995} could be used in this context, but we would anticipate slow convergence and potential sensitivity issues when choosing proposal distributions (particularly for the jump proposals).  Another option is the Dirichlet process framework, although it would not be straightforward to apply this approach given the computational challenges that would result from the lack of conjugacy \citep{neal2000}.

Given these various issues with Bayesian and information-based
criteria (at least in the context of our proposed model), 
we instead use cross-validation with log-probability scores (e.g., see
\citet{vehtari2002, gneiting2007strictly, gelman2013, li2014}) 
for both choosing the number of changepoints and for  model selection
among alternative frameworks. 
We take advantage of the fact that we are working with multiple sequences and use cross-validation at the sequence level, using the log-probability of held-out sequences as our scoring function. Our simulation results (next section) suggest that this approach is feasible even with a relatively small number of sequences (e.g., 10).
When training our
model, we hold out $t$ of our $L$ sequences, referred to as the test
set, and train our model using only $L-t$ of the available sequences, referred to as the training set. Denote the $L-t$ sequences in a
training set by $\bm Y_D$ and the $t$ sequences in the test set by
$\bm Y_{-D}$. To evaluate the quality of the model, we estimate the
logarithm of the probability of observing the test set $\bm Y_{-D}$
given the training set $\bm Y_D$ and the model. We average this log-probability for a number of different train/test set pairs using
cross-validation. The Monte Carlo estimate of the log-probability of
one of the sequences $\bm y^{(l)}$ in the test set is
\begin{equation}
\ln p(\bm y^{(l)}|\bm
Y_{D})=\frac{1}{N}\sum_{i=1}^{N}\frac{1}{M}\sum_{j=1}^{M}\ln p(\bm
y^{(l)}|\bm \tau^{(l)[j]},\bar{\mathbf{Q}}^{[i]},y_0^{(l)}),
\end{equation}
where $\bm \tau^{(l)[j]}\sim p(\bm \tau^{(l)}|T_l,\bar{\bm \theta}^{[i]})$ for
$j=1,...,M$, and $(\bar{\bm \theta}^{[i]},\bar{\mathbf{Q}}^{[i]})\sim p(
\bar{\bm \theta},\bar{\mathbf{Q}}|\bm Y_D)$ for $i=1,...,N$, and where
we use superscript $[\cdot]$ to denote simulated samples. After convergence of the
algorithm and for $N$ independent samples of $\bar{\bm \theta}$ and
$\bar{\mathbf{Q}}$, we simulate $M$ samples from $p(\bm
\tau^{(l)}|T_l,\bar{\bm \theta})$. For each of these $M$ samples, we calculate the value of the logarithm of the data likelihood in \eqref{eq:dataLikelihood}, and then compute the mean. 
For the results in this paper we used $M=1000$ samples, although
sensitivity analysis (not shown) indicates that $M=100$ is sufficient
to obtain consistent estimates of the log-probability score. For $N$,
which is the number of independent samples from the MCMC algorithm we
used $N=1000$ following the discussion in Section S.1 in the supplementary materials on
burn-in period and thinning. For all of our results we use 10-fold
cross-validation, i.e. we partition our data
into 10 test and training sets (unless otherwise stated). 

The total log-probability score is defined to be the sum of the individual scores divided by the sum of the length of the sequences
\begin{equation}
S(\bm Y_{-D}|\bm Y_{D})=\frac{\sum_{l=1}^t \ln p(\bm y^{(l)}|\bm Y_D)}{\sum_{l=1}^tT_l},
\end{equation}
where $\bm y^{(1)},...,\bm y^{(t)}$ are the sequences in the test
set. Using the equation above we can compare different models, either our
proposed model with different numbers of changepoints, or our proposed
model versus an alternative model. When calculating the cross-validated log-probability scores for two models $\mathcal{M}_1$
and $\mathcal{M}_2$ we use the same train/test sets and report the
difference between the scores for each train/test set, i.e., we
calculate $S(\bm Y_{-D}|\bm Y_D,\mathcal{M}_1)-S(\bm Y_{-D}|\bm
Y_D,\mathcal{M}_2)$ for each train/test set.

\section{ANALYSIS AND RESULTS}\label{analysis}
In this section, we analyze one simulated and two real data sets. For the simulated data we generate the sequences and changepoints, allowing us to test our model and compare to alternative methods. Additional simulated data sets are provided in the supplementary material to further explore different aspects of our model. The two real data sets consists of branching patterns for apple trees and daily rainfall over a region Northern India.
%We also analyze a real rainfall data set in this section, where we use our model detect the changepoints of monsoon onset and withdrawal. An additional real-world example, involving segmentation of branching patterns for trees,  is described in Section S.3.4 of the supplement.

%In this section, we first provide a simulated example where we
%generate the data. Additional simulated examples are given in the supplementary materials. This allows us to test our method and its ability
%to correctly place the changepoints, choose the number of
%changepoints, and estimate the model parameters. The simulated
%example give here is followed by the analysis of one real data set, monsoon rainfall. One additional real data set, branching of apple trees, is given in the supplementary materials in Section S.3.4. 

In all of our analyses we compare our method to three baseline models. The first baseline SI (Segmental-Independence) is the same as our proposed model but assumes the observations $\bm y$ are independent within each segment. The second baseline is a standard hidden Markov model (HMM) with a left-to-right transition matrix. The third baseline is the double hidden Markov chain model (dHMM) from \cite{berchtold1999} with a left-to-right transition matrix.
These three baselines differ from our proposed model in that the SI
and HMM models assume independence of the observations $\bm y$ (rather
than Markov dependence), and the HMM and dHMM models assume geometric
distributions on the changepoint locations (rather than a
negative-binomial distribution). For each of the three baselines we
use a Bayesian inference procedure similar to that described in
Section S.1 in the supplementary materials, and we report the results for the models
with the number of changepoints 
corresponding to the highest log-probability score (which for
all the baselines turned out to be the same number of changepoints as in our preferred model). In addition to comparing our model to these baseline models 
%presented in the paper
we also compare our results to those obtained with a model where each sequence is analyzed
independently. In particular, we use the single-sequence model of
\cite{fearnhead2006} for comparison, originally proposed for
real-valued data with independence. See the supplementary materials for these results.

\subsection{\sc Synthetic Data Example: Scenario 1}\label{sec:ex1}
%In this section we analyze data under a simulated scenario. 
%In the first scenario the  sequences all have the same number of changepoints and in the second scenario the sequences have different numbers of changepoints. We use these simulated data sets, where the ground truth is known, to test our model's accuracy and sensitivity in terms of parameter estimation,  accuracy of inferred changepoint locations, and  the number of subsamples required for cross-validation.
%\subsubsection{\sc Scenario 1: sequences with the same number of changepoints}\label{sec51}  
We simulated $L=10$ binary sequences, all of length $200$, from our
proposed  model. For each sequence $l=1,...,10$, we simulated one
changepoint $\tau^{(l)}_1$, and the position of the changepoint was
generated from the truncated negative binomial distribution as
explained in Section \ref{sec:multipeSeq}, using the parameters
$r_1=0.5$ and $b_1=0.8$, such that each segment will have length about
100. The transition matrices used to generate the binary data in each segment had diagonal entries $q_{1,i}$ and $q_{2,i}$, with $i$ denoting the segment. We used $q_{1,1}=q_{2,1}=0.8$ in the first segment, and $q_{1,2}=0.5$ and $q_{2,2}=0.4$ in the second segment.

\begin{figure}
\vspace{-1cm}
\begin{center}
\subfigure[]{ \label{fig:nrOfK_toy}\includegraphics[scale=0.4]{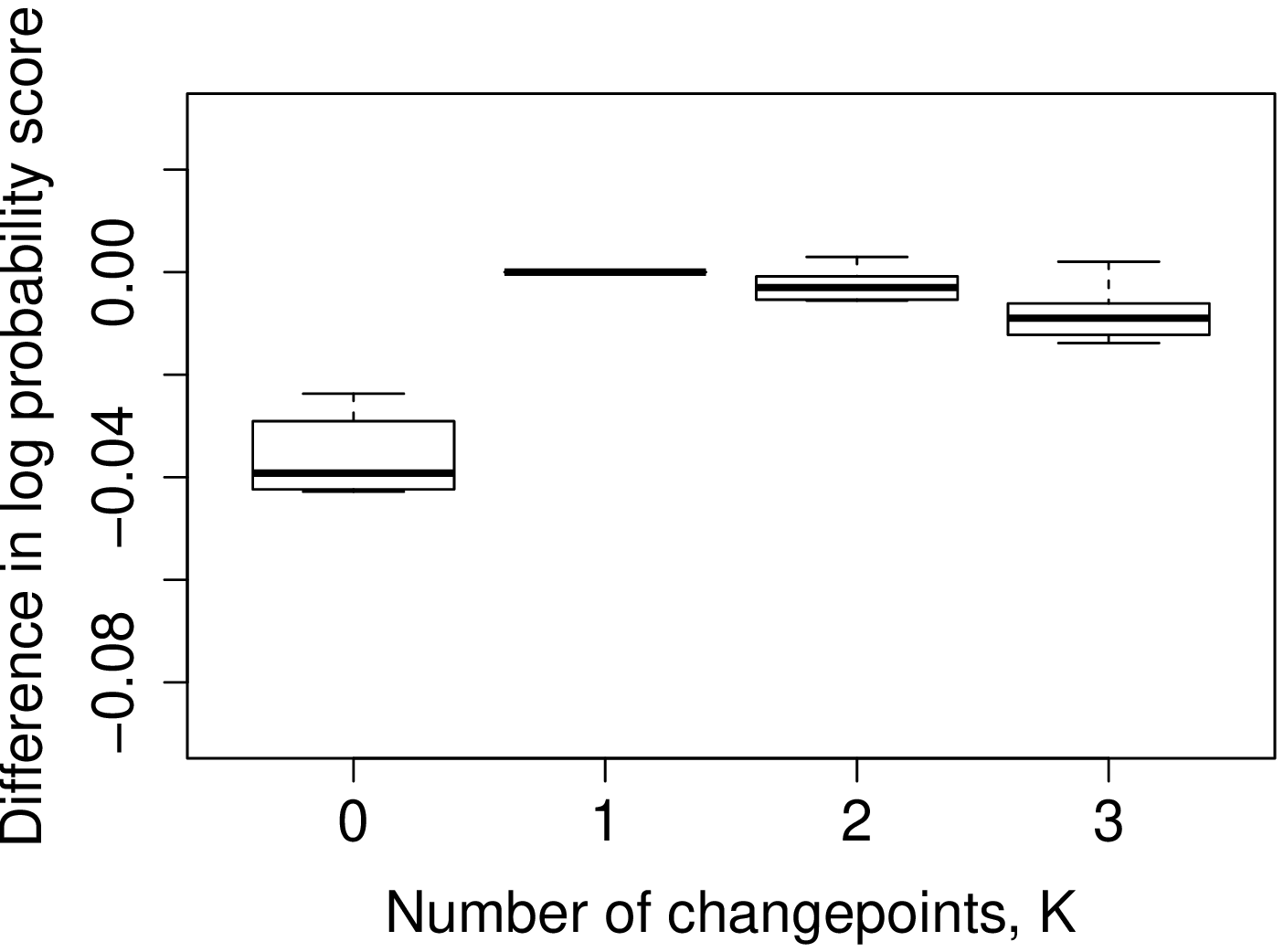}}
\subfigure[]{   \label{fig:baselines_toy}\includegraphics[scale=0.4]{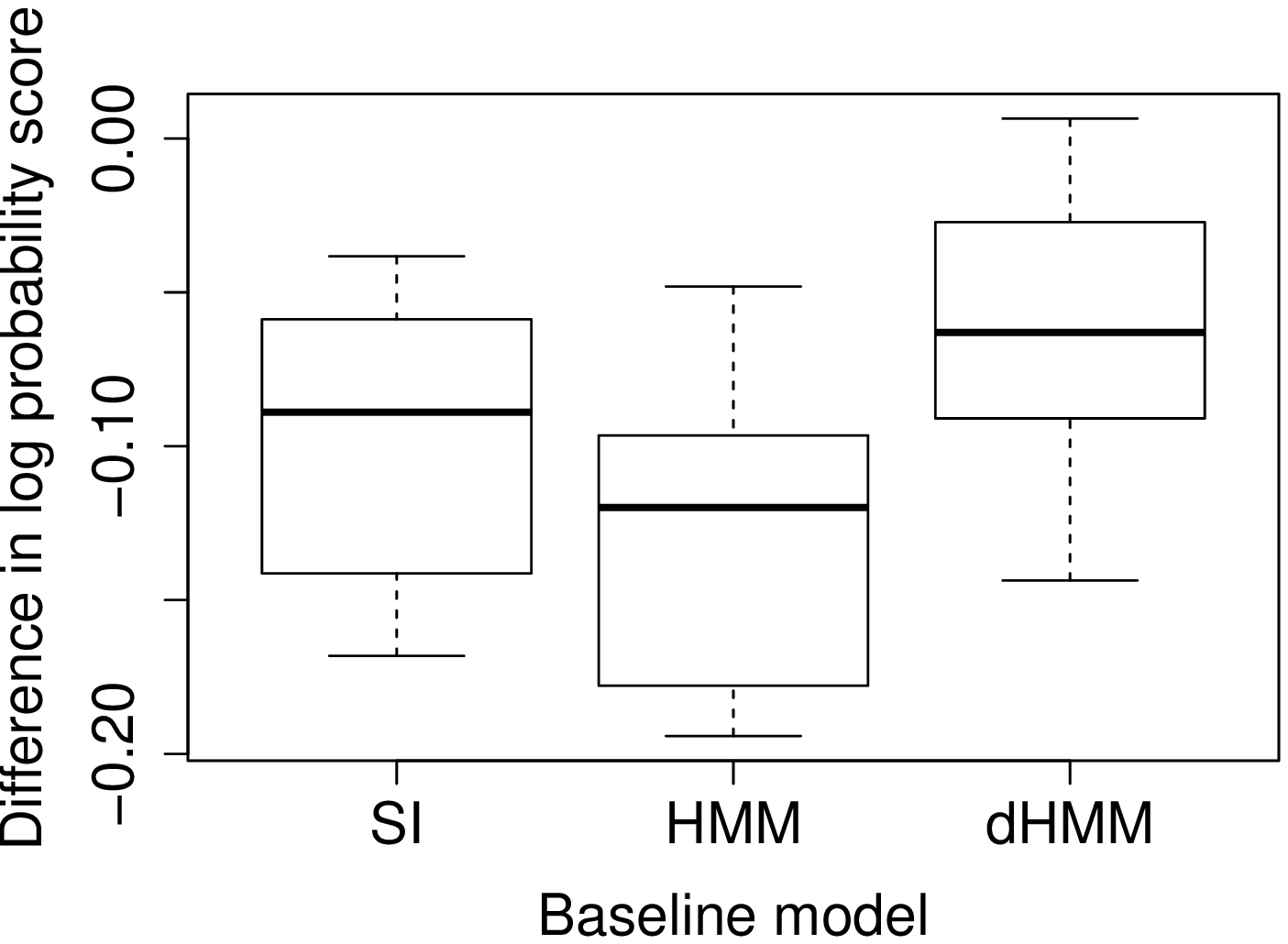}}
   \caption{Each boxplot shows the log-probability scores, across the
     validation sets, for a different model. The y-axis is defined as
     the log-probability score for (a) other numbers of changepoints
     and (b) other baseline models, minus the log-probability score
     for the model with one changepoint.}
\end{center}
\label{fig:boxPlot_toy}
\end{figure}

We used our cross-validation model selection procedure described in Section \ref{sec:modelEvaluation} to determine the number of changepoints. We defined 10 train/test sets by leaving out one of the 10 sequences in each fold, resulting in a ``hold-one-sequence-out'' cross-validation test. The result of this model comparison is shown in Figure \ref{fig:nrOfK_toy}.
It is clear from the figure that the model with a single changepoint
$(K=1)$ is the preferred one. As the number of changepoints increases
beyond a single changepoint the cross-validated log-probability scores become slightly worse. 

The cross-validated log-probability scores of the three
baseline models relative to our model are shown in Figure
\ref{fig:baselines_toy}, and as we can see our model performs the best. All models were
fit using the optimal changepoint number ($K=1$).

We then trained our model with the correct number of changepoints
($K=1$), using all 10 sequences. The estimated parameter values
with 95\% credible intervals (CI) (corresponding to the 2.5\% maximum
and minimum percentiles of the posterior samples after convergence)
are shown in Table S.1 %\ref{tab:est_toy1} 
and Table S.2 %\ref{tab:toy_Q_individual} 
in the supplementary materials. All of the
parameters are well estimated, although there is considerable
uncertainty concerning the estimated value of $b = b_1$. For a more
detailed discussion of the posterior analysis for $b$ and $r$ see the supplementary materials Section S.2.1. %\ref{supp:1}.

\begin{figure}
%\vspace{-1cm}
\begin{center}
\subfigure[]{\label{fig:sequences_toy}\includegraphics[scale=0.45]{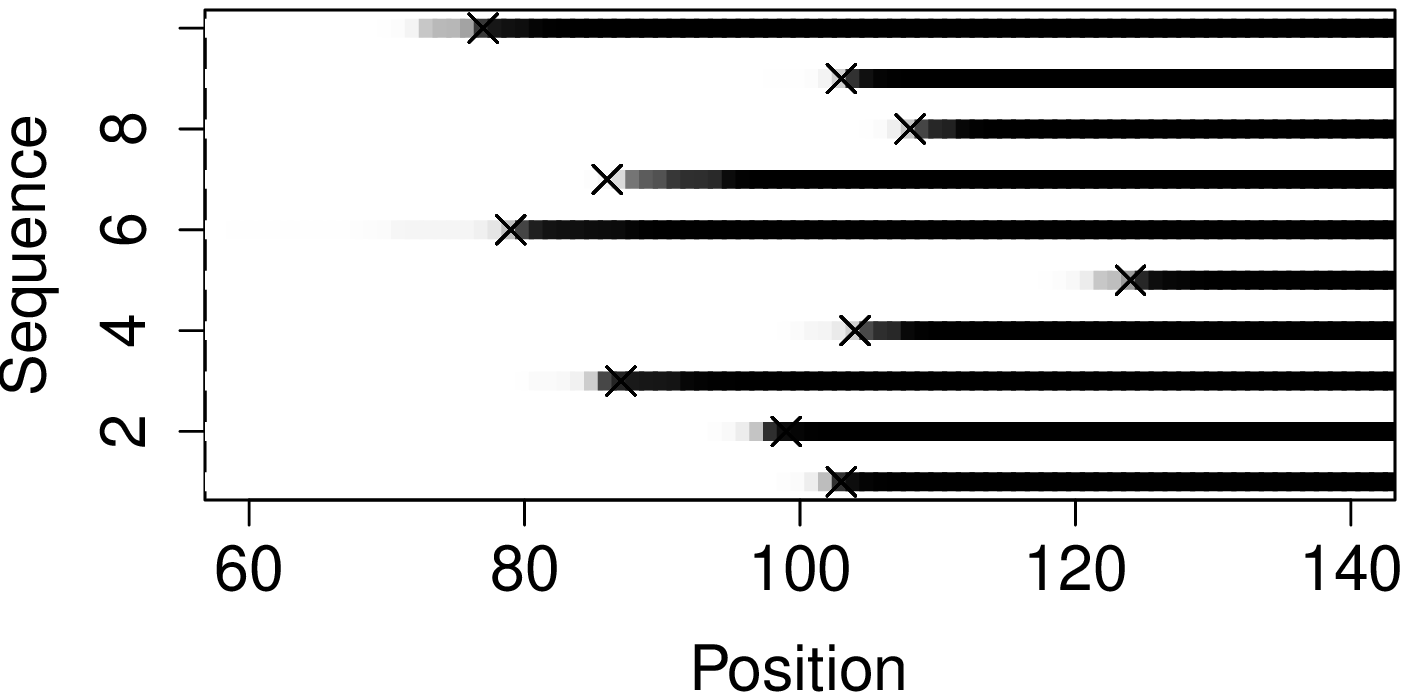}}
\subfigure[\vspace{-0.5cm}]{\label{fig:changepoint_toy}\includegraphics[scale=0.45]{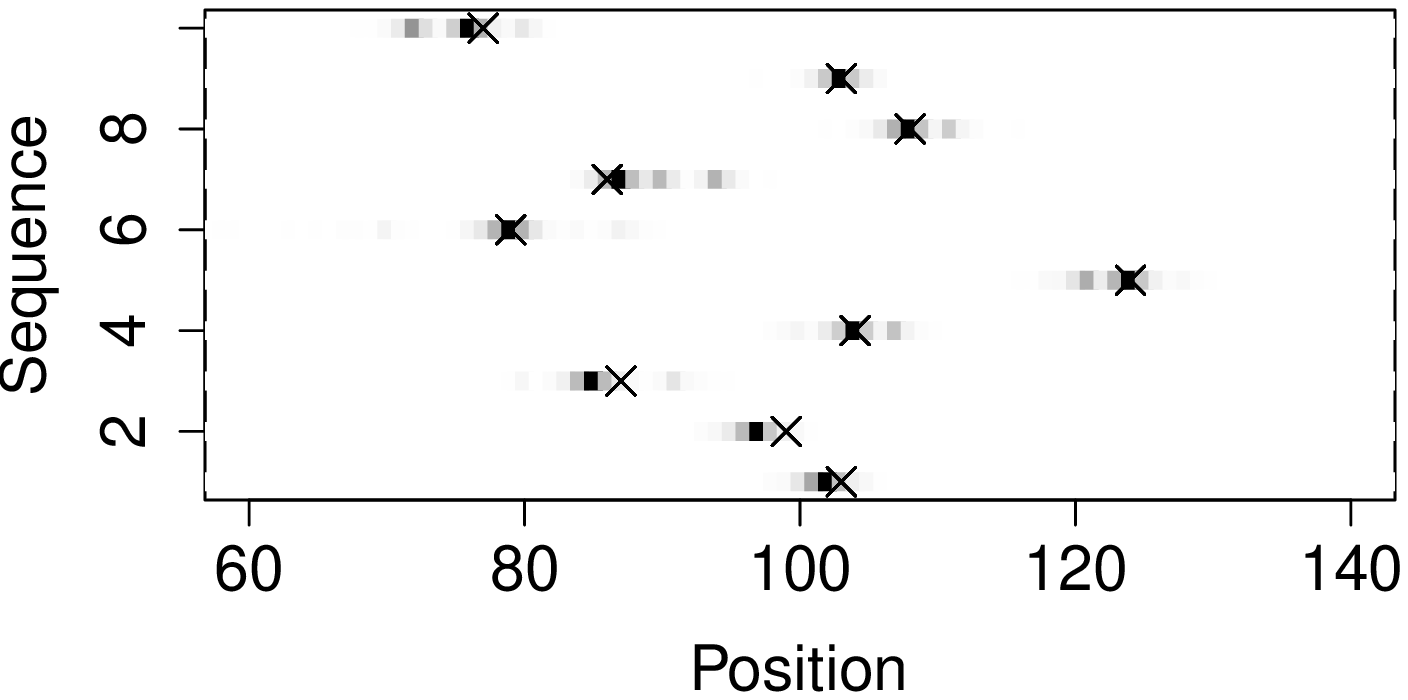}}
        \caption{The estimated marginal probabilities of (a) the
          classification of each observation to segment 2 (probability
          1 is black) and (b) a particular position being a
          changepoint, where the gray scale has been adjusted for each
          sequence so that the position with the maximum probability
          is black and the position with the minimum probability is
          white. The true changepoint locations are marked with
          ($\times$) in both plots. Note that only position 60 to 140
          are shown in each sequence.}
     \label{fig:ch_toy}
\end{center}
\end{figure}
Figure \ref{fig:sequences_toy} shows, for each sequence, the estimated
marginal probability that each observation in that sequence belongs to
segment 2. The true changepoint is marked (with an $\times$) for each sequence and, as we can see, the changepoints are well recovered. In Figure
\ref{fig:changepoint_toy}, we see the marginal probability plot for
the positions of the changepoints. It is worth commenting on the fact
that the marginal probabilities in Figure \ref{fig:changepoint_toy} do
not necessarily vary smoothly as a function of location. That is,
there are observations that are considered to be unlikely candidates
for a changepoint even though the previous and next observations have
a high probability of being a changepoint.

\subsection{\sc Real Data Analysis: Branching of Apple Trees}
\setcounter{footnote}{0}
%\section{REAL DATA ANALYSIS: BRANCHING OF APPLE TREES}
\label{sec:apple}
In this section we analyze the apple tree branching data set\footnote{The data set is available as part of the AMAPmod software \citep{Godin1997} available at \url{http://openalea.gforge.inria.fr/dokuwiki/doku.php}. For more details the reader is referred to \cite{Godin1999}} presented in \cite{guedon2001} and \cite{guedon2003}. \cite{guedon2001} describe how the branching structure of the first annual shoot of the trunk can be useful as a predictor of the adult tree structure and more broadly how branching sequences play an important role in understanding plant development. As stated by \cite{guedon2001} ``the sequences may be viewed as a succession of homogeneous zones or segments where the composition properties do not change substantially within each zone, but change markedly between zones."

\begin{figure}
\vspace{-1cm}
  \centering
    \includegraphics[scale=0.6]{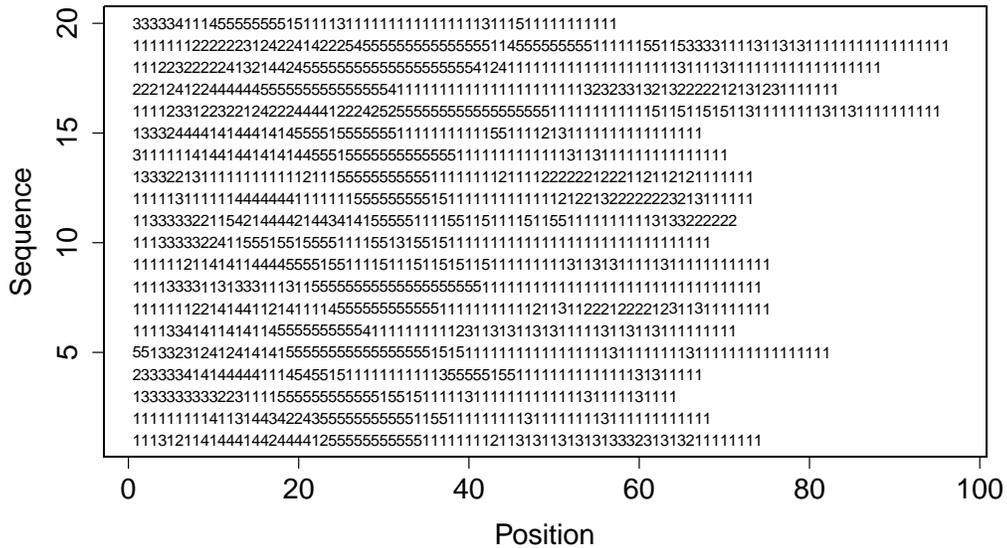}
 \caption{20 apple tree sequences. For this figure sequences 1 to 18 were randomly chosen from the data set, while sequences 19 and 20 are the longest and the shortest sequences, respectively.}
\label{fig:data_apple}
\end{figure}

We analyzed 94 data sequences related to the branching of the main trunk of two-year old apple trees that are left without pruning for one year \citep{guedon2003}, and 20 of these sequences are shown in Figure \ref{fig:data_apple}. The main trunk consists of nodes (metamers or growth units), and each of these nodes can be categorized into one of five types of {\it axillary production} (or {\it states}): (1) latent bud, (2) 1-year-delayed short shoot, (3) 1-year-delayed-long shoot, (4) 1-year-delayed flowering shoot, and (5) immediate shoot. These different  states correspond to nodes without any activity (state 1) and four different types of branching (states 2--5). The data was recorded  by examining the main trunk node by node from the top to the base \citep{Godin1999}. The lengths of the resulting sequences range from 57 to 96 observations as shown in Figure \ref{fig:data_apple}.

Looking at the data in Figure \ref{fig:data_apple} there is clearly some inhomogeneity in the observed sequences. For example, there is a  tendency to have more occurrences of state 5 (immediate shoots) before the middle of each sequence, and the sequences seems to be changing between different states more frequently at the beginning (top of the main trunk) compared to at the end (base of the main trunk).
However, it is not clear from visual inspection how many segments the
data should be divided into, nor where changepoints between segments
should occur. Using exploratory data analysis, \cite{guedon2001} and
\cite{guedon2003} suggested partitioning the sequences into six
segments. As an alternative we can, in a data-driven manner, %use our model to determine the number of segments and the positions of these segments in a data-driven manner. In contrast to earlier work (e.g., \cite{guedon2003}),  we can 
simultaneously determine the appropriate number of segments, the likely positions of these segments, and  estimates of the Markov transition parameters within each segment.

To find the optimal number of changepoints for the model we randomly partition the sequences into ten test sets (and ten corresponding training sets).
The split is done so that four test sets contain ten sequences and six test sets contain nine sequences.
{ In Figure \ref{fig:nrOfK_apple}, we compare
the cross-validated log probability scores of the model with one changepoint ($K=1$) and models with
zero, two, three, four, and five changepoints. We choose the $K=1$
model as the reference in this case because it was the model with the
highest median log-probability score, i.e., the model with a single changepoint is the preferred model.
\begin{figure}
\vspace{-1cm}
\begin{center}
\subfigure[]{ \label{fig:nrOfK_apple}\includegraphics[scale=0.4]{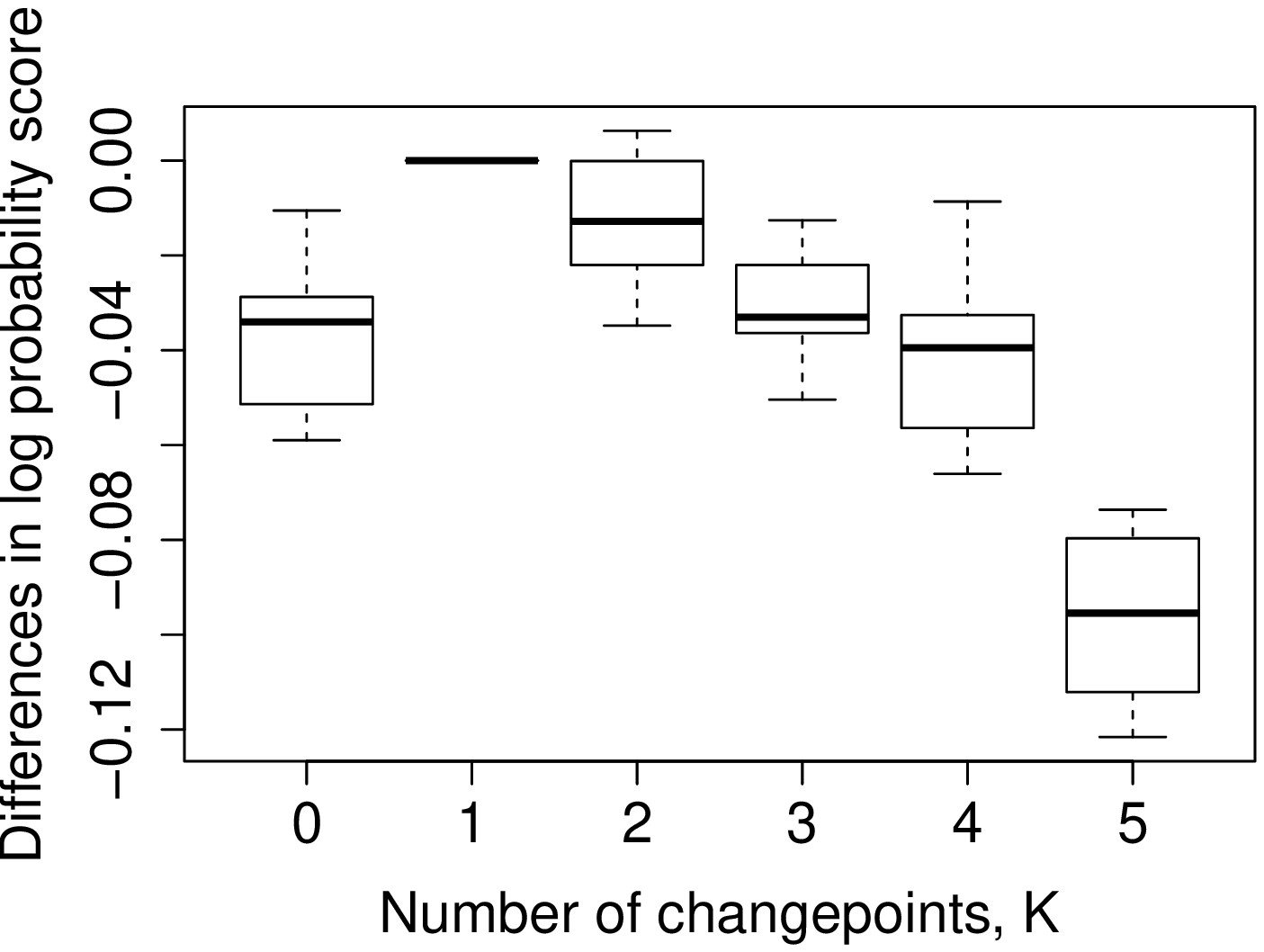}}
\subfigure[]{   \label{fig:baselines_apple}\includegraphics[scale=0.4]{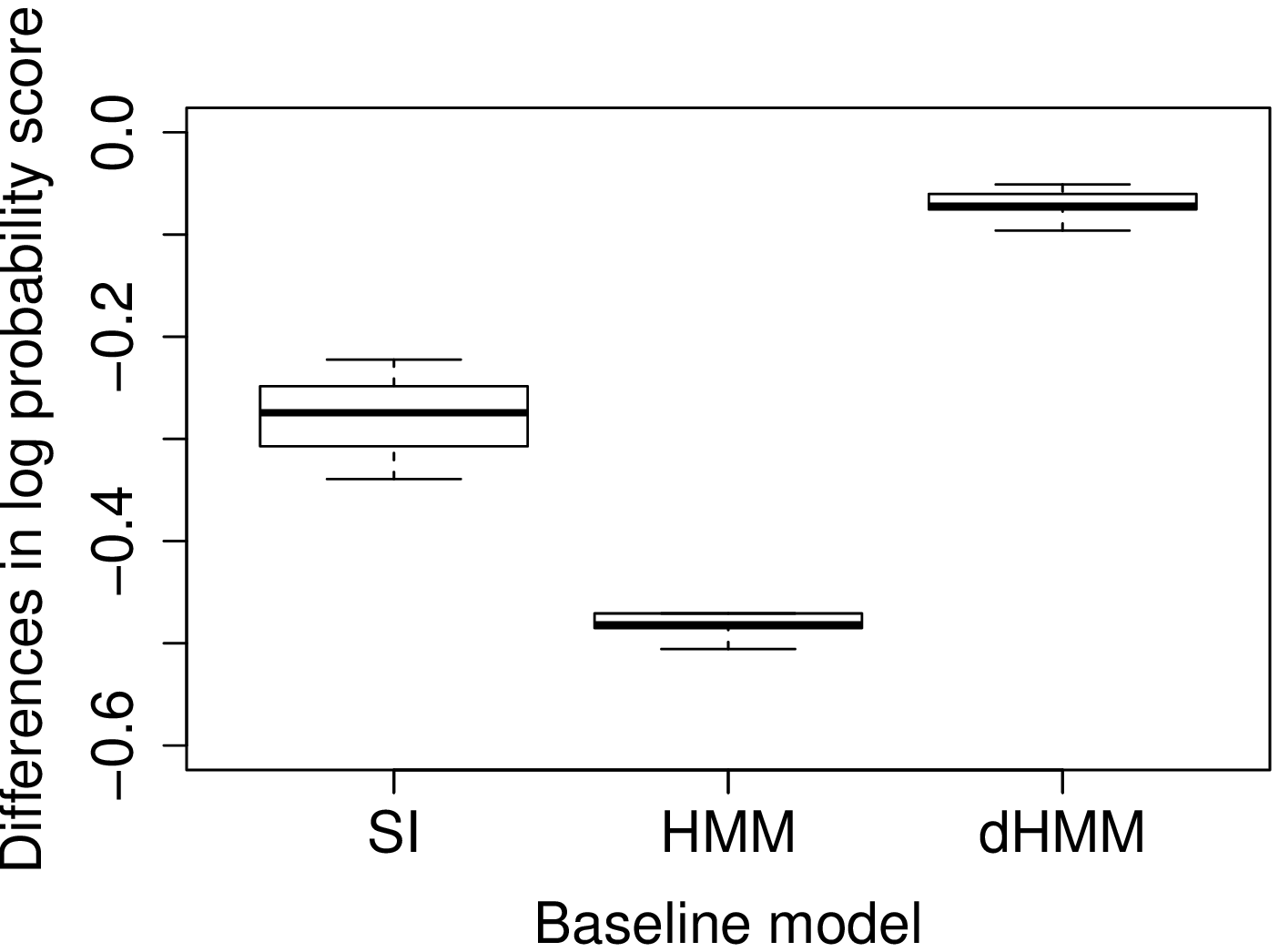}}
   \caption{{Each boxplot shows the log-probability scores, across the
     validation sets, for a different model. The y-axis is defined as
     the log-probability score for (a) other numbers of changepoints
     and (b) other baseline models, minus the log-probability score
     for the model with one changepoint.}}
\end{center}
\label{fig:boxPlot_apple}
\end{figure}

The cross-validated log-probability scores of the baseline
models relative to the preferred
model, are shown in Figure \ref{fig:baselines_apple}. Our model
clearly outperforms the two baseline models using conditionally
independent observations (SI and HMM) supporting the assumption that
there is Markov dependence in the observed sequences. Our model also
performs better than the dHMM, suggesting that the negative binomial
distribution for the changepoint locations is a better choice for this
data set than the  geometric distribution used by the dHMM and
HMM. 
For a comparison between our joint model and the
approach where each sequence is analyzed independently, see the supplementary materials Section S.2.2. }%\ref{supp:4} }

\begin{figure}
\vspace{-1cm}
\begin{center}
\subfigure[]{\label{fig:segments_apple}\includegraphics[scale=0.45]{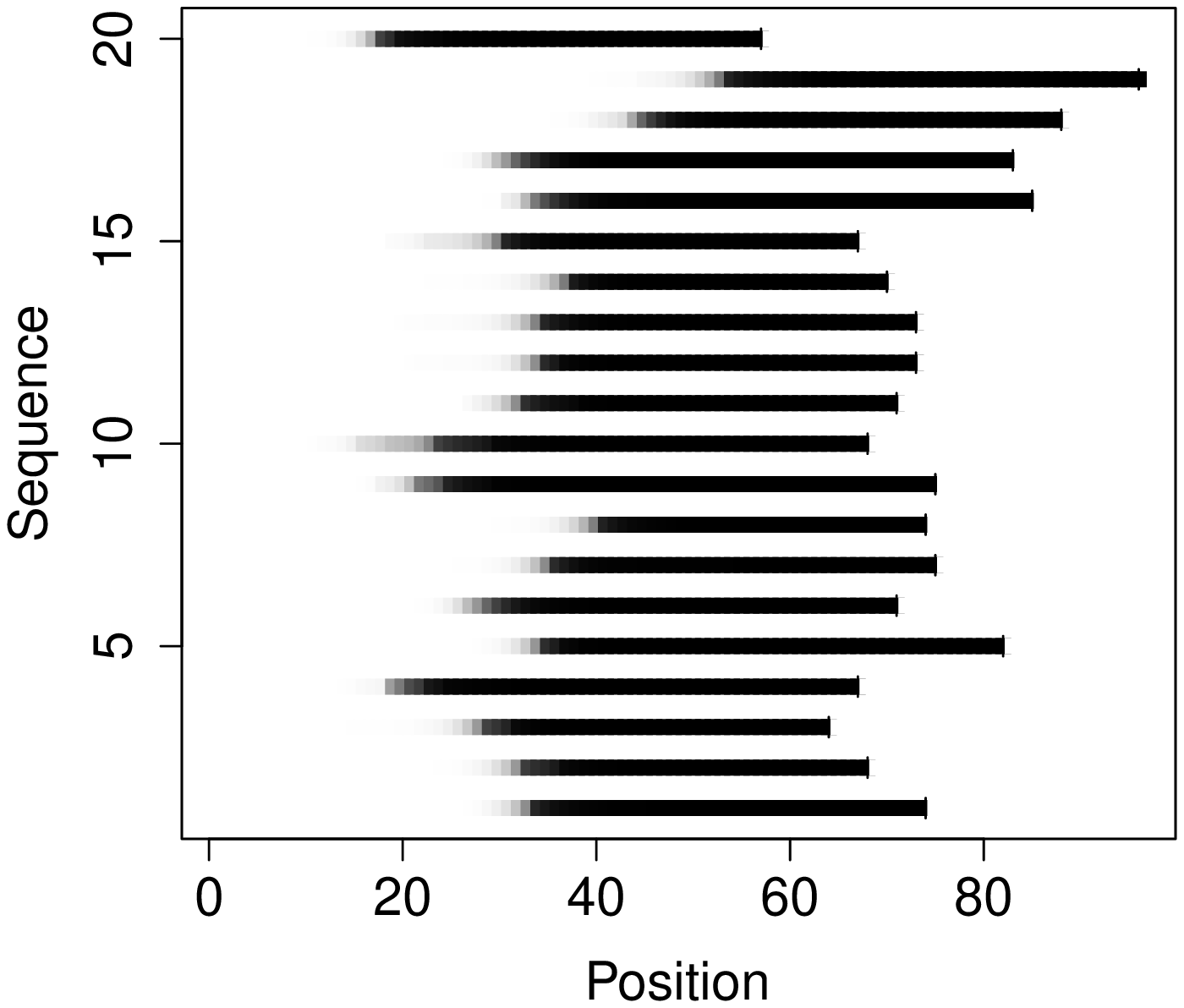}}
\subfigure[]{\label{fig:changepoints_apple}\includegraphics[scale=0.45]{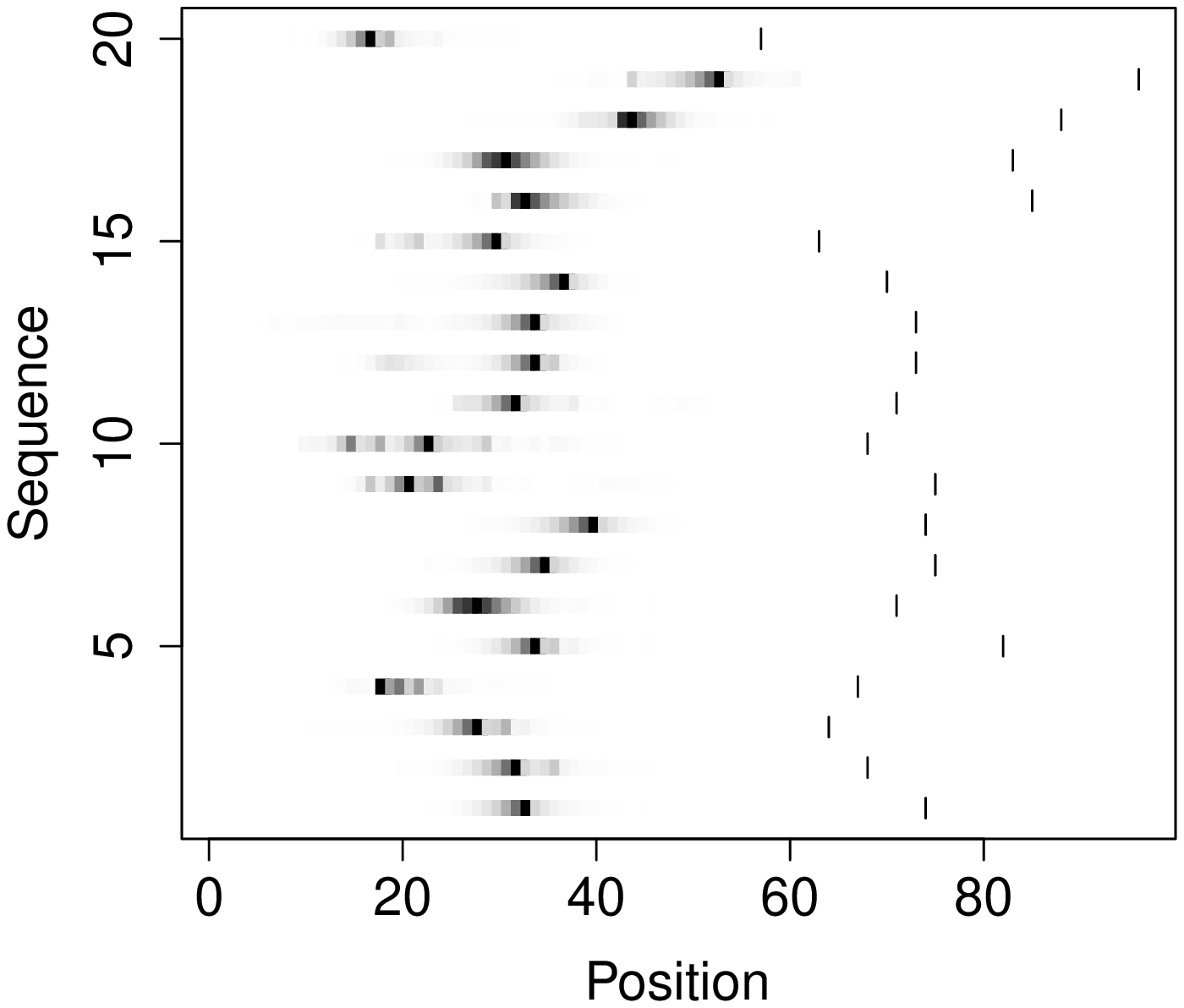}}
        \caption{The estimated marginal probability distribution of (a) the classification of each node to segment 2 (probability 1 is black) and (b) a particular node being a changepoint, where the gray scale has been adjusted for each sequence so that the node with the maximum probability is black and the node with the minimum probability is white. The end position of each observed sequence is marked with a short vertical black line.}
     \label{fig:segmentsAndChangepoints_apple}
\end{center}
\end{figure}

Using our proposed model with a single changepoint, we ran our sampling algorithm on all of the sequences  
in order to infer the position of the changepoint within each sequence
and the transition matrix within each segment. Figure \ref{fig:segments_apple} shows the estimated marginal probability of each node being assigned to segment 2,  for each of the 20 sequences from Figure \ref{fig:data_apple}. The estimated marginal distribution of the position of the changepoint within each of these sequences is shown in Figure \ref{fig:changepoints_apple}.
The results suggest that the changepoint tends to occur at a location roughly one quarter to one half of the length of the
sequence from the start. None of the segments in any of the sequences
where found to be skipped.

\begin{figure}
\vspace{-1cm}
\begin{center}
\subfigure{\label{fig:seqWithEstShortest}\includegraphics[scale=0.4]{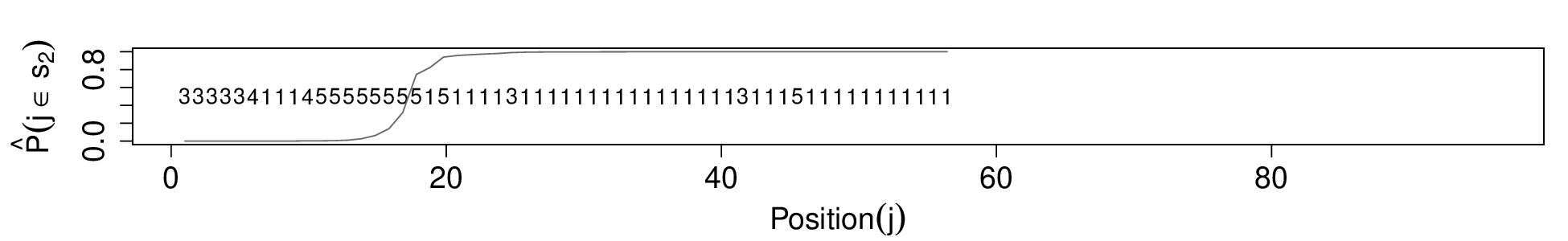}}\\ \vspace{-0.5cm}
\subfigure{\label{fig:seqWithEstLongest}\includegraphics[scale=0.4]{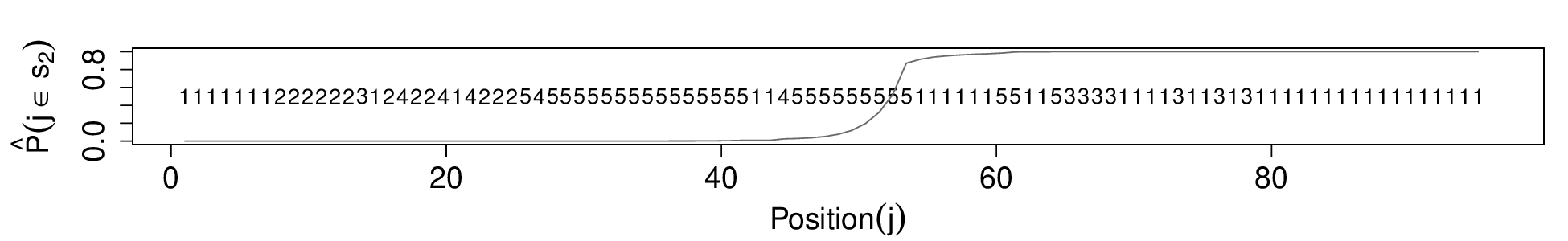}}\\ \vspace{-0.5cm}
\subfigure{\label{fig:seqWithEstNr1}\includegraphics[scale=0.4]{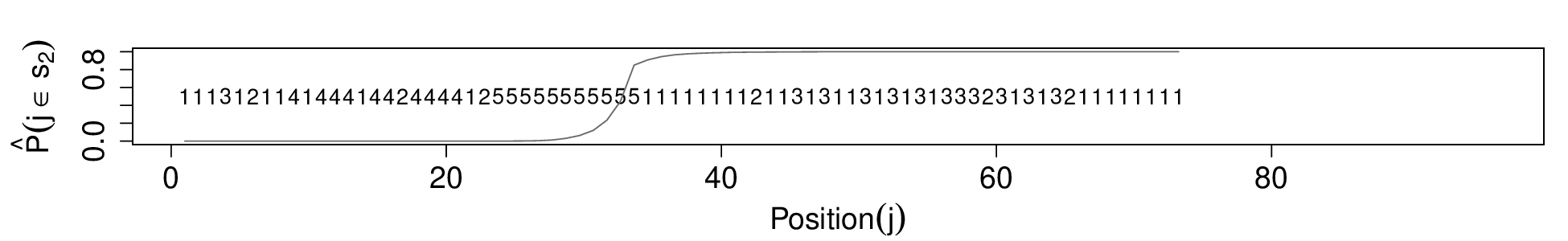}}
        \caption{Sequence 1 (bottom), 19 (middle) and 20 (top) from Figure \ref{fig:data_apple} with marginal estimates of the probability of each position being assigned to segment 2 (as a gray line).}
     \label{fig:seqWithEst}
\end{center}
\end{figure}

In Figure \ref{fig:seqWithEst} we have superposed the observed data over the estimated marginal probability of each node being assigned to segment 2, for three selected sequences, namely the first sequence in Figure \ref{fig:data_apple}, the longest sequence (sequence 19) in the same figure, and the shortest sequence (sequence 20).
From the figure we can see that the changepoints are all placed after the phase where state 5 (immediate shoot) is frequently visited. Investigating the position of the six segments presented in \cite{guedon2001} and \cite{guedon2003}, our first and last segments correspond to the their first three and last three segments, respectively.

{ The estimates of the transition matrices for each segment,
\[\hat{\bm Q}^{(1)}= \left[ \begin{array}{ccccc}
0.54 & 0.07& 0.10&0.20 &0.09 \\
0.21 & 0.37& 0.13&0.21&0.08 \\
0.17 & 0.10& 0.50&0.17&0.06 \\
0.30&0.08&0.08&0.46&0.08\\
0.09&0.03&0.01&0.03&0.84 \end{array} \right],
\hat{\bm Q}^{(2)}= \left[ \begin{array}{ccccc}
0.83 & 0.04& 0.07&0.01&0.05 \\
0.31 &0.36 & 0.22 &0.05&0.06\\
0.53 &0.15 &0.24 &0.02&0.06\\
0.30&0.19&0.07&0.26&0.18\\
0.51&0.04&0.04&0.03&0.38   \end{array} \right], \]
are consistent with the biology of tree growth. In segment 1, for
example, the self-transition and transitions into state 4
("flowering") are much higher than in segment 2. Since segment 2
corresponds to the lower part of the trunk, this makes sense since
flowering is much less likely to occur towards the bottom of the trunk
(segment 2) relative to the top (segment 1). We can also observe that
the marginal probability of state 1 is much higher in segment 2 than
in segment 1 (via its self-transition and in-transition
probabilities). State 1 is the non-branching state, so it is
biologically reasonable to expect less branching in the lower part of
the trunk (segment 2). The estimate for the changepoint parameters $r_1$ and $b_1$ are $0.416$ and $0.652$, respectively, with 95\% credible
interval $[0.389,0.442]$ and $[0.456,0.867]$, respectively.}

\subsection{\sc Real Data Analysis: Monsoon Rainfall}\label{sec:rain}

To illustrate the applicability %The second data set we use to illustrate the applicability 
of the model to a real-world problem we analyze the Indian monsoon data described earlier in Section 1.
The onset and withdrawal of the annual summer monsoon is of critical importance in India since it directly impacts agricultural production, water resource management, and hydroelectricity production \citep{Lima09}. There is no precise definition of the monsoon season, but there is a general understanding that the onset is the time of consistent and substantial increase in rainfall over a regional area and the withdrawal is the time that marks the return to a dry period \citep{fasullo, joseph}. In terms of understanding the climatological variability of the monsoon over  time, a first step is to label the changepoints of the onset and withdrawal in the historic record \citep{joseph}, enabling (for example) prediction of 
onset and withdrawal as a function of exogenous variables such as large-scale atmospheric quantities  \citep{pai-nair}.
Our proposed model provides a framework to not only detect but also to quantify our uncertainty about the estimation of the onset and withdrawal dates and our results can be viewed as an alternative to other approaches that use non-Markov models for this purpose (e.g., \cite{Stern82, Lima09}).

Finite-state Markov models have long been used to characterize daily
rainfall occurrence in climatological applications \citep{gab62,
  katz74}. Rainfall occurrence is typically defined as any amount more
than 0.2mm or 0.01in. A second cutoff can be used to distinguish light
and heavy rainfall events (anything over 20mm is considered a heavy
rainfall event). In our analysis below  the daily rainfall amount is
assigned to one of three discrete categories: no/light rainfall with
0-0.2mm ($y_j=1$), medium rainfall with 0.2-20mm ($y_j=2$), or heavy
rainfall with $>$20mm ($y_j=3$) \citep{katz77}.

Our data set consists of daily rainfall measurements over a 31 year period (1979--2010) from a weather station located  at latitude 24.65 and longitude 77.32 in the monsoon region of India\footnote{The data were obtained from the U.S. National Centers for Environmental Prediction (NCEP) Climate Prediction Center (CPC) Global Summary of the Day (GSOD) Observations, \url{http://rda.ucar.edu/datasets/ds512.0}}. The data is plotted in Figure \ref{fig:rainfallExample} with each year plotted as a sequence. Rainfall amounts were categorized into three states as described above and the 153 missing observations (1.35\% of the data) were imputed during inference as described in Section S.1 in the supplementary materials. We denote each day of the year as $j=1,\ldots,365$ (leap days are removed). We assume that the first and last segments in each year have the same Markov transition matrix, reflecting the fact that the end of one sequence on December 31 is contiguous in time with the start of the next sequence on January 1. This constraint rules out the possibility of a single changepoint in the model, and thus, the number of possible changepoints we can consider is K=0, 2, 3, 4, $\ldots$.

\begin{figure}
\begin{center}
\vspace{-1cm}
\subfigure[]{ \label{fig:nrOfK_rainfall}\includegraphics[scale=0.4]{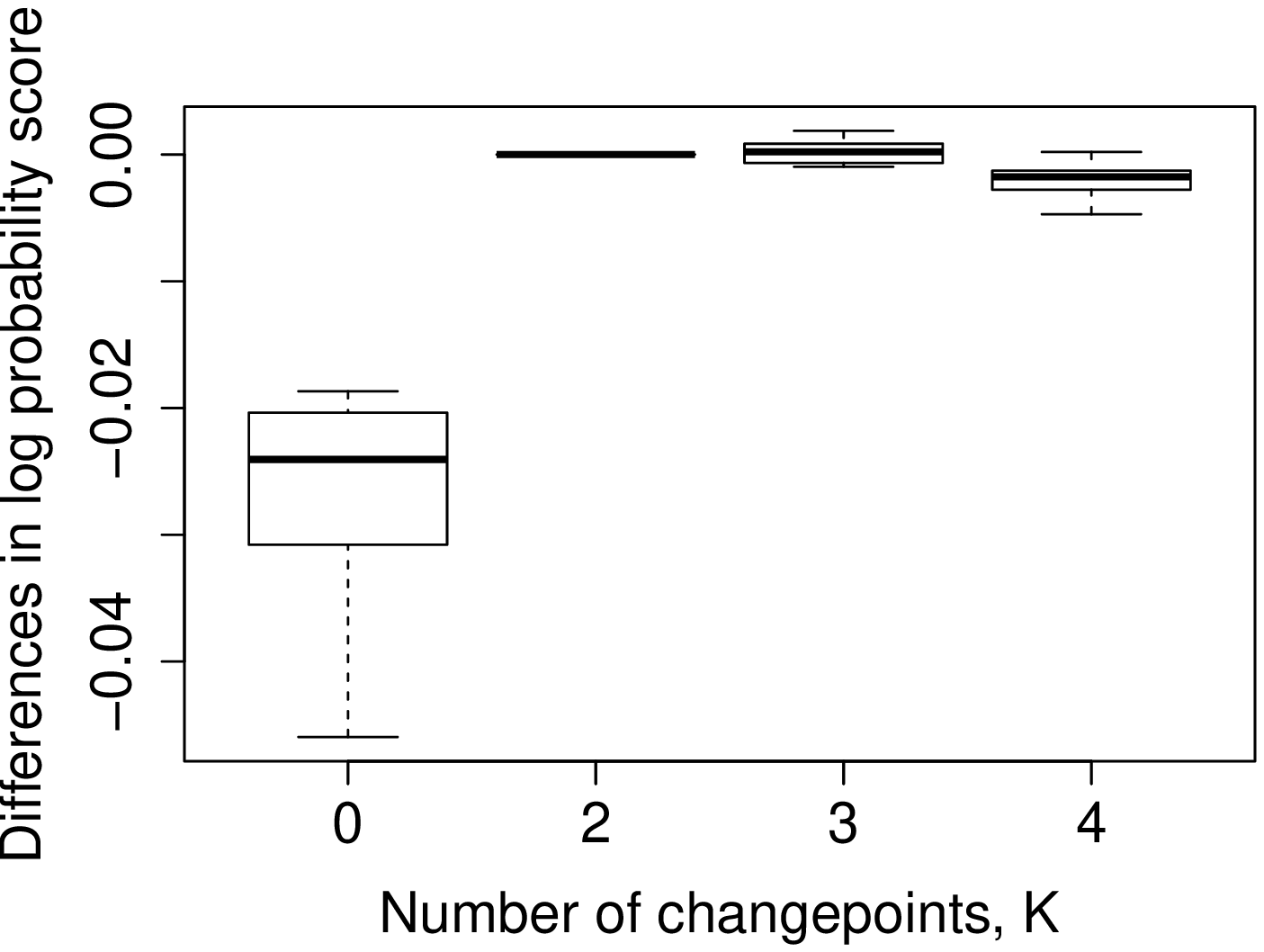}}
\subfigure[]{   \label{fig:baselines_rainfall}\includegraphics[scale=0.4]{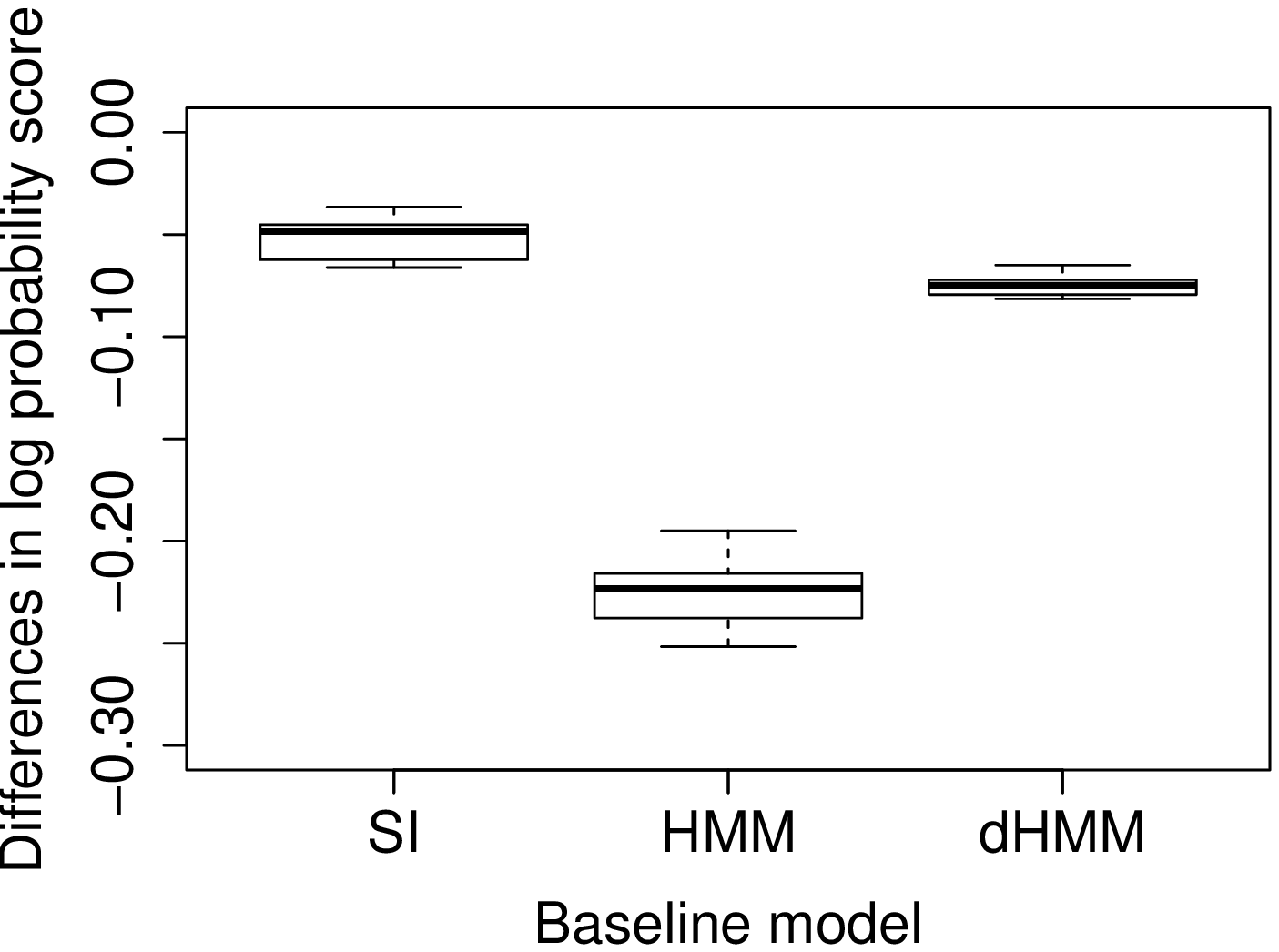}}
   \caption{Each boxplot shows the log-probability scores, across the
     validation sets, for a different model. The y-axis is defined as
     the log-probability score for (a) other numbers of changepoints
     and (b) other baseline models, minus the log-probability score
     for the model with two changepoints.}
\end{center}
\end{figure}

To determine the optimal number of changepoints using cross-validation
we randomly partitioned our sequences (or years) into 10 training/test sets where nine of the test sets contain three sequences and one has four sequences. Figure \ref{fig:nrOfK_rainfall} compares the cross-validated log
probability scores of the model with two changepoints, to models with  zero,
three, and four changepoints. The models with two and three
changepoints are very close in performance and outperform models with
zero or four changepoints. We select the model with two changepoints ($K=2$) as our
preferred model given that it is simpler and it corresponds to our
physical intuition about the monsoon phenomenon (i.e. onset and
withdrawal of the monsoon season). 

Figure \ref{fig:baselines_rainfall} compares the model with two changepoints with the baseline models SI, HMM, and dHMM. %, in the same manner described earlier in Section \ref{sec:apple}).
%As with the tree branching data set
As we can see, our proposed model significantly outperforms the three baselines.
A comparison of the joint model and the approach based
  on individual sequences is provided in Section S.2.3 %\ref{supp:4} 
in the supplementary materials.

Figure \ref{fig:rainfall_segMarginals2} shows the estimates of the marginal probability of each day belonging to the monsoon season, providing a visual interpretation of the estimated interannual variability in the dates of the Indian monsoon onset and withdrawal across 31 years. The marginal distribution across all 31 years is shown at the top of the figure with each of the years plotted per row below.  Most days in July and August, across all years, are highly probable to be classified as monsoon days (black) and the days at the beginning and end of the monsoon season have less certainty of being in the monsoon (gray).
\begin{figure}
\begin{center}
\vspace{-1cm}
\subfigure[]{\label{fig:rainfall_segMarginals2}\includegraphics[scale=0.31]{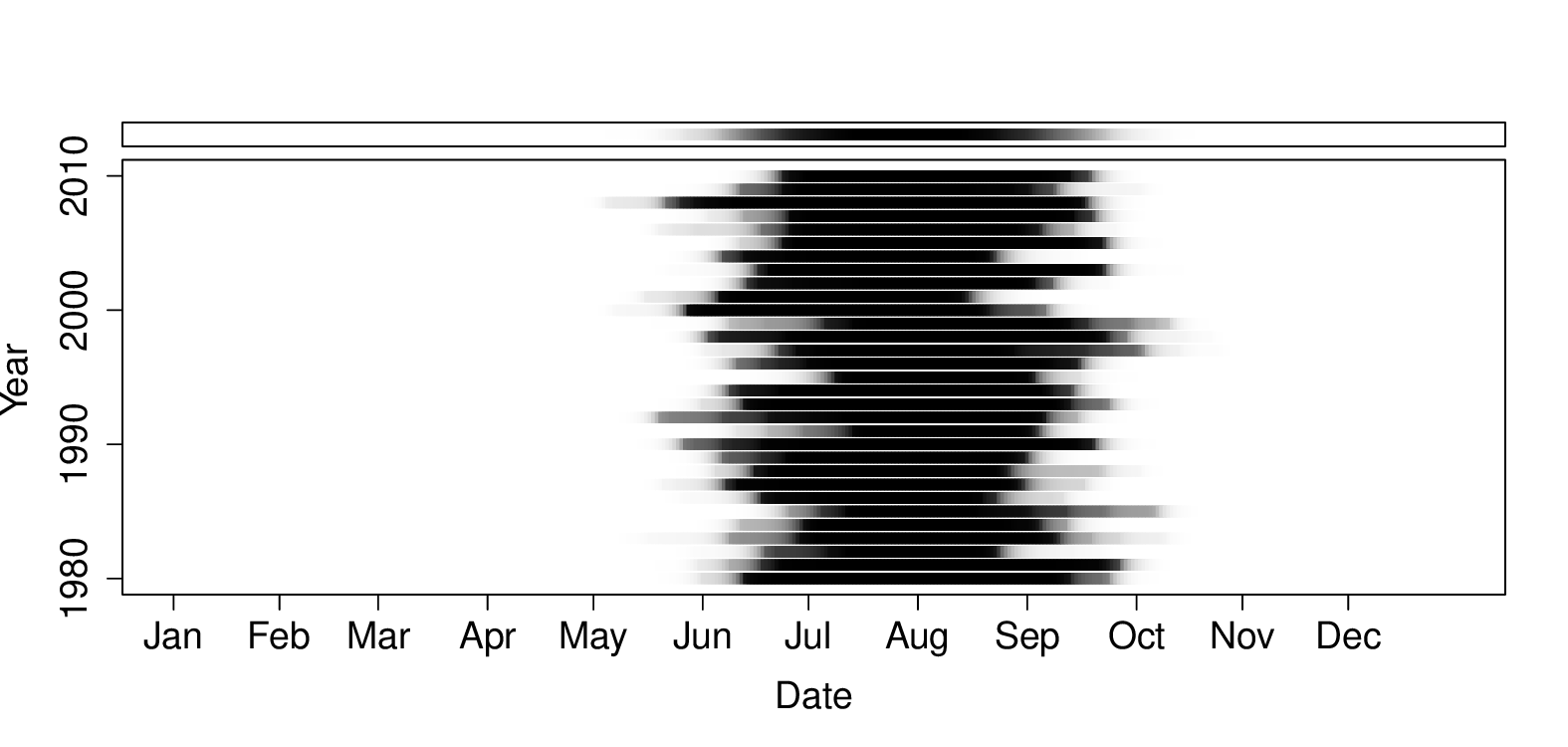}} %\vspace{-1.2cm}
\subfigure[]{\label{fig:rainfall_changeMarginals2}\includegraphics[scale=0.31]{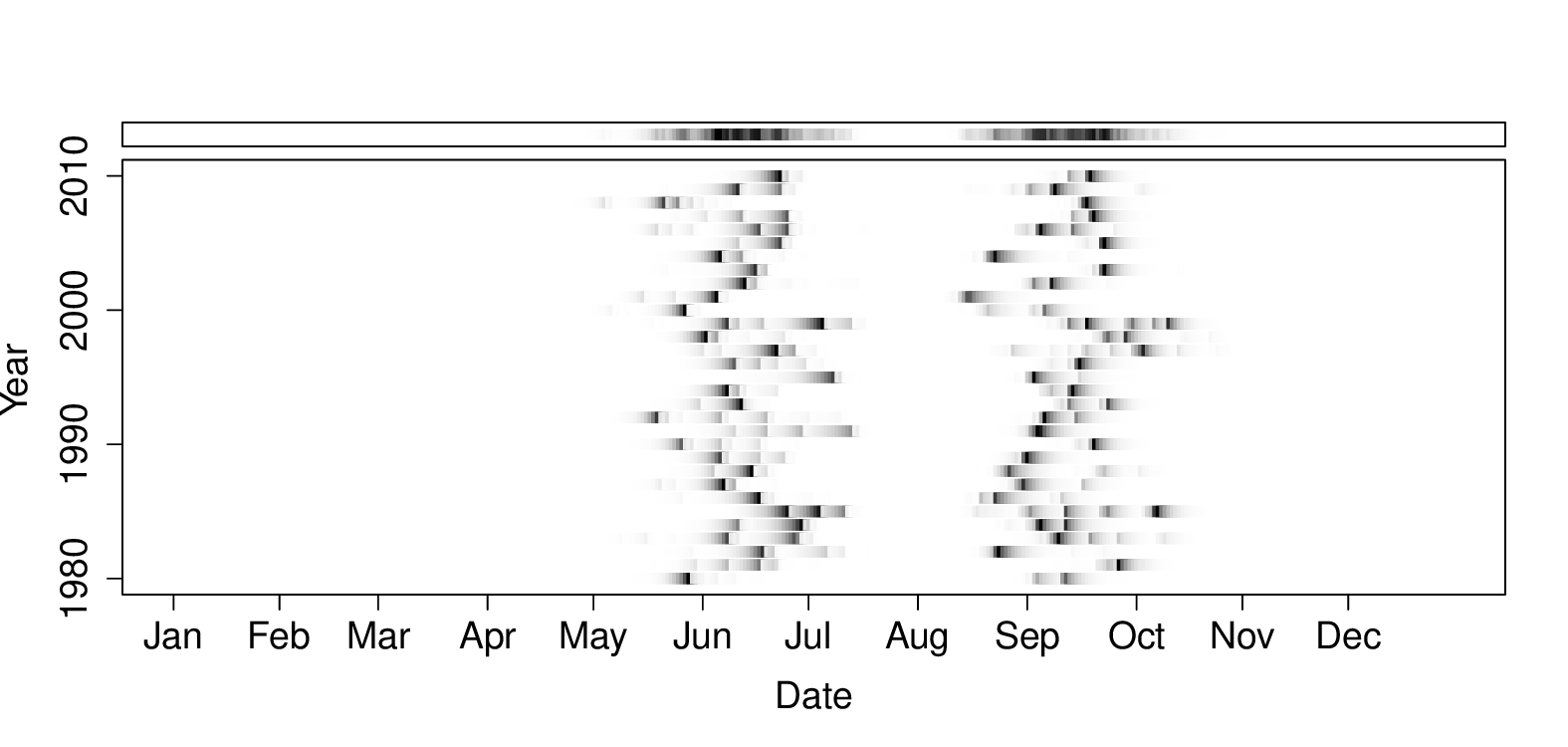}}
        \caption{The estimated marginal probability distribution for
          (a) the classification of each day to the monsoon season
          (probability 1 is black) and (b) a particular day being a
          changepoint, where the gray scale has been adjusted for each year so that the day with the maximum probability is black and the day with the minimum probability is white.}
     \label{fig:rainfall}
\end{center}
\end{figure}

The probability of each individual day being a changepoint (onset or
withdrawal of the monsoon) can also be obtained from the model. Figure
\ref{fig:rainfall_changeMarginals2} shows the probability of a
particular day being a changepoint in each of the 31 years. Days that
are darker gray are more likely to be the monsoon onset or withdrawal
date. The top line shows the marginal over all years. We see that the
highest probability of onset is in June and the highest probability of
withdrawal is in September.

The estimated transition matrices for our model with two changepoints, the first for the dry (non-monsoon) season and the second for the monsoon season are as follows
\[\hat{\bm Q}^{dry}= \left[ \begin{array}{ccc}
0.916 & 0.026& 0.002 \\
0.742 & 0.232 & 0.027 \\
0.423 & 0.367& 0.210 \end{array} \right],
\hat{\bm Q}^{monsoon}= \left[ \begin{array}{cccc}
0.720 & 0.216& 0.069 \\
0.439 & 0.436& 0.125 \\
0.258 & 0.489& 0.253 \end{array} \right],\]
with rows/columns 1, 2, 3, corresponding to no rainfall, light rainfall, and heavy rainfall categories respectively.
The monsoon and non-monsoon seasons are distinctly different in terms
of their Markov transition probabilities, with transitions into light
and heavy rainfall categories being significantly higher in the
monsoon season compared to the dry season.
The estimates for the changepoint parameters $r_1$, $b_1$, $r_2$ and $b_2$ are $0.446$, $0.810$, $0.703$ and $0.885$, respectively, with 95 \% credible intervals $[0.431,0.461]$, $[0.550,0.985]$, $[0.687,0.719]$ and $[0.693,0.993]$, respectively.

Upon examining the model
with three changepoints ($K=3$), we found it added an additional changepoint in the middle of the monsoon season. However the estimated transition matrices for these two segments did not differ significantly from one another.

Our results are broadly consistent with other findings. For example, \cite{fasullo} analyzed data from a station further south in India, estimating that the monsoon onset dates at this location range  from May 19 to June 20 with the median being June 6, and the withdrawal dates range  from August 12 to September 27 with the median being September 4.
These dates are consistent with our estimates at a more northerly location, given that the monsoon approaches from the south in India, and thus, we would expect our calculated onset and withdrawal dates to be somewhat later than those of \cite{fasullo}. We obtain a mean estimate of June 13 for the monsoon onset with 95\% probability interval (PI) of [May 23, July 4] and the mean date for the monsoon withdrawal is September 13 with an estimated 95\% PI of [August 24, October 3]. These estimates are based on the data shown at the top of Figure \ref{fig:rainfall_changeMarginals2} where the 95\% PI is computed by summing up the estimated marginal probabilities symmetrically around the mean until a 95\% interval is reached.

\section{DISCUSSION AND CONCLUSIONS}\label{sec:discussion}

We introduced a piecewise homogeneous Markov chain model where changepoint positions are modeled by a discrete distribution, in particular, a truncated version of the negative binomial distribution. The model is constructed to handle multiple sequences of variable length where each sequence moves between the underlying Markov chains in the same order.
We show
that the changepoints are well recovered on synthetic data, resulting in accurate estimates for
the parameters used to define our model. To illustrate the utility of the model on real-world
data sets we applied the model to an apple tree branching data set and to daily rainfall data
collected for 31 years in Northern India. In the apple tree data the model suggests that a single
changepoint for each sequence is best, and for the rainfall data the model is able to detect
the onset and withdrawal of the monsoon season. In both cases the proposed model produced
inferences of parameters and changepoint locations that were scientifically interpretable. In
addition the model systematically outperformed alternative approaches such as hidden Markov
and double-hidden Markov models.

% We show  that the changepoints are well recovered on synthetic data, resulting% in accurate estimates for the parameters used to define our model. To illustra%te the utility of the model on real-world data sets we applied the model to dai%ly rainfall data collected for 31 years in Northern India. %, and in addtion to% an apple tree branching data set (see the supplementary materials Section S.3.%4). In the apple tree data the model suggests that a single changepoint for eac%h sequence is best, and 
%For this data set %the rainfall data 
%the model is able to detect the onset and withdrawal of the monsoon season, and% %In both cases the proposed model
%produced inferences of parameters % and changepoint locations 
%that were scientifically interpretable. In addition the model  outperformed alt%ernative approaches such as hidden Markov and double Markov models.

Our Bayesian framework allows different sequences to draw strength from each other both when finding the changepoints and for parameter estimation. Estimates of uncertainty about both parameters and latent variables, which arise naturally from our MCMC inference algorithm, provide an appealing interpretation of the uncertainty regarding position of the changepoints. This is of particular interest for example in analyzing the onset and withdrawal of the Indian monsoon season. Uncertainty quantification is an increasingly important component of climate data analysis, and the type of Bayesian approach used in this paper can provide a useful data-driven alternative to more traditional methods such as using threshold values \citep{Lima09} or relying entirely on definitions based on prior knowledge \citep{fasullo}.

\begin{table}
\centering
\begin{tabular}{ c| c c c c}
Data set& $K=0$ & $K=1$ & $K=2$ & $K=3$\\
\hline
  Simulated Data 1& 0.002 & 0.015 & 0.065 &0.138 \\
  Monsoon Data& 0.054 & 0.163 & 0.432 & 0.795 
\end{tabular}
\vspace{0.2cm}
  \caption{Computational time per iteration in seconds for two of the
    data sets discussed in the paper, with code in Matlab on a computer with 2.80 GHz CPU.}
\label{tab:compTime}
\end{table} 
Comparisons of the computation times for two of the data examples in the paper are shown 
in Table \ref{tab:compTime}, comparing the times for the smallest data
set (the first simulated data set) and the largest data set (the
monsoon data).  The numbers shown are for one MCMC iteration when
fitting the model to the full data set (we used parallelization for
our cross-validation runs). We see that as the number of changepoints
increase the method is more computationally expensive, but overall the
method is relatively fast even on the larger data set. Additional speedups could be obtained (for example) by parallelizing the analysis across sequences, coding the algorithm in a more efficient language than Matlab, and so on.

There are a number of additional extensions to the model that may be
potentially useful to explore.  For example, a useful direction would
be the development of conditional models $p(\bm y|\bm x)$, where each
sequence $\bm y = (y_1,\ldots,y_T)$ is accompanied by a
sequence $\bm x = (\bm x_1,\ldots,\bm x_T)$ of exogenous variables and
where each $\bm x_t$ could  be multivariate, $t = 1\ldots, T$. The
exogenous variables $\bm x$ could influence the $\bm y$'s directly
and/or the locations of the changepoints. Another direction would be
models for handling high-dimensional and/or real-valued observation
sequences, where one could assume the existence of a categorical
latent sequence $\bm z = (z_1,\ldots, z_T)$ as a
low-dimensional representation of the observed sequence
data $\bm y$, extending the general  approach presented here to piecewise homogeneous hidden Markov chains.

\section*{ACKNOWLEDGMENTS}
This work was supported in part by  US Department of Energy award DOE-SC0006619 (TH and PS), US Office of Naval Research under MURI grant N00014-08-1-1015 (TH and PS), and US National Science Foundation award IIS-1320527 and a Google Faculty Award (PS).

\section*{SUPPLEMENTARY MATERIALS}

\begin{description}

\item[Additional Supporting Information:] Supplementary materials containing additional results, simulations, sensitivity analysis, etc. is given at the end of this document.

%\item[Software implementation of the changepoint Markov model:]  The
%  Matlab code for implementing the changepoint Markov model and the
%  simulated data sets used in the paper (GNU zipped tar file).

\end{description}

\bibliographystyle{jasa}
\bibliography{bibl}	% For bibtex

\clearpage

\renewcommand{\baselinestretch}{1.0} \small\normalsize
{\begin{center}{\LARGE \textbf{Supplementary materials to the paper: Bayesian Detection of Changepoints in Finite-State Markov Chains for Multiple Sequences} \\}
	\vspace{1cm}
{\Large Petter \textsc{Arnesen}, Tracy \textsc{Holsclaw} and Padhraic \textsc{Smyth}	}
\let\thefootnote\relax\footnote{Petter Arnesen is a PhD student with the Department of Mathematical Sciences,
Norwegian University of Science and Technology, Trondheim 7491, Norway
(email: \textit{petterar@math.ntnu.no}).
Tracy Holsclaw is a Postdoctoral Scholar with the Department of Statistics, University of California, Irvine, CA (email: \textit{tholscla@ams.ucsc.edu}). Padhraic Smyth is a Professor with the
Department of Computer Science and the Department of Statistics, University of California, Irvine, CA (email: \textit{smyth@ics.uci.edu}).}
\end{center}}

\renewcommand{\baselinestretch}{1.9} \small\normalsize
\setcounter{section}{0}
\setcounter{subsection}{0}
\setcounter{table}{0}
\setcounter{figure}{0}
\renewcommand{\thesection}{S.\arabic{section}}
\renewcommand{\thesubsection}{\thesection.\arabic{subsection}}
\renewcommand{\thetable}{S.\arabic{table}}
\renewcommand{\thefigure}{S.\arabic{figure}}
%\renewcommand{\thesubfigure}{S.\arabic{subfigure}}

%{\color{blue}
\section{SAMPLING ALGORITHM FOR POSTERIOR INFERENCE}\label{sec:algorithm}
For notational convenience we assume in this section that the number of observed sequences is $L=1$. The generalization to sampling with multiple sequences is straightforward  given the assumption of conditionally independent sequences from Section 2.4. %\ref{sec:multipeSeq}. 
We use the Metropolis-Hasting algorithm in all of our sampling updates.

The changepoints $\tau_i$ are sampled by proposing a new position
$\tilde{\tau}_i=\tau_i+1$ or $\tilde{\tau}_i=\tau_i-1$, both with
probability 0.5, with obvious adjustments if there are zero-length
segments involved in the update. Letting $\bm \tau=(\tau_1,...,\tau_i,...,\tau_{K})$ be the current segmentation and $\tilde{\bm\tau}=(\tau_1,...,\tilde{\tau}_i,...,\tau_{K})$ be the proposed new segmentation, this yields the following acceptance probability
\begin{equation}
\mbox{acc}(\tilde{\bm\tau}|\bm\tau)=\min \left \{1,\frac{p(\bm
    y|\tilde{\bm
      \tau},\bar{\mathbf{Q}},y_0)p(\tilde{\bm\tau}|T,\bar{\bm\theta})q(\bm\tau|\tilde{\bm
      \tau})}{p(\bm
    y|\bm \tau,\bar{\mathbf{Q}},y_0)p(\bm \tau|T,\bar{\bm
    \theta})q(\tilde{\bm \tau}|\bm \tau)} \right \},
\end{equation}
where $q(\tilde{\bm \tau}|\bm\tau)$ is the probability of proposing
$\tilde{\bm \tau}$ given that the current state is $\bm \tau$.

To update $r_i$ we adopt the following strategy. First we propose a
new value $\tilde{r}_i$ from a proposal distribution of the form $(\tilde{r}_i,1-\tilde{r}_i)|r_i\sim Dir(a_r(r_i,1-r_i))$,
where $a_r$ is a tuning parameter that controls how close the proposed
value $\tilde{r}_i$ will be to the current value $r_i$. The same
strategy is used when proposing new values for $b_i$, using a separate
tuning parameter $a_b$. To update the values in the transition matrix, $\bm Q^{(i)}$, we
propose to change one of the $n$ rows. For example, $\bm Q^{(i)}_{k,\cdot}$ is updated by a proposal distribution
of the form $\tilde{\bm Q}^{(i)}_{k,\cdot}|\bm Q^{(i)}_{k,\cdot}\sim
Dir(a_Q\bm Q^{(i)}_{k,\cdot})$, where $a_Q$ is a tuning parameter
controlling the acceptance rates for such updates. For the results reported in this paper we set tuning parameter values to $a_r = 1000,
a_b = 100,$ and $a_Q = 1000$. These values were chosen to obtain fast
convergence and good mixing of the MCMC chain, and we found that the algorithm is not particularly sensitive to these values. In particular, a five-factor change in these values does not significantly effect the sampling.

We define one iteration of our algorithm to be a single proposed
update of $b_i$ and $r_i$ for one value of $i=1,...,K$, a single
proposed update for one of the rows in one of the transition matrices,
and $V$ proposed updates of each of the changepoints in each
sequence. We found empirically that using $V > 1$ updates leads to
faster convergence and better mixing of the MCMC sampler  (we used
$V=5$ for the results in this paper). Results are obtained from
100,000 iterations of the MCMC chain after removing a burn-in period of 10,000 iterations. In addition, we do thinning by using only every 100th iteration in order to obtain independent samples.

In practice it is common to have missing data (for example for the
rainfall data discussed in the paper) motivating the use of a
systematic approach for handling such missing observations. We
marginalize over missing observations in our sampling algorithm
whenever we need to evaluate the data likelihood in (3). %\eqref{eq:dataLikelihood}. 
The calculation is carried out using a
version of the recursive algorithm given in \cite{Rue2007}. Let
$m=\{j|\text{$y_j$ is missing}\}$, and let $y_j^*$ denote the actual
value of a non-missing variable $y_j$. We write
$p(y_j=y_j^*|y_{j-1},\bm \tau,\bar{\mathbf{Q}})$ when evaluating the probability of that event. When marginalizing over missing values the data likelihood of the non-missing values $\bm{y}_{-m}$ becomes
\begin{equation}
p(\bm{y}_{-m}|\bm \tau,\bar{\mathbf{Q}},y_0)=\sum_{y_j:j\in m}p(\bm y|\bm\tau,\bar{\mathbf{Q}},y_0).
\end{equation}
To evaluate this let $z_0(y_1)=p(y_1|y_0)$, and for $j=1,...,T-1$ we define recursively
\begin{eqnarray*}
z_j(y_{j+1})=
\begin{cases}
p(y_{j+1}|y_j=y_j^*)z_{j-1}(y_j=y_j^*) \hspace{1cm}  \mbox{$y_j$ is not missing}\\
\sum_{y_j}p(y_{j+1}|y_j)z_{j-1}(y_j)  \hspace{2.05cm}  \mbox{$y_j$ is missing.}
\end{cases}
\end{eqnarray*}
To complete the calculation we find
\begin{eqnarray*}
p(\bm{y}_{-m}=\bm{y}_{-m}^*|\bar{\mathbf{Q}},\bm \tau,y_0)=
\begin{cases}
z_{T-1}(y_T=y_T^*) \hspace{1cm} \mbox{$y_T$ is not missing}\\
\sum_{y_T}z_{T-1}(y_T)  \hspace{1cm} \ \mbox{$y_T$ is missing.}
\end{cases}
\end{eqnarray*}

\section{ADDITIONAL POSTERIOR RESULTS FOR THE EXAMPLES IN THE PAPER}

\subsection{\sc Synthetic data: Scenario 1}
\begin{figure}
%\vspace{-1cm}
\begin{center}
\subfigure[]{\label{fig:tauDistVSb}\includegraphics[scale=0.45]{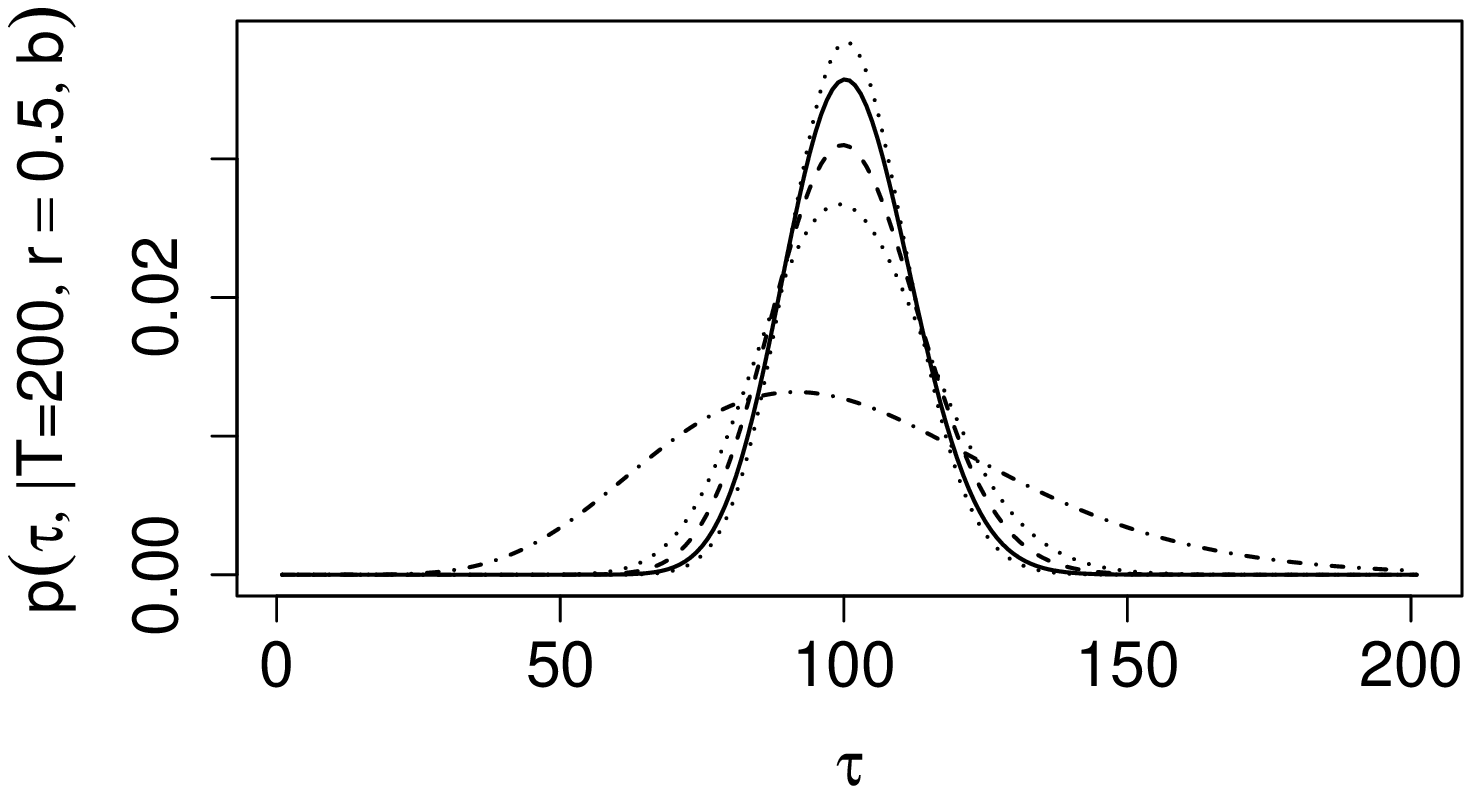}}
\subfigure[\vspace{-0.5cm}]{\label{fig:tauDistVSr}\includegraphics[scale=0.45]{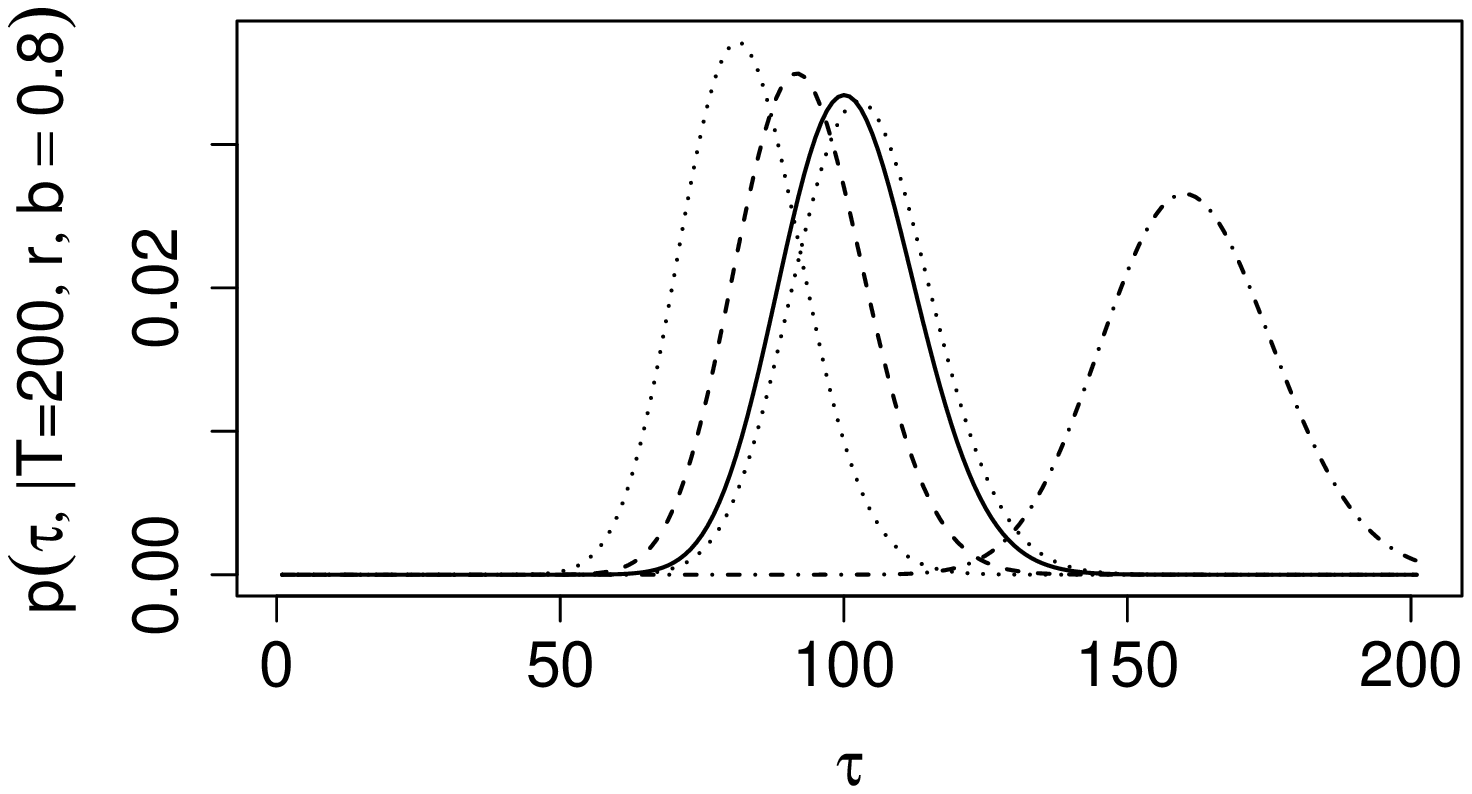}}
        \caption{(a) Sensitivity of the probability distribution $p(\bm
          \tau|T=200,r=0.5,b)$. The solid line corresponds to $b=0.8$
          (true value), dashed line to $b=0.615$ (estimated value),
          and the dotted lines correspond to $b=0.444$ and $b=0.927$
          (limits of the 95\% CI). The dash-dotted line corresponds to
          $b=0.1$.  (b) Sensitivity of the probability distribution $p(\bm \tau|T=200,r,b=0.8)$. The solid line corresponds to $r=0.5$
          (true value), dashed line to $r=0.458$ (estimated value),
          and the dotted lines correspond to $r=0.406$ and $r=0.513$
          (limits of the 95\% CI). The dash-dotted line corresponds to
          $r=0.8$.}
     \label{fig:tauDist}
\end{center}
\end{figure}

Parameters $b$ and $r$ are correlated with one another thus making their values interdependent. Because of this interdependence we investigate below the marginals and the sensitivity of the model relative to these parameters. Both parameters are well estimated, although there is considerable uncertainty concerning the estimated value of $b$ (see Table \ref{tab:est_toy1}). Figure \ref{fig:tauDist}(a) shows, however, that the distribution of the changepoints, $p(\bm \tau|T=200,r=0.5,b)$, is not particularly sensitive to the value of $b$ within the credibility interval. The parameter $r$ primarily controls the position of the changepoints. Sensitivity analysis for this parameter is shown in Figure \ref{fig:tauDist}(b).  The relatively high posterior uncertainty for these two parameters is somewhat due to the small sample size of only 10 sequences. We also generated a data set with $L=100$ sequences and found that the estimate for $b$ is $0.712$ with a 95\% CI of $(0.555,0.902)$, and as expected the extra sequences improve the accuracy and precision of the estimate.

\def\arraystretch{0.5}
\begin{table}
\centering
\begin{tabular}{ c| c c c }
&True &Est.& 95 \% CI\\
\hline
  $r_1$& 0.5 & 0.458 & [0.406,0.513] \\
  $b_1$& 0.8 & 0.615 & [0.444,0.927] \\
\end{tabular}
\vspace{0.2cm}
  \caption{Parameter estimate and 95\% CI.}
\label{tab:est_toy1}
\end{table}

\begin{figure}
\vspace{-1cm}
\begin{center}
\subfigure[]{ \label{fig:nrOfK_toy_L5_cross}\includegraphics[scale=0.33]{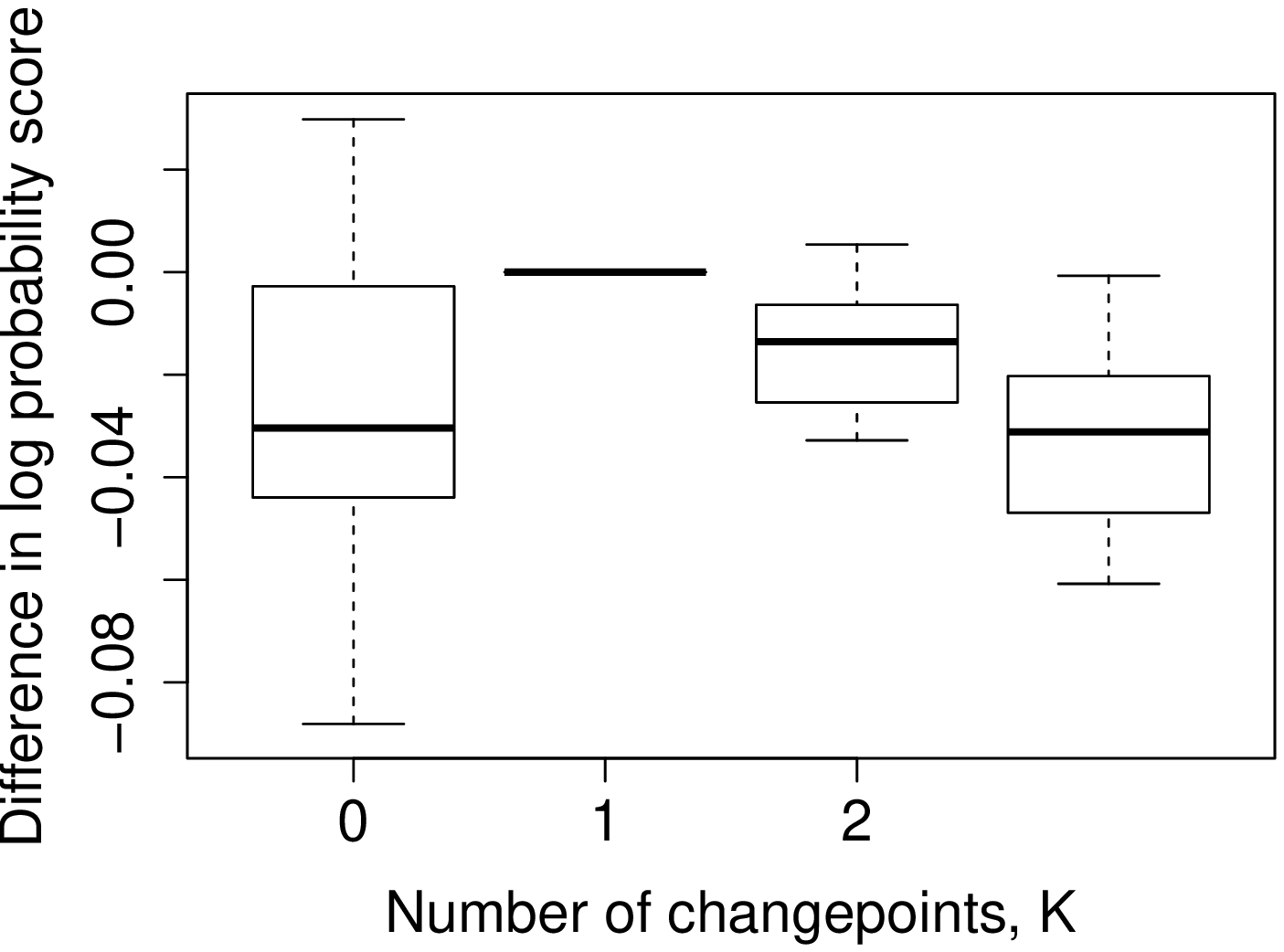}}
\subfigure[]{   \label{fig:nrOfK_toy_L20}\includegraphics[scale=0.33]{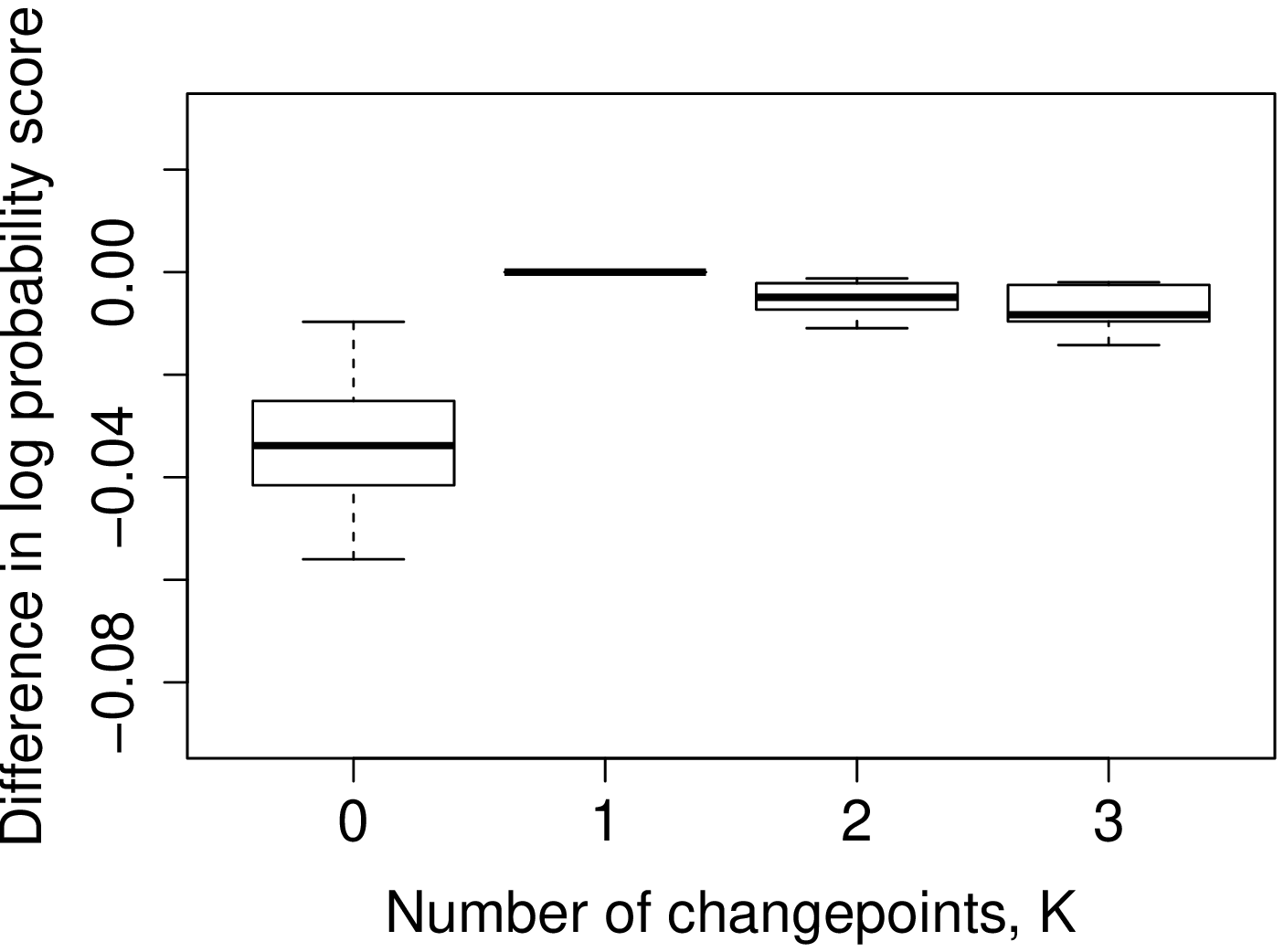}}
\subfigure[]{   \label{fig:nrOfK_toy_L100}\includegraphics[scale=0.33]{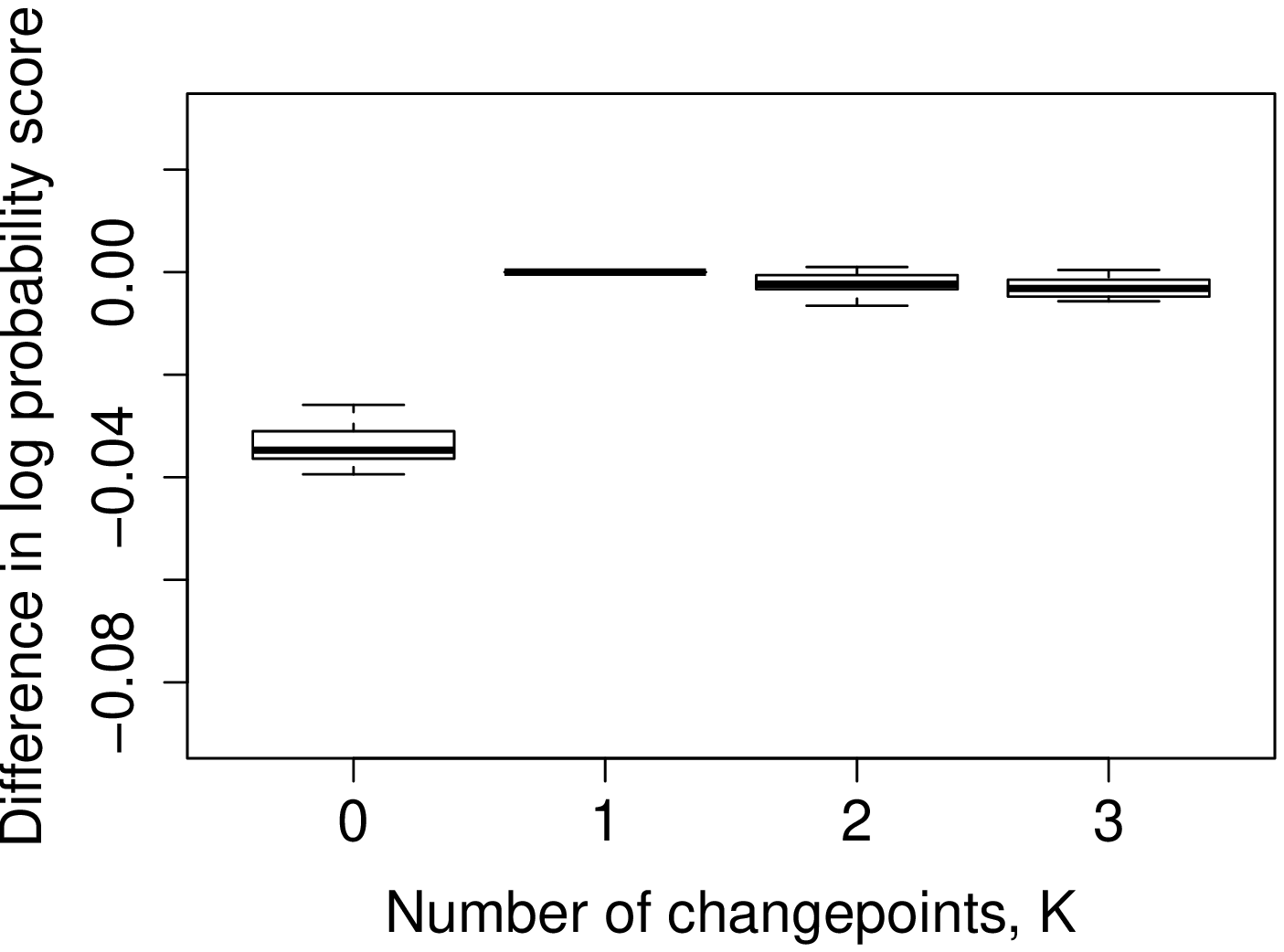}}
   \caption{Each boxplot shows the log-probability scores, across the
     validation sets, for a different model. The y-axis is defined as
     the log-probability score for other numbers of changepoints minus the log-probability score
     for the model with one changepoint, and the three plots correponds
     to different number of sequences $L$: (a) L=5, (b) L=20 and (c)
     L=100.}
	\label{fig:boxPlot_toy_diffL1}
\end{center}
\end{figure} 
We also varied $L$ in this scenario to test the sensitivity of the model to the number of sequences. 
We generated data sets with $L=5$, $L=20$, and $L=100$ and display the results in Figure \ref{fig:boxPlot_toy_diffL1}. For the data sets with $L=20$ and $L=100$ the results are similar to Figure 2(a) %\ref{fig:nrOfK_toy} 
with $L=10$. 
For $L=5$ the cross-validation test tends to prefer the correct solution ($K=1$)  (where we generated 15 folds in
total, comprised of validation sets with either a single sequence or a
pair of sequences),
although $K=0$ (no changepoint) is preferred in some cases, likely due
to limited availability of data to support a more complex model. 

{
In addition to comparing our model to the baseline models presented in the paper
we also compare our results to those obtained with a model where each sequence is analyzed
independently. In particular, we use the single-sequence model of
Fearnhead (2006) for comparison, originally proposed for
real-valued data with independence. We adapt this model by
conditioning on the number of changepoints and extending to a Markov
assumption. In the Fearnhead model, the prior assumes a geometric
distribution to the length of each segment. By fixing the number of changepoints this simply becomes a
uniform distribution on the position of the changepoints. Note that
this is a special case of our model applied to single sequences when
$b_i$ approaches zero for all $i$. Figure
\ref{fig:ch_toy_individual} shows the results from running this model
with a fixed (correct) number of changepoints, and the results can be
compared to those of Figure 3 in the paper. %\ref{fig:ch_toy}.   
\begin{figure}
%\vspace{-1cm}
\begin{center}
\subfigure[]{\label{fig:toy_segments_individual}\includegraphics[scale=0.45]{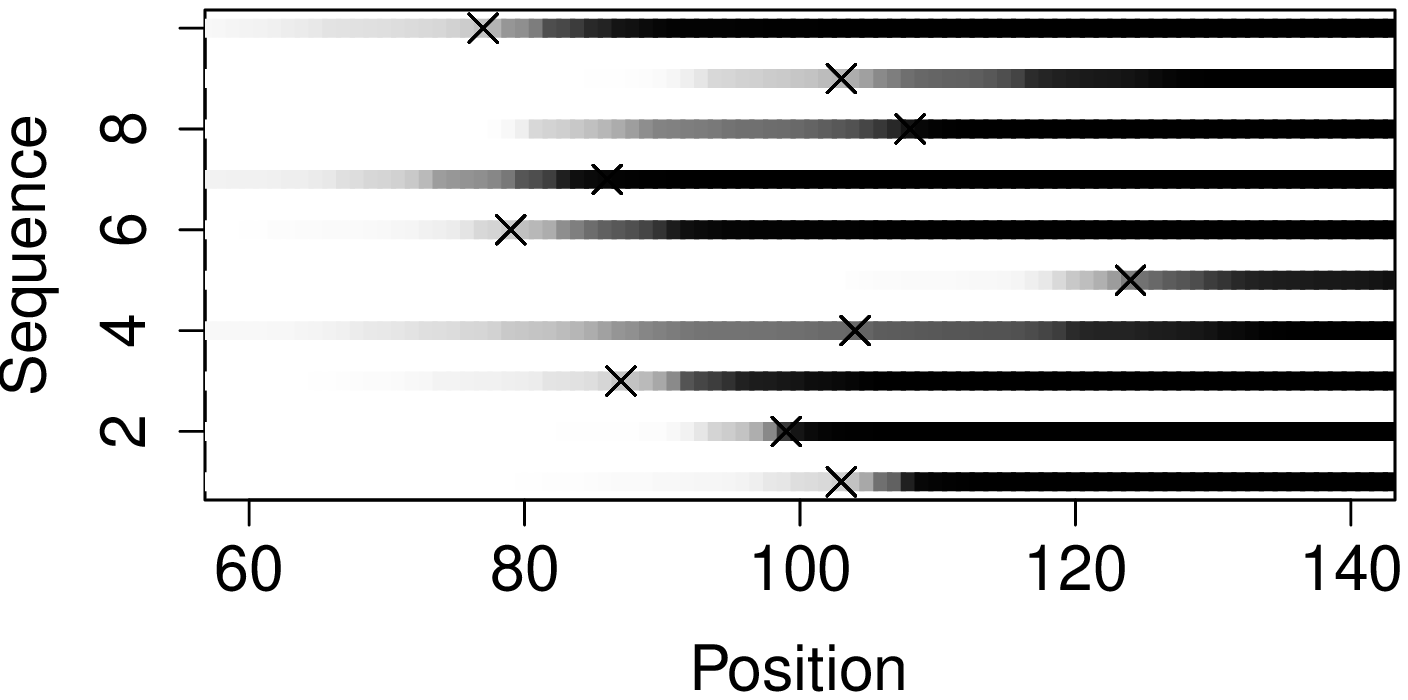}}
\subfigure[\vspace{-0.5cm}]{\label{fig:toy_changepoints_individual}\includegraphics[scale=0.45]{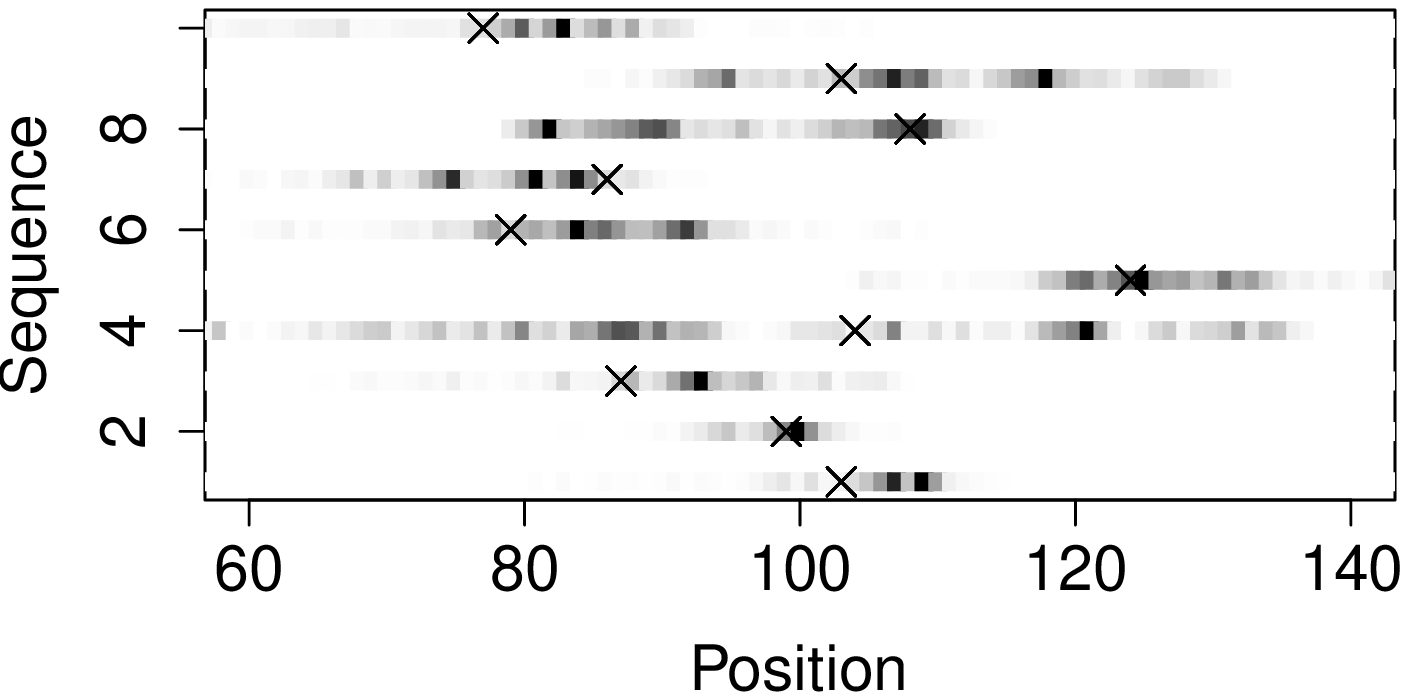}}
        \caption{The estimated marginal probabilities for individual
          sequence model of (a) the
          classification of each observation to segment 2 (probability
          1 is black) and (b) a particular position being a
          changepoint, where the gray scale has been adjusted for each
          sequence so that the position with the maximum probability
          is black and the position with the minimum probability is
          white. The true changepoint locations are marked with
          ($\times$) in both plots. Note that only position 60 to 140
          are shown in each sequence.}
     \label{fig:ch_toy_individual}
\end{center}
\end{figure}
Comparing these figures illustrates the strength of our joint
modeling approach. Since our model is able to draw strength across
sequences we achieve more accurate estimates of the position of
the changepoints
compared to the single sequence model. Similar results can be shown for
the estimate of the transition matrices in each of the segments (see 
Table \ref{tab:toy_Q_individual}).
%the supplementary materials Section \ref{supp:4} for these results.
% }
%
%When comparing this scenario to the single sequence model we get (in
%addition to Figure \ref{fig:ch_toy_individual} in the paper)
%the results shown in 
% for the estimates of the transition probabilities. 
\def\arraystretch{0.5}
\begin{table}
\centering
\begin{tabular}{ c|  c c c c} 
Seq.&$\hat{q}_{1,1}$(95 \%CI)& $\hat{q}_{2,1}$(95 \%CI)& $\hat{q}_{1,2}$(95 \%CI) &$\hat{q}_{2,2}$(95 \%CI)\\
\hline
1 & 0.856(0.757,0.907) & 0.711 (0.580,0.824) & 0.462 (0.351,0.582) &
0.190 (0.113,0.300)\\
2 & 0.770(0.654,0.840) & 0.725 (0.608,0.829) & 0.392 (0.295,0.497) &
0.389 (0.284,0.506)\\
3 & 0.842(0.751,0.910) & 0.760 (0.628,0.876) & 0.504 (0.393,0.614) &
0.430 (0.323,0.530)\\
4 & 0.820(0.722,0.895) & 0.844 (0.738,0.928) & 0.601 (0.446,0.717) &
0.447 (0.316,0.565)\\
5 & 0.764(0.682,0.849) & 0.739 (0.645,0.816) & 0.506 (0.389,0.646) &
0.484 (0.392,0.620)\\
6 & 0.807(0.689,0.890) & 0.701 (0.576,0.798) & 0.565 (0.464,0.658) &
0.393 (0.295,0.504)\\
7 & 0.866(0.777,0.928) & 0.713 (0.557,0.840) & 0.475 (0.366,0.606) &
0.371 (0.276,0.483)\\
8 & 0.802(0.690,0.884) & 0.842 (0.741,0.917) & \ 0.651 (0.5440,0.746) &
0.326 (0.216,0.472)\\
9 & 0.846(0.746,0.898) & 0.621 (0.395,0.763) & 0.513 (0.390,0.629) &
0.480 (0.340,0.606)\\
10 & 0.811(0.668,0.929) & 0.885 (0.789,0.947) & 0.512 (0.395,0.627) &
0.379 (0.271,0.475)\\
\hline
Joint&0.829(0.792,0.852)&0.772(0.742,0.810)&0.514 (0.475,0.550)&0.382(0.344,0.423)
\end{tabular}
\vspace{0.2cm}
  \caption{Parameter estimate and 95\% CI for the transition
    probabilities estimated based on individual sequences (Seq. 1-10), and for our
    joint model approach (Joint).}
\label{tab:toy_Q_individual}
\end{table} 
In this table we clearly see that the estimates for the transition
probabilities for the individual sequences have a much higher
uncertainty than the estimates for our joint model approach (shown in
the last row).

\subsection{\sc Real Data Analysis: Branching of apple Trees}
In this section we present results from running the individual
sequence approach on the apple tree data set. In Figure
\ref{fig:segmentsAndChangepoints_apple_individual} we show the
estimates for the position of the changepoints using a model with $K=1$,
and this is to be compared with the result from our joint model shown
in Figure 6 in the paper. % \ref{fig:segmentsAndChangepoints_apple}. 
\begin{figure}
\vspace{-1cm}
\begin{center}
\subfigure[]{\label{fig:segments_apple_individual}\includegraphics[scale=0.45]{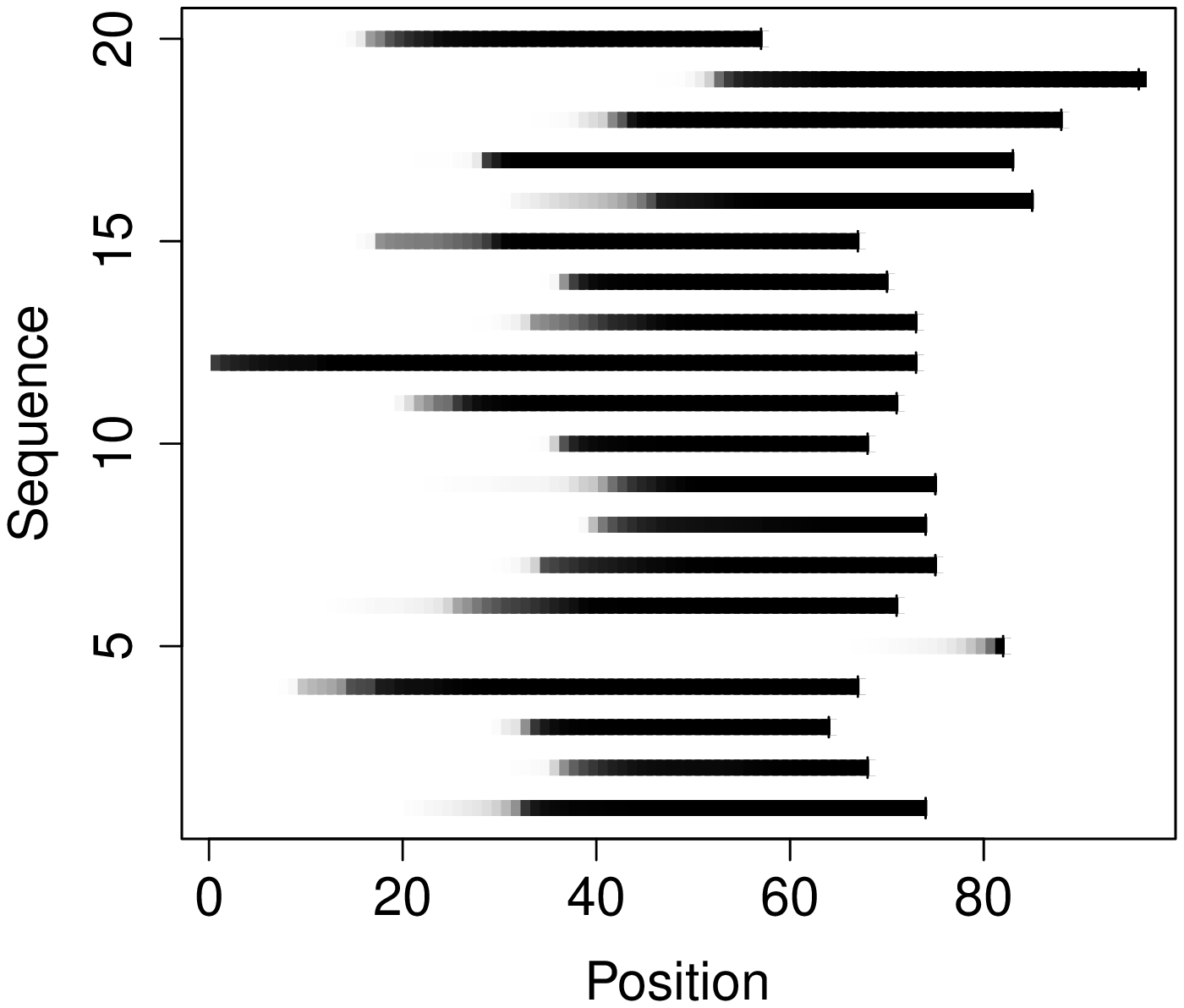}}
\subfigure[]{\label{fig:changepoints_apple_individual}\includegraphics[scale=0.45]{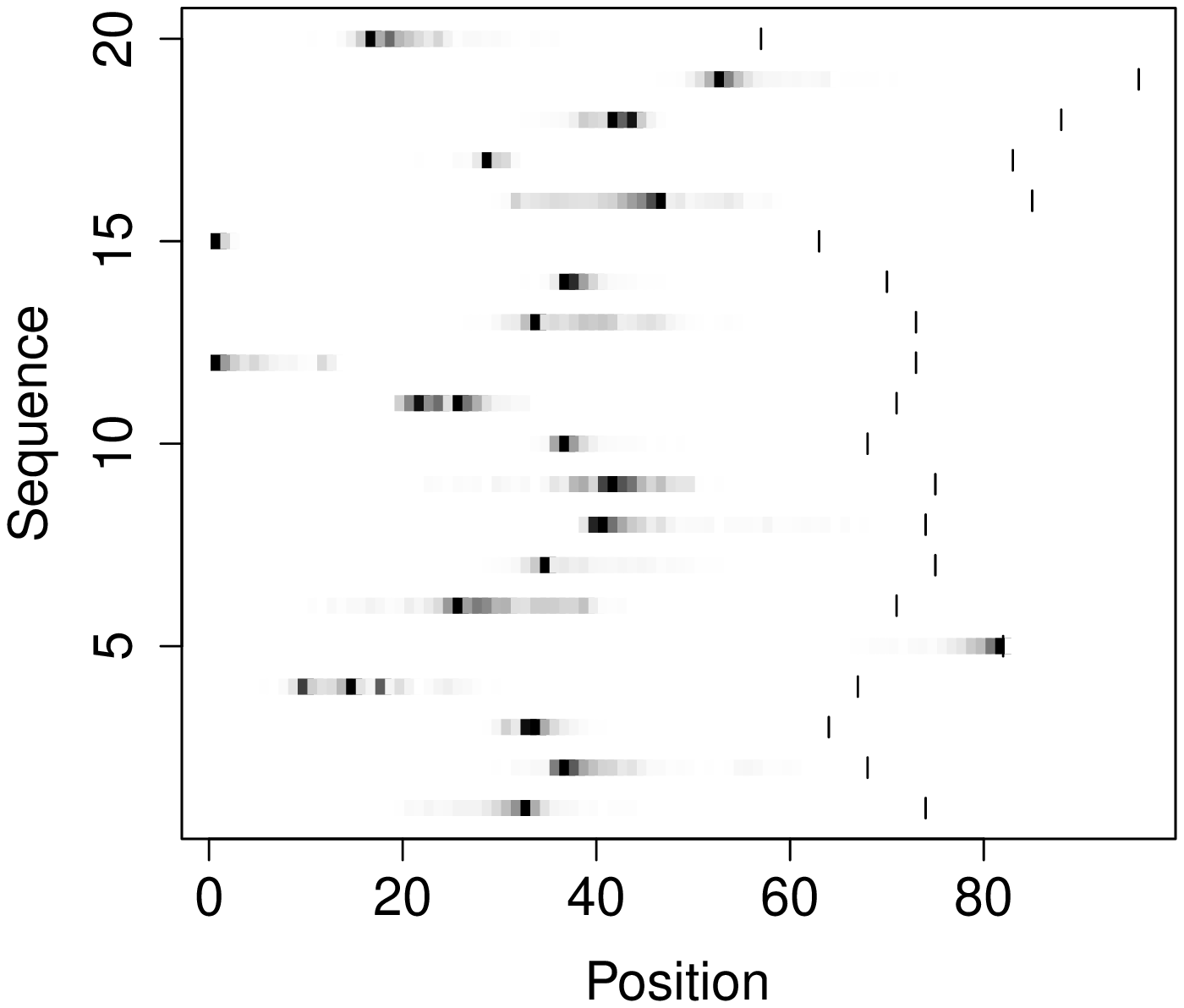}}
        \caption{The estimated marginal probability distribution of (a) the classification of each node to segment 2 (probability 1 is black) and (b) a particular node being a changepoint, where the gray scale has been adjusted for each sequence so that the node with the maximum probability is black and the node with the minimum probability is white. The end position of each observed sequence is marked with a short vertical black line.}
     \label{fig:segmentsAndChangepoints_apple_individual}
\end{center}
\end{figure}
From these figures we see that the individual sequence approach
approximately
%, and for the most of the sequences, 
finds the same changepoints as the joint model approach,
although with somewhat higher uncertainty. Note that for some of the
sequences, for instance sequences 5 and 12 in Figure
\ref{fig:segmentsAndChangepoints_apple_individual}, the individual
sequence approach converges into a state with a zero length segment, which is somewhat
surprising as there seems to be no significant difference in the
structure of these two sequences compared to the others (see Figure 4 in the paper). This result is not present in
our joint approach due to the sharing of information between the
sequences. Also, because this is categorical data with five states the
uncertainty in the estimates of the transition probabilities
% very high
for the
individual sequence approach becomes
quite high due to lack of data, compared to our joint approach.

\subsection{\sc Real Data Analysis: Monsoon Rainfall}
%\clearpage
We also applied the individual sequence approach to the rainfall data
presented in Section 4.3. %\ref{sec:rain} 
in the paper. We assumed for this
experiment that $K=2$ and the results are shown in Figure \ref{fig:rainfall_individual}.
\begin{figure}
\begin{center}
\vspace{-1cm}
\subfigure[]{\label{fig:rainfall_segMarginals}\includegraphics[scale=0.31]{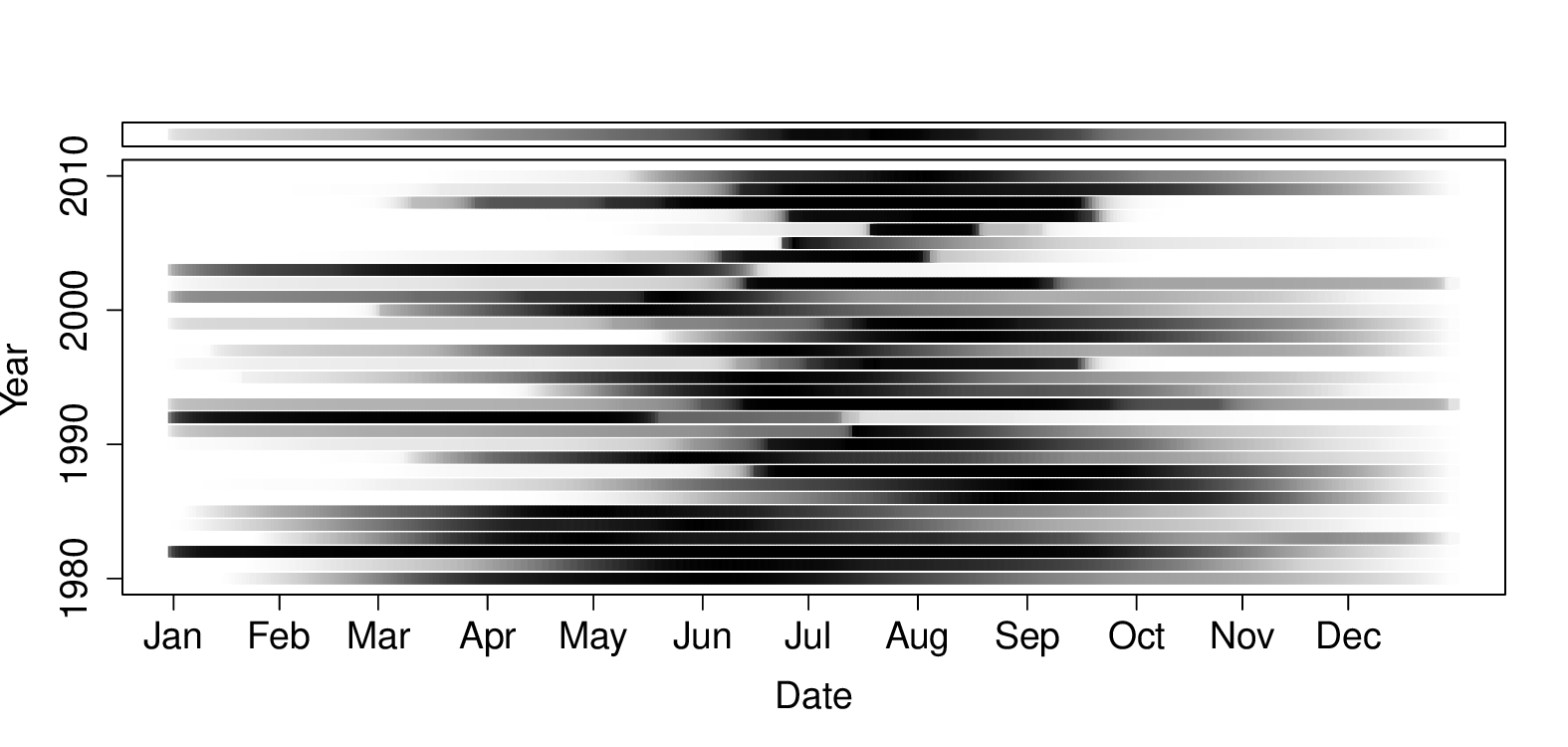}}% \vspace{-1.2cm}
\subfigure[]{\label{fig:rainfall_changeMarginals}\includegraphics[scale=0.31]{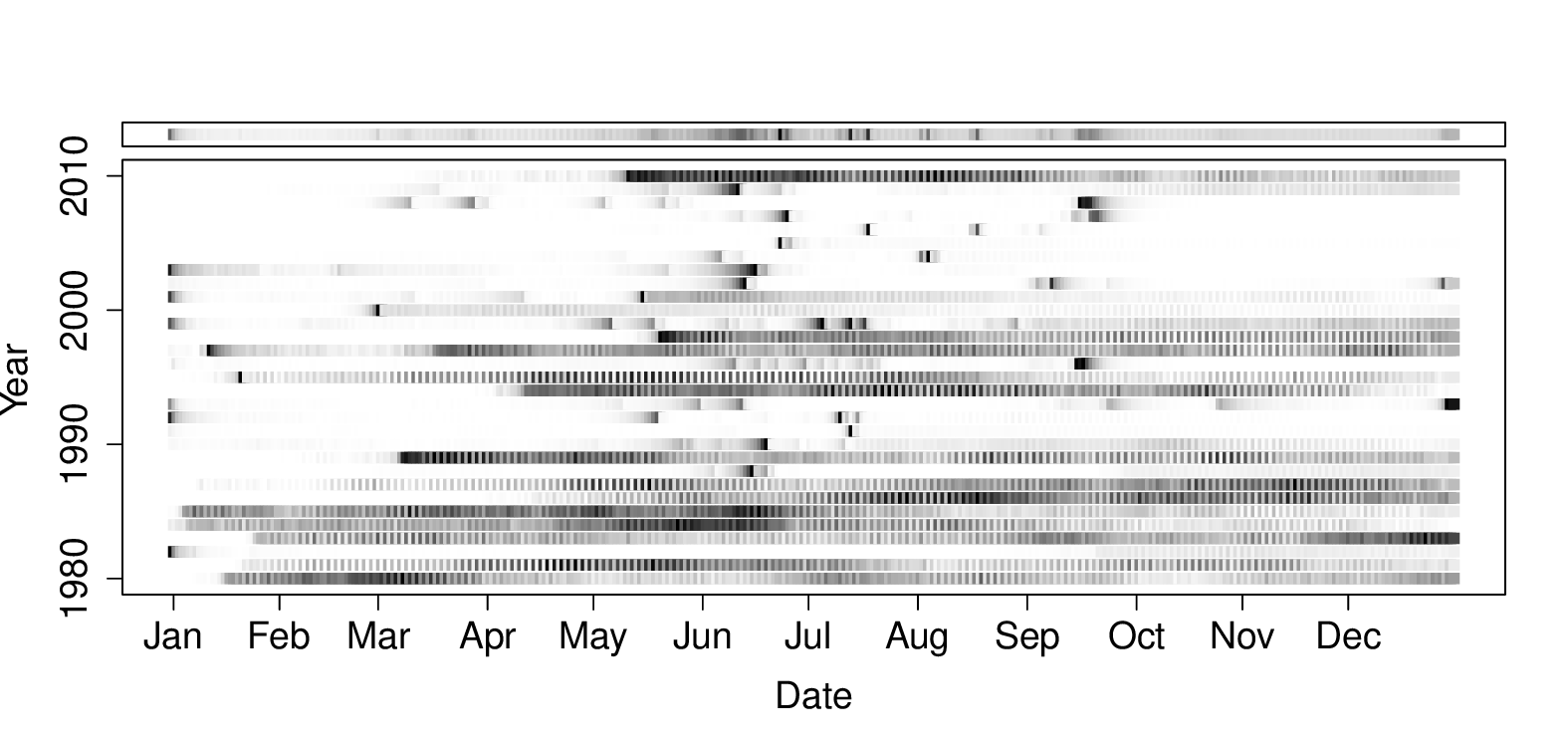}}
        \caption{The estimated marginal probability distribution for
          (a) the classification of each day to the monsoon season
          (probability 1 is black) and (b) a particular day being a
          changepoint, where the gray scale has been adjusted for each year so that the day with the maximum probability is black and the day with the minimum probability is white.}
     \label{fig:rainfall_individual}
\end{center}
\end{figure}
This figure clearly shows very high uncertainty regarding the position of
the changepoints, and in most of these cases these changepoints are completely unrelated to the actual  onset and offset of the monsoon season. This figure should be
compared to the results using our joint approach shown in Figure 9 %\ref{fig:rainfall} 
in the main paper.

\section{ADDITIONAL SYNTHETIC AND REAL DATA ANALYSIS EXAMPLES}

%\section{POSTERIOR DISTRIBUTION OF THE SYNTHETIC DATA ANALYSIS}
%\label{supp:1}

%Here we discuss the posterior distributions for several of the parameters of the model and also the sensitivity of the model to some of these parameters. 

%\subsection*{\sc Scenario 1}

\subsection{\sc Synthetic Data Example: Scenario 2}
%\section{Scenario 2: sequences with different number of changepoints}\label{sec52}
In this scenario, we generate binary data from our model for $L=30$
sequences with length $T_l=200$. The first 20 sequences are generated
with two changepoints and the last 10 sequences are generated with only one
changepoint. The 10 sequences with one changepoint are generated using the same parameters as described in Section 4.1 in the paper. The 20 sequences with two changepoints have similar first and last segments (as in Section 4.1) plus an additional middle segment where $(r_2,b_2)=(0.75,0.8)$ and $q_{1,2}=q_{2,2}=0.2$. 

\begin{figure}
\vspace{-1cm}
\begin{center}
\subfigure[]{ \label{fig:nrOfK_mixedData}\includegraphics[scale=0.4]{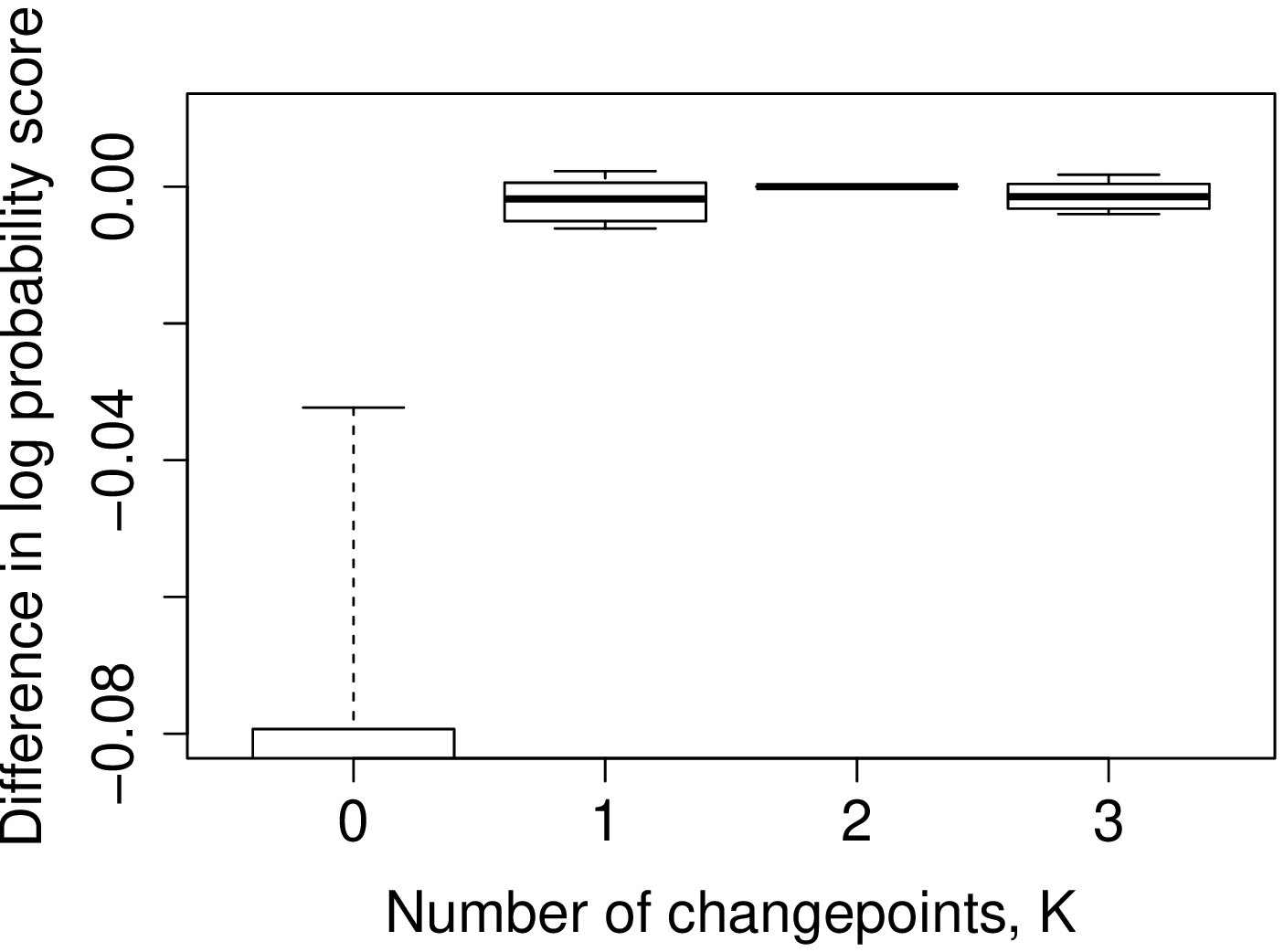}}
\subfigure[]{   \label{fig:baselines_mixedData}\includegraphics[scale=0.4]{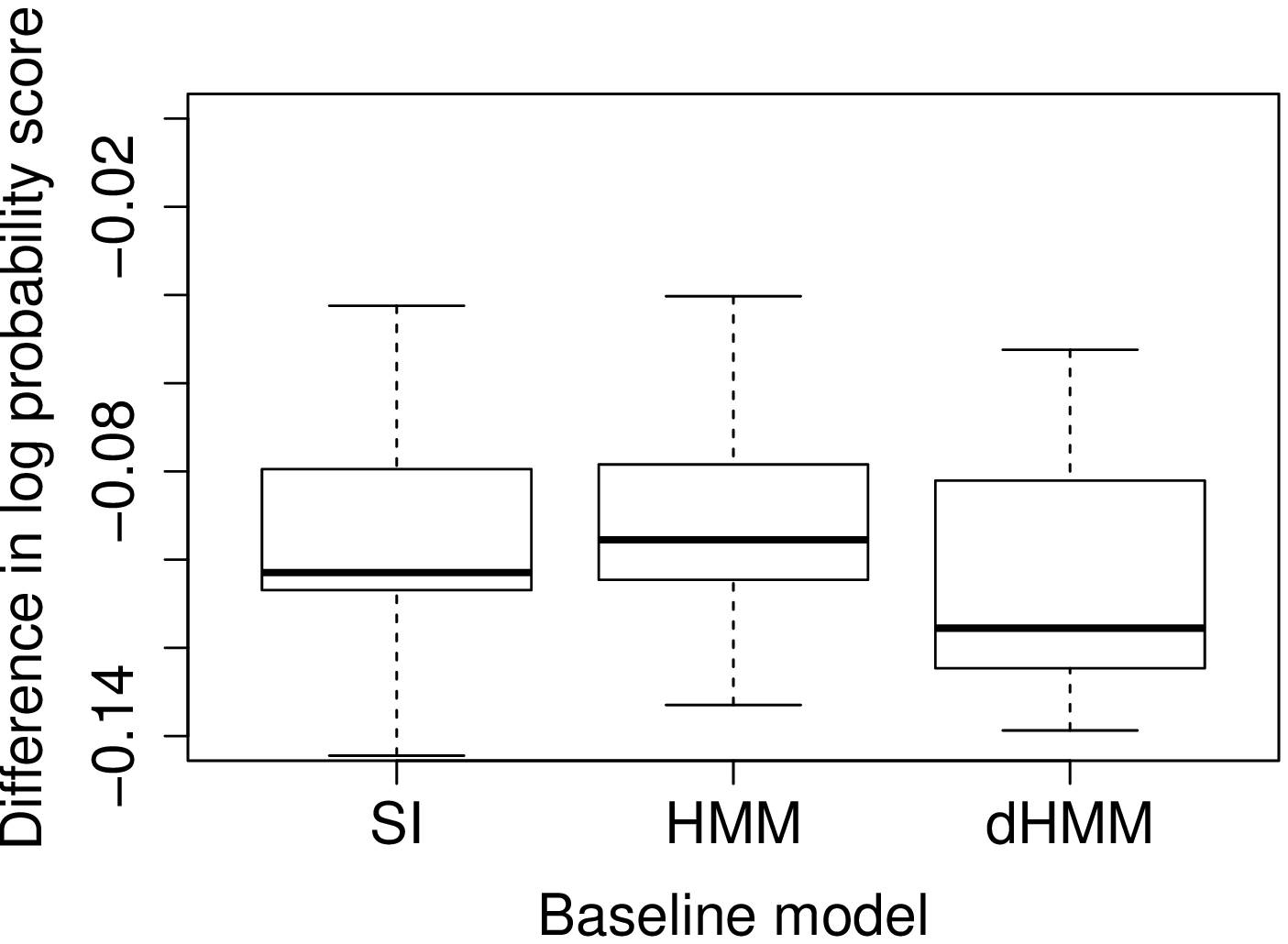}}
   \caption{Each boxplot shows the log-probability scores, across the
     validation sets, for a different model. The y-axis is defined as
     the log-probability score for (a) other numbers of changepoints
     and (b) other baseline models, minus the log-probability score
     for the model with two changepoints.}
	\label{fig:boxPlot_toy}
\end{center}
\end{figure}
We used 10-fold cross validation  to find the optimal number of
changepoints. In this scenario $K=2$ effectively represents the true
maximal number of changepoints in any sequence, where sequences with fewer
changepoints can skip the ``extra'' segments. Thus, the interpretation
of a single ``overall best $K$" has less meaning in this
context. Nonetheless, using $K=2$ changepoints as the reference,
Figure S.\ref{fig:nrOfK_mixedData} shows that two changepoints ($K=2$) are preferred using the cross-validation score, which is consistent with the fact that 20 out of the 30 sequences were generated with two changepoints.  Figure S.\ref{fig:baselines_mixedData} shows that our model outperforms the other baseline models, all fit with $K=2$.

\begin{figure} 
\vspace{-1cm}
\begin{center}
\subfigure[]{ \label{fig:seg1_mixedData}\includegraphics[scale=0.4]{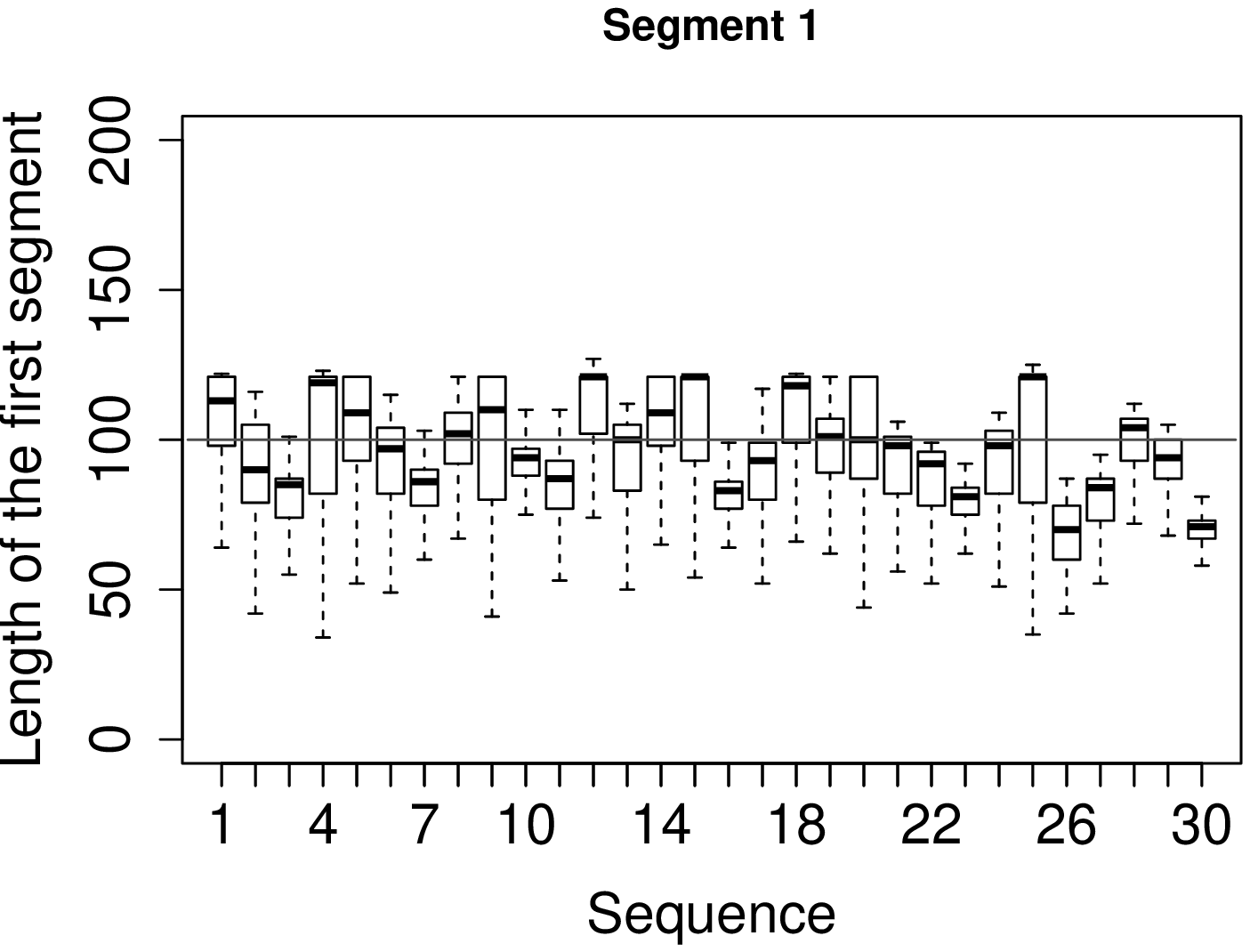}}
\subfigure[]{   \label{fig:seg2_mixedData}\includegraphics[scale=0.4]{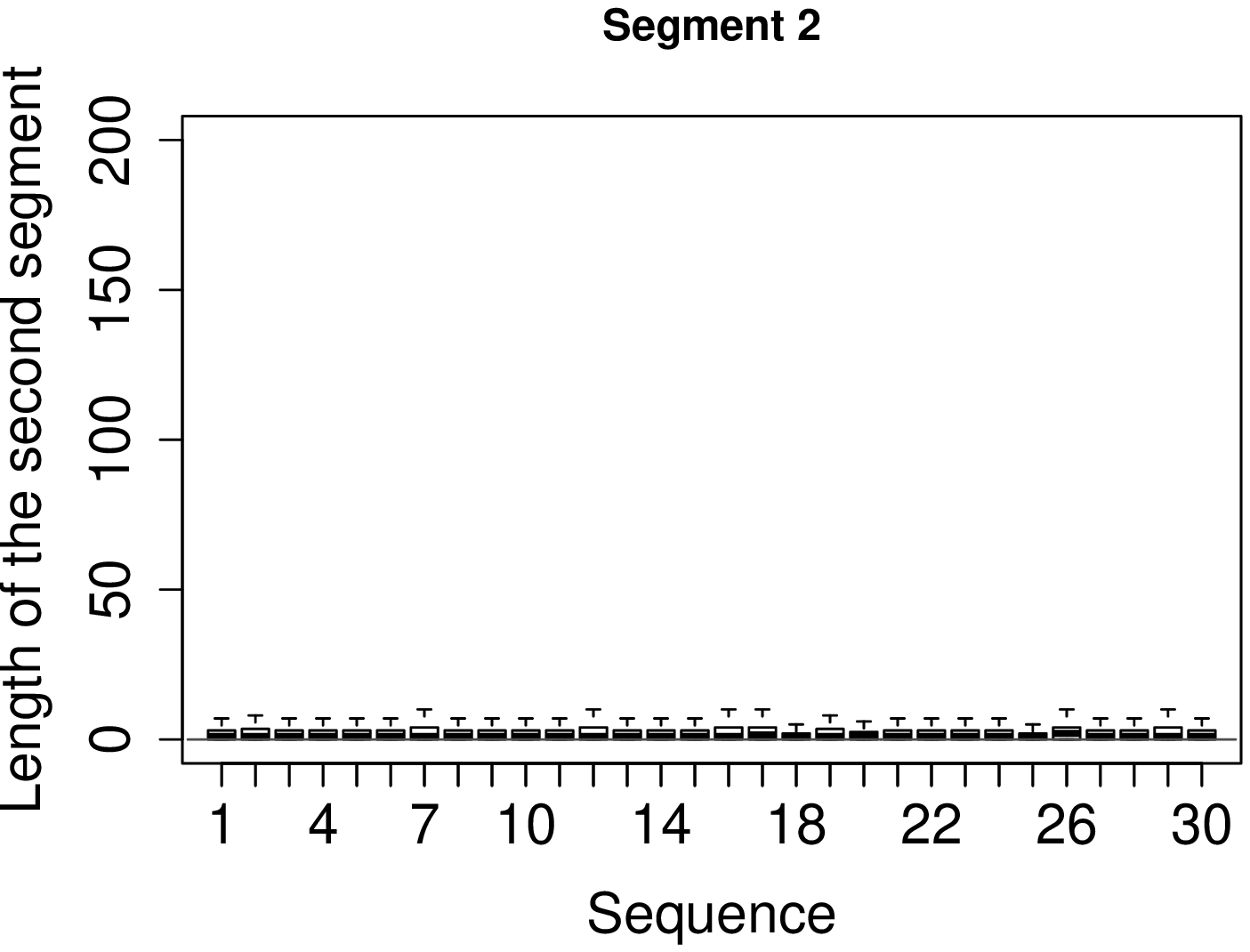}}\\\vspace{-0.8cm}
\subfigure[]{ \label{fig:seg3_mixedData}\includegraphics[scale=0.4]{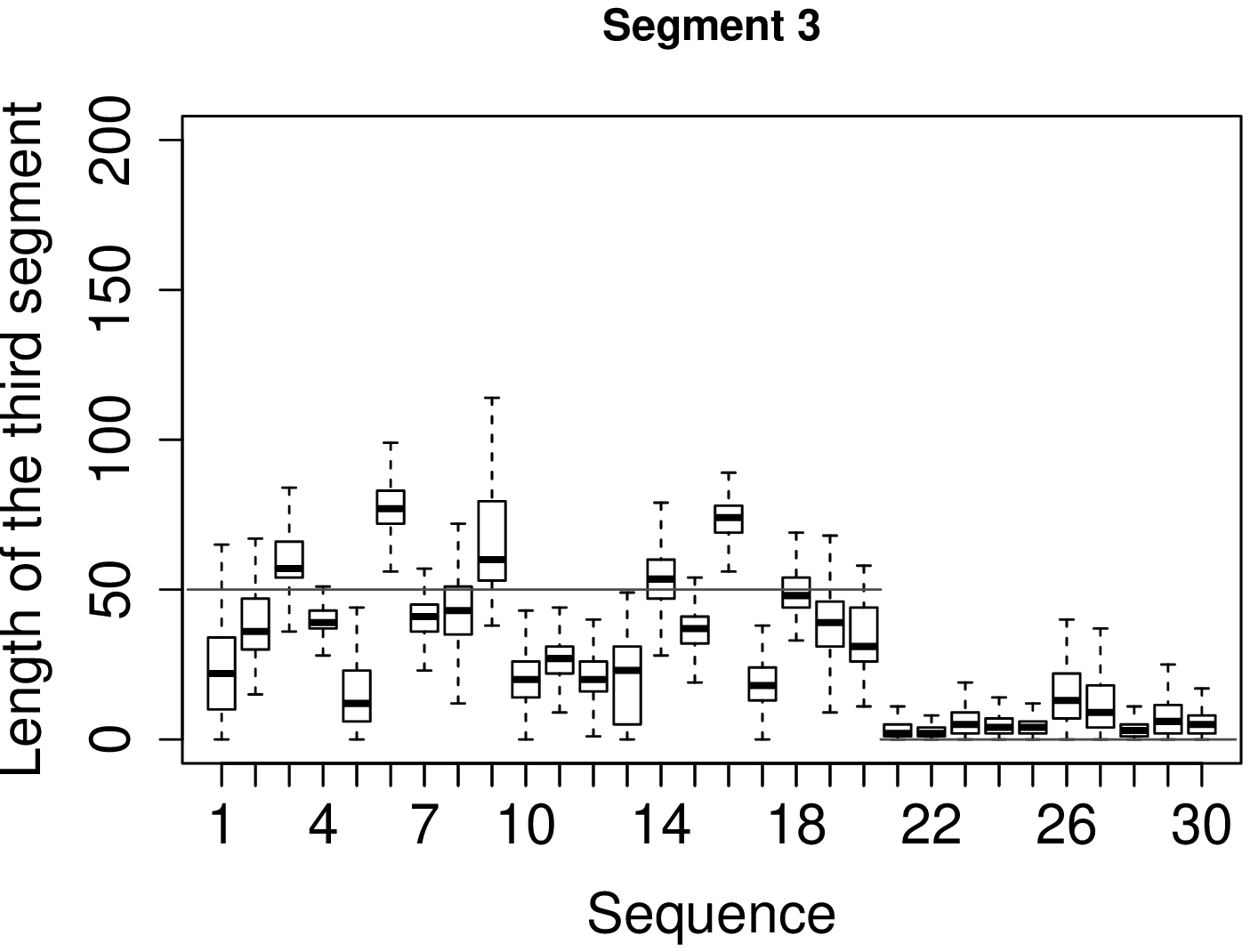}}
\subfigure[]{   \label{fig:seg4_mixedData}\includegraphics[scale=0.4]{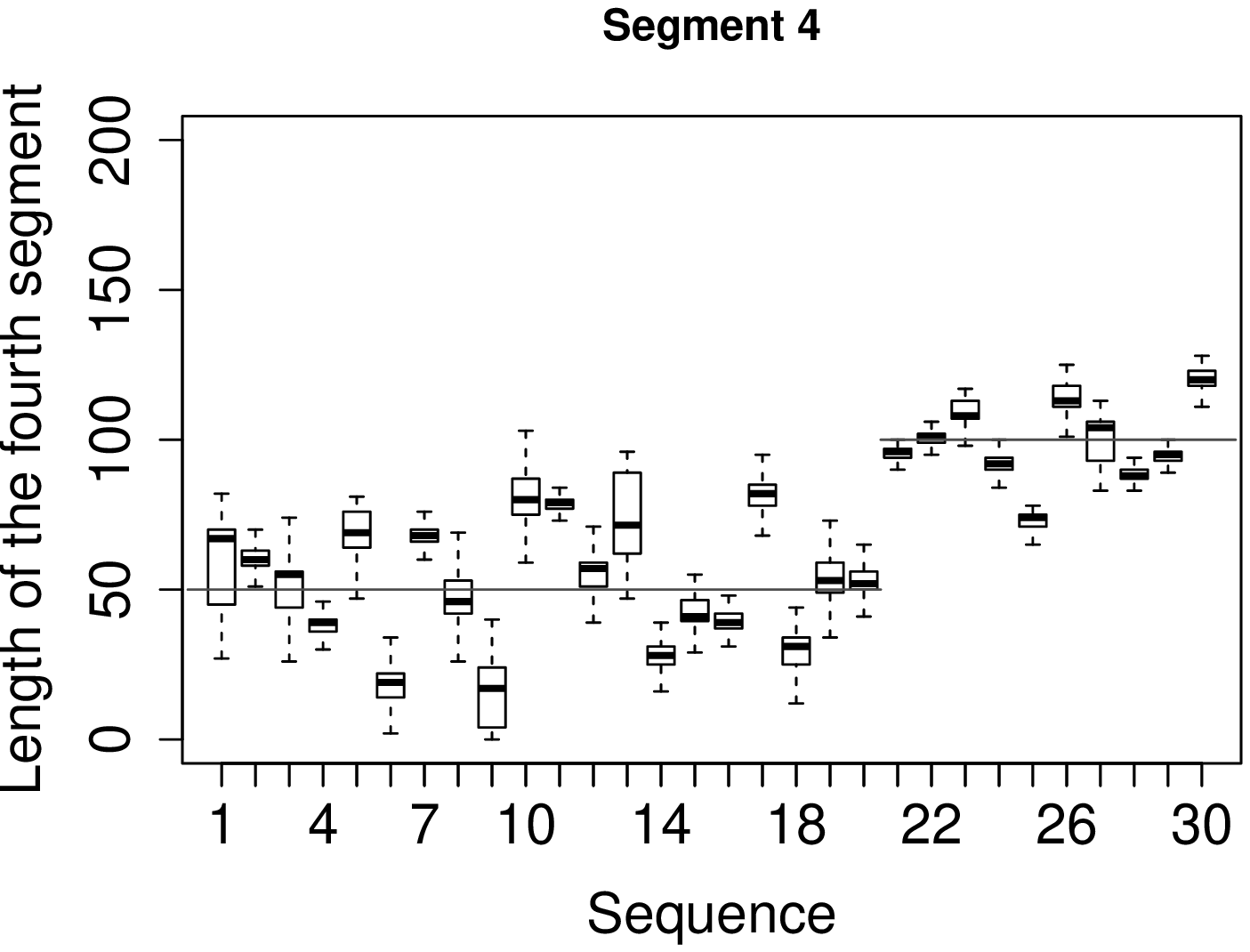}}
   \caption{Distribution of the length of each segments in each
     sequence in the $K=3$ solution. a)-d) are segments 1 to 4
     respectively. Each sequence has a box plot for the posterior
     estimate of the length of its segment. For simplicity we show the
     expected length of each segment as a solid gray line, noting that
     for each simulated sequence the actual segment length will vary
     around this expected length.}
		\label{fig:segments_mixedData10}
\end{center}
\end{figure}

To illustrate how our model skips segments, we examine in detail the solutions found by the model when it is fit to all 30 sequences and allowed to use $K=3$ changepoints, even though the data was generated with only $K=1$ or $K=2$ changepoints. 
Figure \ref{fig:segments_mixedData10} shows the results of this experiment, with each of the four panes corresponding to one of the four possible segments that the model could infer. The  estimated
posterior distribution of the length of each segment is shown in the
form of a box plot (the $y$-axis) for each of the 30 sequences (the
$x$-axis). Recall that sequences 1 to 20 were generated with $K=2$ changepoints
and sequences 21 to 30 with $K=1$ changepoint. The expected length of each segment given the true changepoint model is shown as a gray line. From the figure we see that the second segment (Figure \ref{fig:segments_mixedData10}(b)) is skipped for all sequences, i.e., the model finds no evidence for an additional segment. Similarly, in Figure \ref{fig:segments_mixedData10}(c) we see that sequences 21 to 30 infer very short or zero length segments for the third segment, and as a consequence the fourth segment is much longer in these sequences (Figure \ref{fig:segments_mixedData10}(d)).
Thus, even with misspecification of the value of $K$, and  when
sequences have variable numbers of changepoints, our model appears to
perform well. 

%The posterior distributions of other parameters of the
%model are discussed in the supplementary materials Section \ref{supp:1}. 

For 
%the $K=3$ solution, we discuss 
the posterior distributions of the
parameters, 
 we see some summary statistics 
%of those distributions 
in Table \ref{tab:est_mixedData}. 
\def\arraystretch{0.5}
\begin{table}
\centering
\begin{tabular}{ c|  c c } 
&Est.& 95 \% CI\\
\hline
  $r_1$ & 0.382 & [0.225,0.512] \\
  $b_1$ & 0.618 & [0.242,0.832] \\
  $r_2$ & 0.412 & [0.245,0.538] \\
  $b_2$ & 0.858 & [0.510,0.901] \\
 $r_3$ & 0.732 & [0.556,0.904] \\
  $b_3$ & 0.004 & [0.001,0.012] \\ 
  $q_{1,1}$& 0.779 & [0.768,0.804] \\
 $q_{2,1}$& 0.803 & [0.786,0.821] \\
 $q_{1,2}$& 0.722 & [0.365,0.846] \\
 $q_{2,2}$& 0.708 & [0.322,0.834] \\
 $q_{1,3}$& 0.216 & [0.194,0.239] \\
 $q_{2,3}$& 0.184 & [0.156,0.209]\\
 $q_{1,4}$& 0.496 & [0.443,0.546]\\ 
 $q_{2,4}$& 0.369 & [0.312,0.422] 
\end{tabular}
\vspace{0.2cm}
  \caption{Parameter estimate and 95\% CI.}
\label{tab:est_mixedData}
\end{table}
For these statistics we find many similarities with the true parameters of the data. For instance, the 95\% CIs of $r_1$ and $b_1$   include the true parameter of the first changepoint for both types of sequences. Next, $r_2$ and $b_2$ correspond to the changepoint leading up to the first (often) zero length segment, so it is reasonable that $r_2$ is close to $r_1$. Next, we see a high uncertainty in the position of the third changepoint, $b_3=0.004$, which makes sense because this is the changepoint leading up to the third changepoint, which is approximately of length 50 for the 20 first sequences and (often) zero for the last 10 sequences. For the transition matrix parameters we see high uncertainty for the estimates of $q_{1,2}$ and $q_{2,2}$ which makes sense because this segment is (often) zero for all the sequences. All other transition matrix parameters are well recovered. 
 
We also investigated both the $K=2$ and the $K=1$ solution for this
data set. For the $K=2$ case, one segment is skipped for the last 10
sequences, as expected. For the $K=1$ solution the position of the
single changepoint in the last 10 sequences gets well estimated, while
in the first 20 sequences the position of the first changepoint is the
one that is recovered accurately and the model merges the second and
third segments. 
%{Broadly similar results were obtained in an experiment with a mix of 20 $K=1$ sequences and 10 $K=2$ sequences (see Section S.3 in the supplementary materials).}

%\subsection*{\sc Scenario 2}

\subsection{\sc Synthetic Data Example: Scenario 3}

%\section{SYNTHETIC DATA EXAMPLE: SCENARIO 3}

In this section we present an experiment similar to what is done in the previous section (Section S.3.1),
% Section
%\ref{sec52} in the paper where 
however we switch the number of sequences
with $K=1$ changepoints with the number of sequences with $K=2$
changepoints. 
%In particula, our data consists of 20 sequences simulated from the model in Section 4.1 and 10
%sequences from the model with $K=2$ changepoints given in Section
%\ref{sec52}. 
In Figure \ref{fig:boxPlot_mixedData_v2} we see the
result from our cross
validation procedure for different numbers of changepoints, and with comparisons
to the baseline models. We observe $K=2$ to be the
superior choice. This is slightly surprising considering that the example in
Section S.3.1 gave a less significant difference between the
$K=1$ and $K=2$ solution even though in that case there were more
sequences with $K=2$ changepoints. We believe that this is likely to result from
differences between these two random data sets.   
\begin{figure}
\vspace{-1cm}
\begin{center}
\subfigure[]{ \label{fig:nrOfK_mixedData_v2}\includegraphics[scale=0.4]{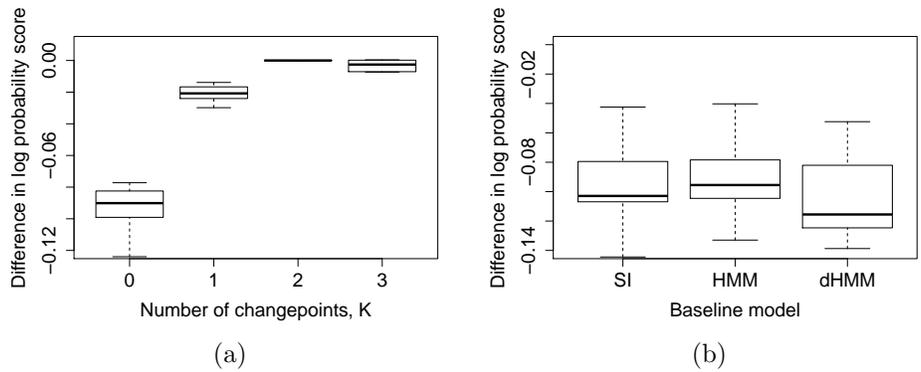}}
\subfigure[]{   \label{fig:baselines_mixedData_v2}\includegraphics[scale=0.4]{baselines_mixedData_rev.ps}}
   \caption{Each boxplot shows the log-probability scores, across the
     validation sets, for a different model. The y-axis is defined as
     the log-probability score for (a) other numbers of changepoints
     and (b) other baseline models, minus the log-probability score
     for the model with two changepoints.}
	\label{fig:boxPlot_mixedData_v2}
\end{center}
\end{figure}

In Figure \ref{fig:segments_mixedData10_v2} we see the
posterior estimation of the length of each segments when we assume a
model with $K=3$ changepoints, again a similar procedure as in Section S.3.1. %\ref{sec52}. 
\begin{figure} 
\vspace{-1cm}
\begin{center}
\subfigure[]{ \label{fig:seg1_mixedData_v2}\includegraphics[scale=0.4]{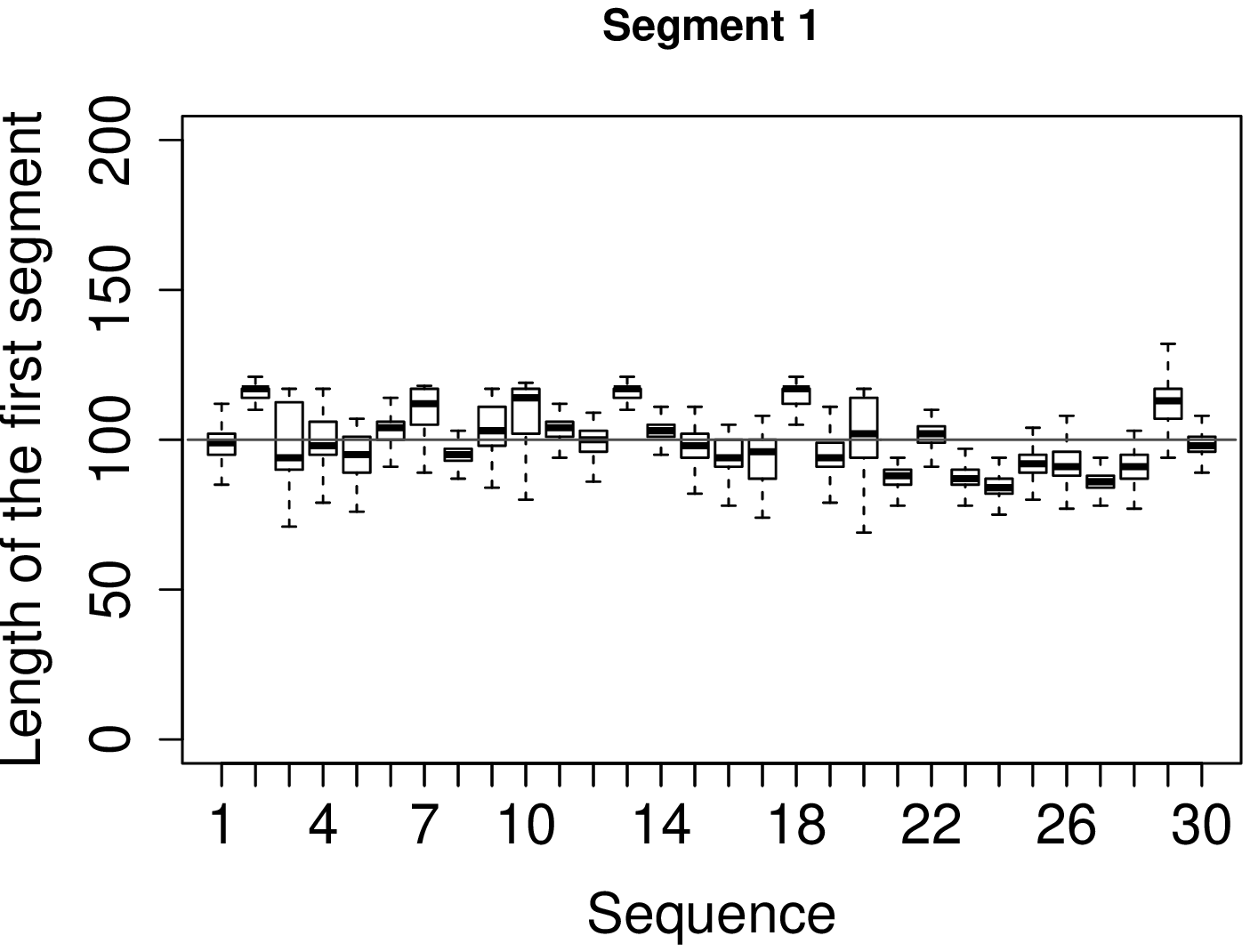}}
\subfigure[]{   \label{fig:seg2_mixedData_v2}\includegraphics[scale=0.4]{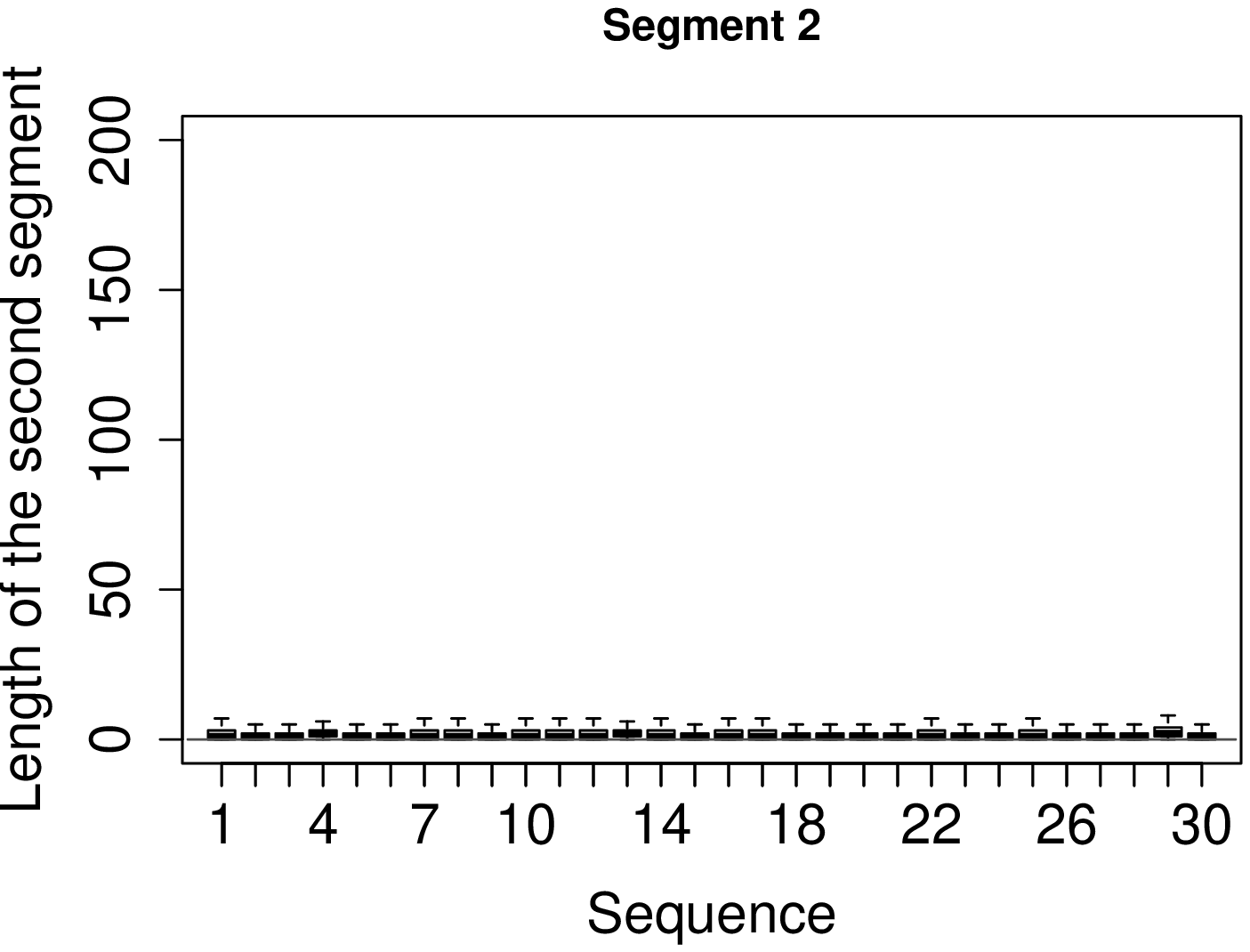}}\\\vspace{-0.8cm}
\subfigure[]{ \label{fig:seg3_mixedData_v2}\includegraphics[scale=0.4]{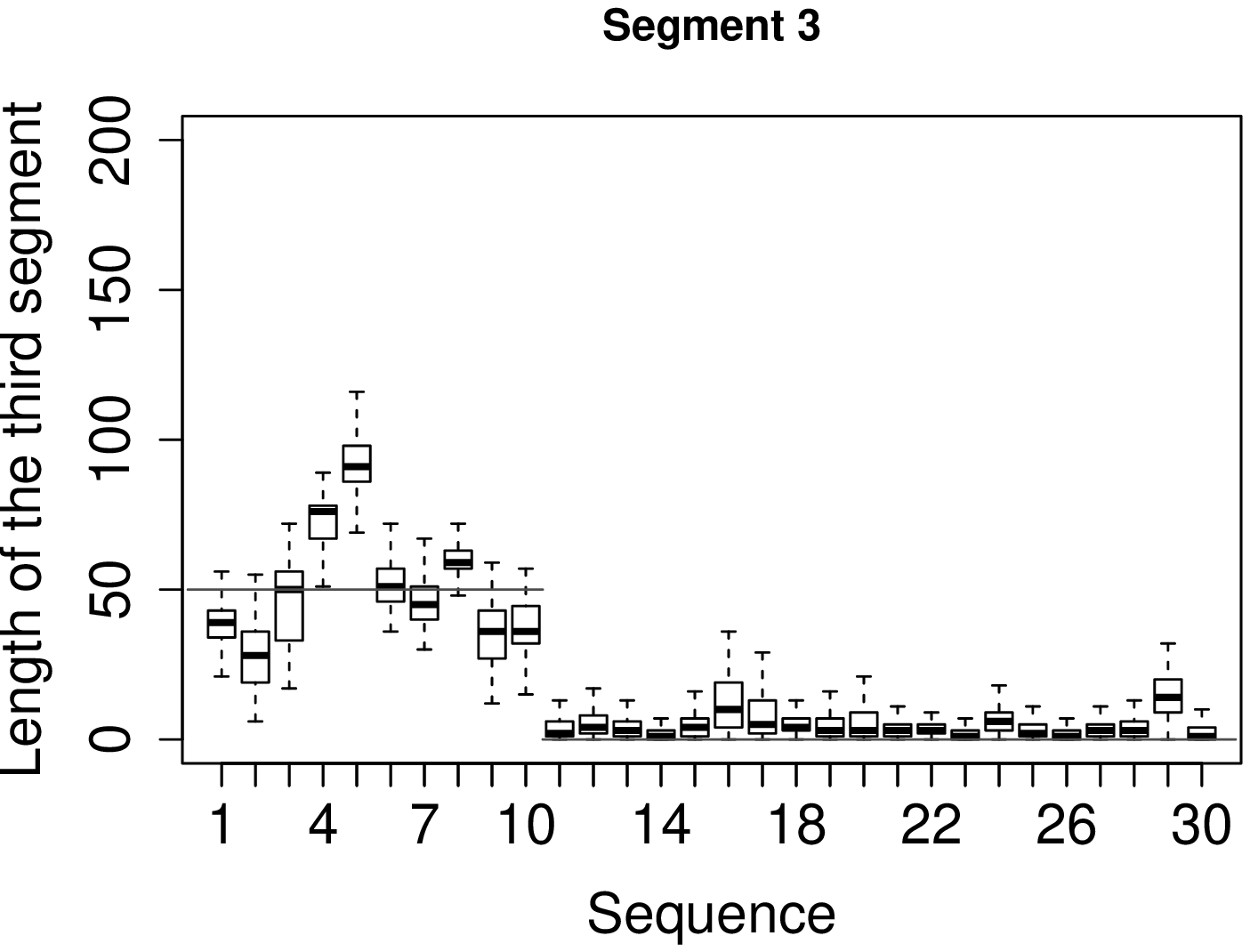}}
\subfigure[]{   \label{fig:seg4_mixedData_v2}\includegraphics[scale=0.4]{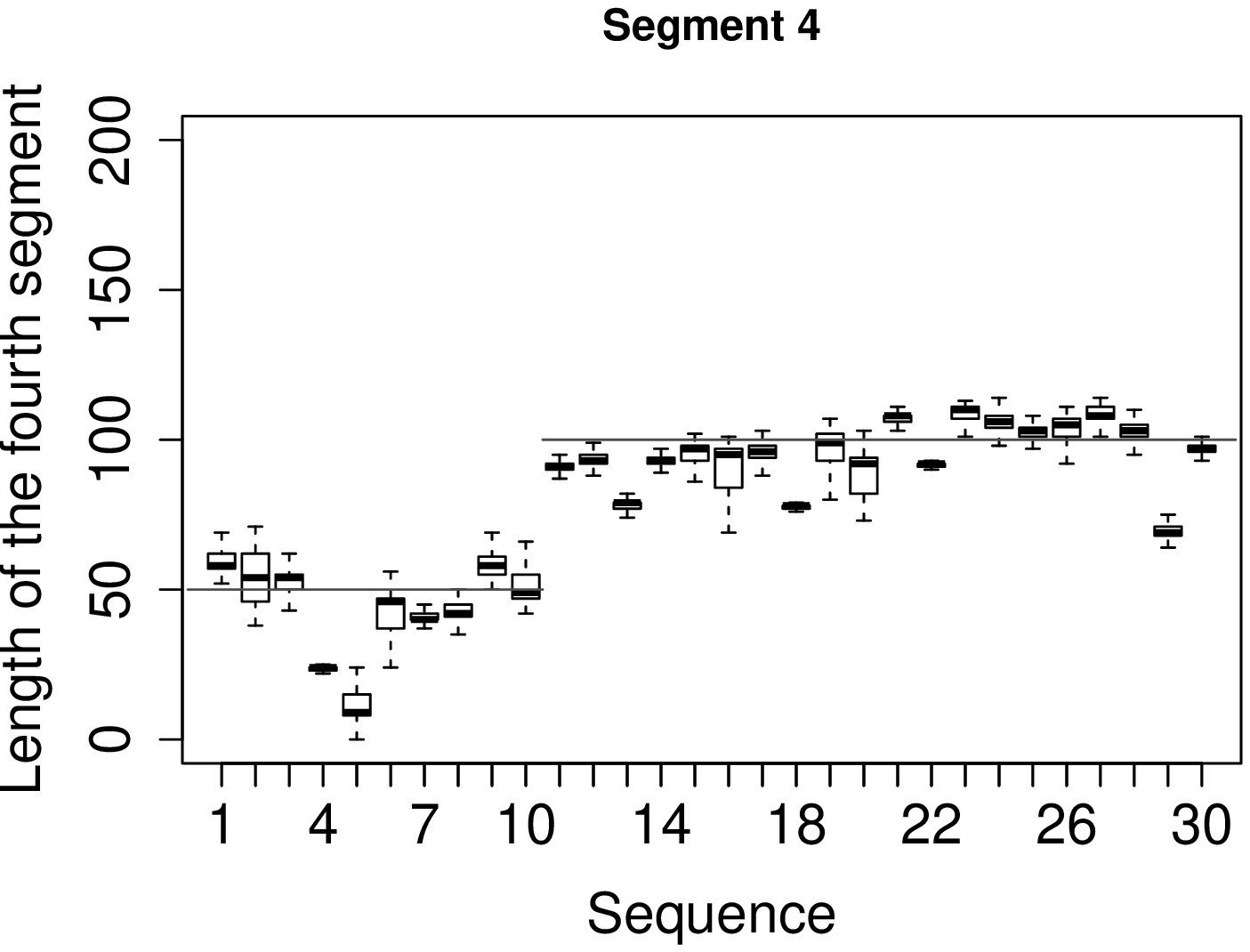}}
   \caption{Distribution of the length of each segments in each
     sequence in the $K=3$ solution. a)-d) are segments 1 to 4
     respectively. Each sequence has a box plot for the posterior
     estimate of the length of its segment. For simplicity we show the
     expected length of each segment as a solid gray line, noting that
     for each simulated sequence the actual segment length will vary
     around this expected length.}
		\label{fig:segments_mixedData10_v2}
\end{center}
\end{figure}
Again the zero length segments can be observed.

\subsection{\sc Synthetic Data Example: Scenario 4 and 5}

%\section{SYNTHETIC DATA EXAMPLE: SCENARIO 4 and 5}
\label{supp:2}
In this section we briefly present an experiment that illustrates the
performance of our procedure when the number of possible categories for $y_j$
increases, i.e. the dimension of $Q^{(i)}$ increases. In particular we
keep the changepoints simulated for the 10 sequences in Section
4.1 %\ref{sec51} 
in the paper, but simulate instead $\bf{y}$, for
scenario 4, by using the following transition matrices
\[\bm Q^{(1)}= \left[ \begin{array}{ccc}
0.8 & 0.1& 0.1 \\
0.1 & 0.8 & 0.1 \\
0.1 & 0.1& 0.8 \end{array} \right],
\bm Q^{(2)}= \left[ \begin{array}{ccc}
0.333 & 0.333& 0.333 \\
0.350 & 0.300& 0.350 \\
0.375 & 0.375& 0.250 \end{array} \right],\]
and for scenario 5 the matrices 
\[\bm Q^{(1)}= \left[ \begin{array}{cccc}
0.800 & 0.033& 0.033 &0.033\\
0.033 & 0.800 & 0.033&0.033 \\
0.033 & 0.033& 0.800 &0.033\\
0.033 & 0.033& 0.033 &0.800\end{array} \right],
\bm Q^{(2)}= \left[ \begin{array}{cccc}
0.250 & 0.250& 0.250&0.250 \\
0.267 & 0.200& 0.267 &0.267\\
0.283 & 0.283& 0.150 &0.283\\
0.300 & 0.300& 0.300 &0.100\end{array} \right],\]
which are chosen to somewhat imitate the transition matrices in
Section 4.1 %\ref{sec51} 
but now with 3 and 4 categories, respectively. Figure
\ref{fig:boxPlot_catData} shows the result when running our cross-validation procedure on these two data sets. In both of these two
multi-category examples we are able to detect the correct number of
changepoints.   
\begin{figure}
\vspace{-1cm}
\begin{center}
\subfigure[]{ \label{fig:nrOfK_cat3Data}\includegraphics[scale=0.4]{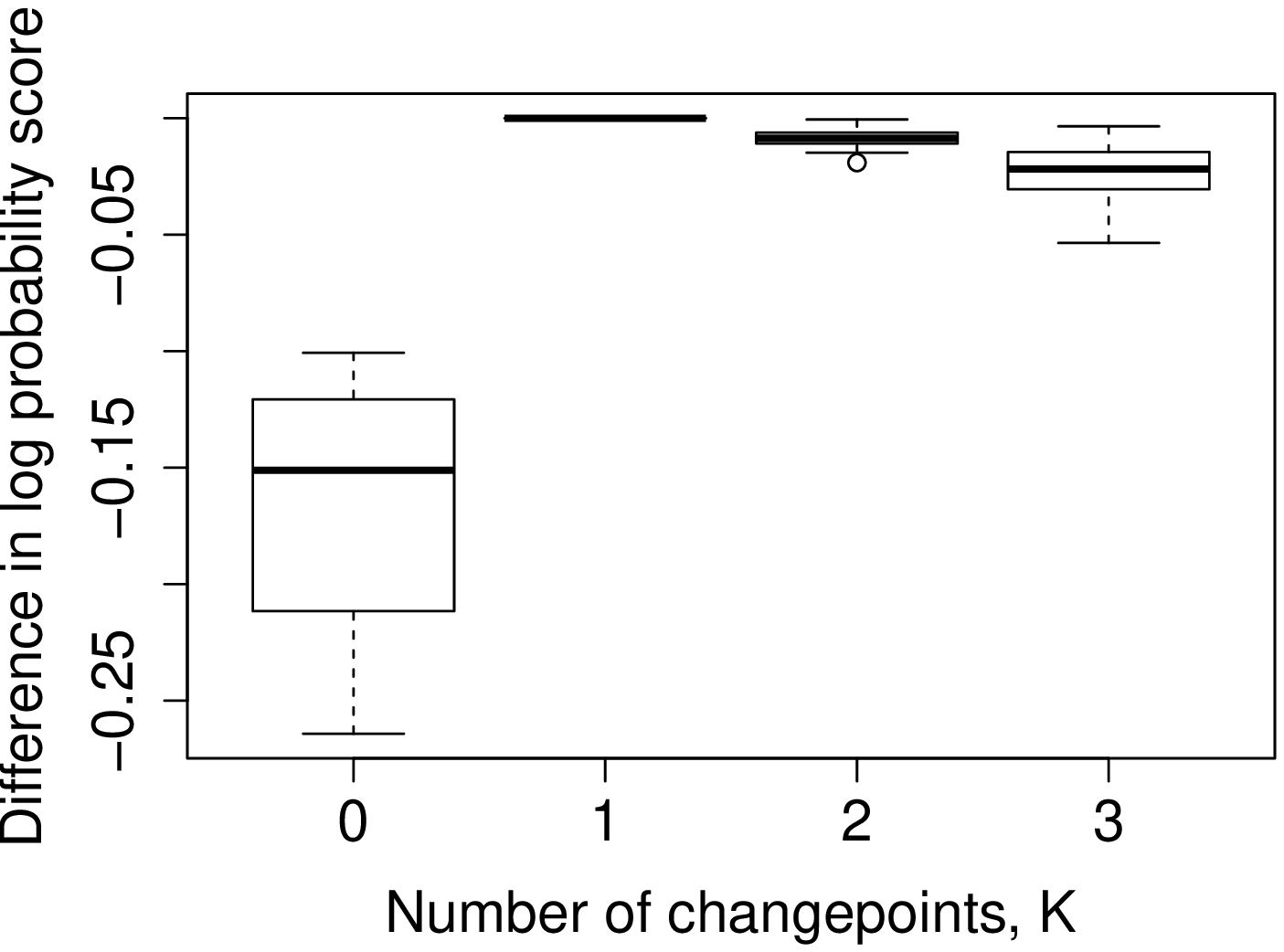}}
\subfigure[]{   \label{fig:nrOfK_cat4Data}\includegraphics[scale=0.4]{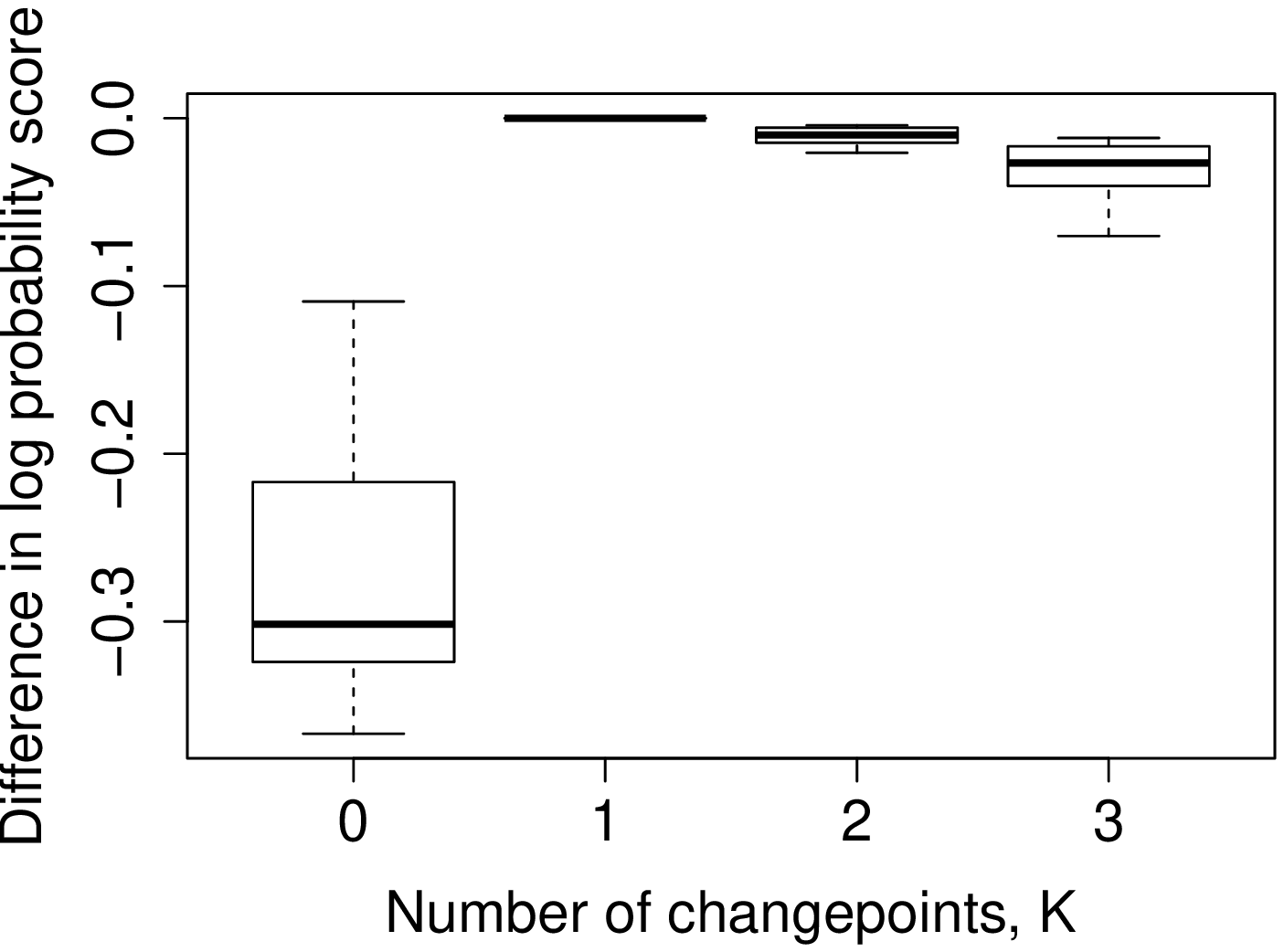}}
   \caption{Each boxplot shows the log-probability scores, across the
     validation sets, for a different model. The y-axis is defined as
     the log-probability score for (a) other numbers of changepoints
     for scenario 4
     and (b) other numbers of changepoints
     for scenario 5.}
	\label{fig:boxPlot_catData}
\end{center}
\end{figure}
We next investigate the distribution of the
position of the changepoints under these two scenarios, see Figure \ref{fig:segmentsAndChangepoints_catData}.
\begin{figure}
\vspace{-1cm}
\begin{center}
\subfigure[]{\label{fig:cat3Data_segments}\includegraphics[scale=0.45]{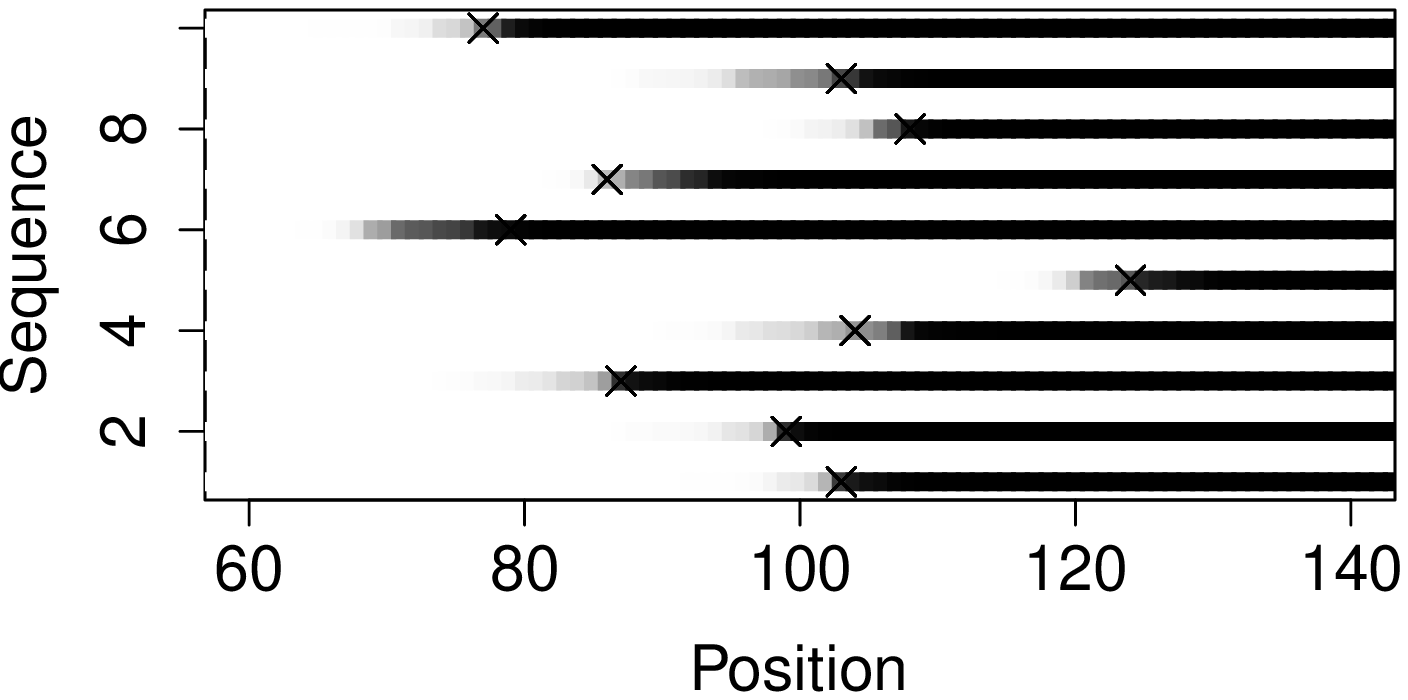}}
\subfigure[]{\label{fig:cat3Data_changepoints}\includegraphics[scale=0.45]{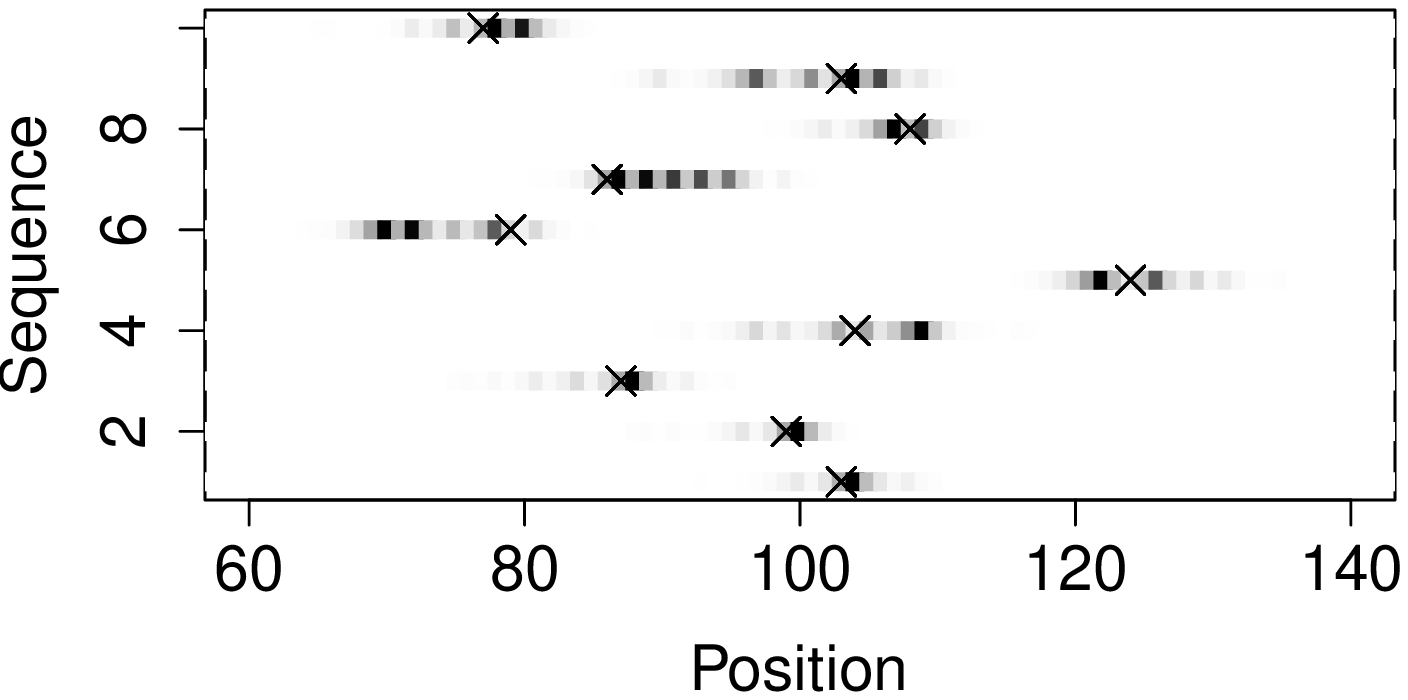}}\\
\subfigure[]{\label{fig:cat4Data_segments}\includegraphics[scale=0.45]{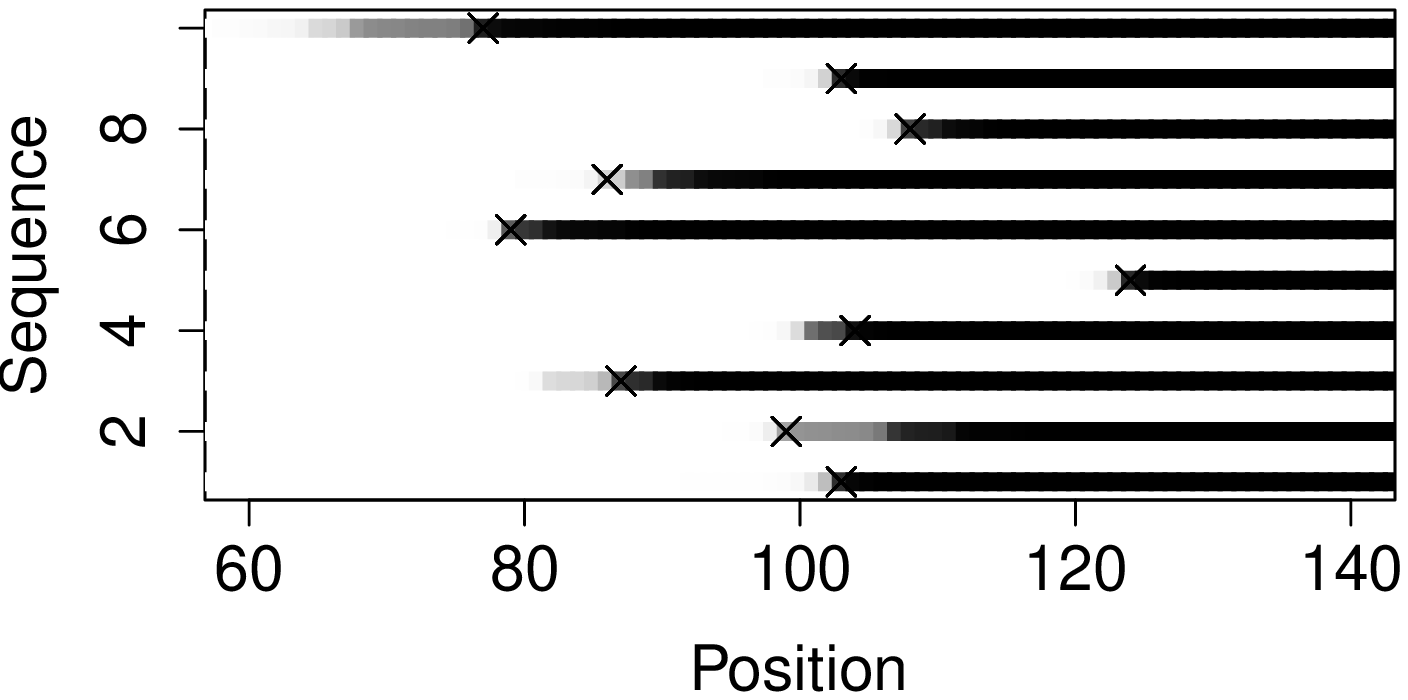}}
\subfigure[]{\label{fig:cat4Data_changepoints}\includegraphics[scale=0.45]{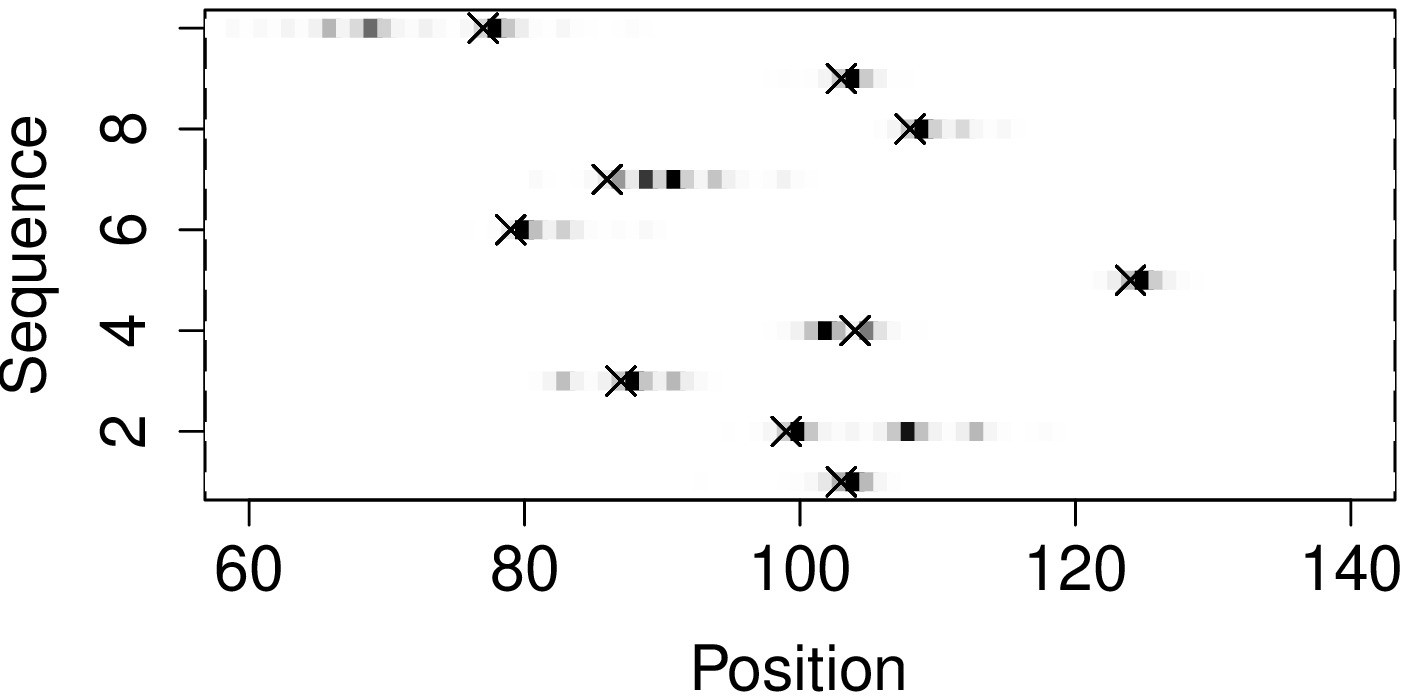}}
        \caption{The estimated marginal probability distribution of
          (a) the classification of each node to segment 2
          (probability 1 is black) for scenario 4 and (b) a particular
          node being a changepoint for scenario 4, where the gray
          scale has been adjusted for each sequence so that the node
          with the maximum probability is black and the node with the
          minimum probability is white. In (c) and (d) similar results
        are shown for scenario 5 respectively.}
     \label{fig:segmentsAndChangepoints_catData}
\end{center}
\end{figure}
Comparing these results to Figure 3 %\ref{fig:ch_toy} 
in the paper, we see that the
uncertainty regarding the position of the changepoints increases due to less data,
although the results are still relatively accurate. Similar results can be
seen in the estimates for the model parameters $\bm{Q}^{(1)}$, $\bm{Q}^{(2)}$, $r_1$ and $b_1$ (not shown).

%\section{COMPARING DATA ANALYSIS TO SINGLE SEQUENCE METHOD}\label{supp:4}
%
%In this section we provide details of the comparison of our joint sequence
%model  to the single sequence model described in the
%paper, see Section \ref{sec51}. 

%\subsection*{\sc Synthetic data: Scenario 1}

%\bibliographystyle{jasa}
%\bibliography{bibl_sup}	% For bibtex

%\clearpage

%\appendix
%\renewcommand{\thesubsection}{\Alph{}}

\end{document}